\documentclass[twocolumn]{aastex63}
\usepackage[utf8]{inputenc}
\usepackage{hyperref}
\usepackage{float}
\usepackage{verbatim}
\usepackage{natbib}
\usepackage[caption=false]{subfig}
\usepackage{multirow}
\usepackage{dcolumn}
\usepackage{rotating}
\usepackage{appendix}
\newcolumntype{d}[1]{D{.}{.}{#1}}

\usepackage{xcolor}

\newcommand\Nquads{52~}
\newcommand\kms{\ifmmode{\rm km\thinspace s^{-1}}\else km\thinspace s$^{-1}$\fi}

\begin{document}

\title{\Nquads Eclipsing Quadruple Star Candidates Discovered in {\em TESS} Full Frame Images}

\correspondingauthor{Veselin B. Kostov}
\email{veselin.b.kostov@nasa.gov}
\author[0000-0001-9786-1031]{Veselin B. Kostov}
\affiliation{NASA Goddard Space Flight Center, 8800 Greenbelt Road, Greenbelt, MD 20771, USA}
\affiliation{SETI Institute, 189 Bernardo Ave, Suite 200, Mountain View, CA 94043, USA}
\author[0000-0003-0501-2636]{Brian P. Powell}
\affiliation{NASA Goddard Space Flight Center, 8800 Greenbelt Road, Greenbelt, MD 20771, USA}
%
\author[0000-0003-3182-5569]{Saul A. Rappaport}
\affiliation{Department of Physics, Kavli Institute for Astrophysics and Space Research, M.I.T., Cambridge, MA 02139, USA}
\author[0000-0002-8806-496X]{Tam\'as Borkovits}
\affiliation{Baja Astronomical Observatory of University of Szeged, H-6500 Baja, Szegedi út, Kt. 766, Hungary}
\affiliation{HUN-REN -- SZTE Stellar Astrophysics Research Group,  H-6500 Baja, Szegedi út, Kt. 766, Hungary}
\affiliation{Konkoly Observatory, Research Centre for Astronomy and Earth Sciences, H-1121 Budapest, Konkoly Thege Miklós út 15-17, Hungary}
%
\author[0000-0002-5665-1879]{Robert Gagliano}
\affiliation{Amateur Astronomer, Glendale, AZ 85308, USA}
\author{Mark Omohundro}
\affiliation{Citizen Scientist, c/o Zooniverse, Department of Physics, University of Oxford, Denys Wilkinson Building, Keble Road, Oxford, OX13RH, UK}
\author[0000-0003-3988-3245]{Thomas L. Jacobs}
\affiliation{Amateur Astronomer, Missouri City, Texas, USA}
\author[0000-0002-2607-138X]{Martti~H.~Kristiansen}
\affil{Brorfelde Observatory, Observator Gyldenkernes Vej 7, DK-4340 T\o{}ll\o{}se, Denmark}
\author[0000-0002-5286-0251]{Guillermo Torres}
\affiliation{Center for Astrophysics $\vert$ Harvard \& Smithsonian, 60 Garden St, Cambridge, MA, 02138, USA}
%
\author[0000-0001-7756-1568]{Gerald Handler}
\affiliation{Nicolaus Copernicus Astronomical Center, Polish Academy of Sciences, ul. Bartycka 18, 00-716, Warszawa, Poland}
%
%
\author[0000-0002-5034-0949]{Allan R. Schmitt}
\affiliation{Citizen Scientist, 616 W. 53rd. St., Apt. 101, Minneapolis, MN 55419, USA}
%
\author{Hans M. Schwengeler}
\affiliation{Citizen Scientist, Planet Hunter, Bottmingen, Switzerland}
%
\author{Tibor Mitnyan}
\affiliation{Baja Astronomical Observatory of University of Szeged, H-6500 Baja, Szegedi út, Kt. 766, Hungary}
\affiliation{HUN-REN -- SZTE Stellar Astrophysics Research Group,  H-6500 Baja, Szegedi út, Kt. 766, Hungary}
\affiliation{Department of Experimental Physics, University of Szeged, H-6720, Szeged, D\'om t\'er 9, Hungary}
%
\author{Ivan A. Terentev}
\affiliation{Citizen Scientist, Planet Hunter, Petrozavodsk, Russia}
%
\author[0000-0002-8527-2114]{Daryll M. LaCourse}
\affiliation{Amateur Astronomer, 7507 52nd Place NE Marysville, WA 98270, USA}
%
\author[0000-0001-7246-5438]{Andrew Vanderburg}
\affiliation{Center for Astrophysics $\vert$ Harvard \& Smithsonian, 60 Garden St, Cambridge, MA, 02138, USA}
%
%
\author[0000-0002-4959-8598]{Svetoslav D. Alexandrov}
\affiliation{Citizen Scientist, Eclipsing Binary Patrol Collaboration}
\affiliation{Institute of Plant Physiology and Genetics, Bulgarian Academy of Sciences, Acad. G. Bontchev Str., BI.23, Sofia 113, Bulgaria}
%
\author[0000-0001-8953-9149]{Cledison Marcos da Silva}
\affiliation{Citizen Scientist, Eclipsing Binary Patrol Collaboration}
\author[0000-0001-7701-6818]{Marco Z. Di Fraia}
\affiliation{Citizen Scientist, Eclipsing Binary Patrol Collaboration}
\author[0000-0001-7062-7632]{Aline U. Fornear}
\affiliation{Citizen Scientist, Eclipsing Binary Patrol Collaboration}
%
\author{Marc Huten}
\affiliation{Citizen Scientist, Eclipsing Binary Patrol Collaboration}
%
\author{Davide Iannone}
\affiliation{Citizen Scientist, Eclipsing Binary Patrol Collaboration}
\author[0000-0003-3464-1554]{Julien S. de Lambilly}
\affiliation{Citizen Scientist, Eclipsing Binary Patrol Collaboration}
%
\author{Sam Lee}
\affiliation{Citizen Scientist, Eclipsing Binary Patrol Collaboration}
%
\author{Jerome Orosz}
\affiliation{Department of Astronomy, San Diego State University, 5500 Campanile Drive, San Diego, CA 92182, USA}
%
\author[0009-0000-2799-7291]{Rafael Rodrigues}
\affiliation{Citizen Scientist, Eclipsing Binary Patrol Collaboration}
%
\author{Allan Tarr}
\affiliation{Citizen Scientist, Eclipsing Binary Patrol Collaboration}
%
\author[0000-0003-2381-5301]{William Welsh}
\affiliation{Department of Astronomy, San Diego State University, 5500 Campanile Drive, San Diego, CA 92182, USA}

\begin{abstract}
We present the discovery of \Nquads eclipsing quadruple star candidates detected in {\em TESS} Full Frame Image {\sc eleanor} data by machine learning and citizen scientists. The uniformly-vetted and -validated targets exhibit two sets of eclipses following two distinct periods, representing quadruple systems with a 2+2 hierarchical configuration. Detailed photocenter measurements confirmed that both sets of eclipses originate within $\sim0.1-0.2$ pixels ($\sim2-4$ arcsec) of the corresponding target, and ruled out resolved nearby field stars. The catalog includes a number of systems producing prominent eclipse timing variations and/or apsidal motion, a quadruple with an outer period of ${\sim 1,400}$ days, and even a 2+2 quadruple in a likely wide quintuple with a resolved co-moving star. Additionally, two systems have complete astrometric solutions for the outer orbits from Gaia. 
We provide the measured ephemerides, eclipse depths and durations, overall statistical properties, and highlight potentially interesting systems that merit further investigations.
\end{abstract}

\accepted{AJ October 2025}


\section{Introduction}\label{sec:intro}

Multiple stellar systems are a fascinating natural outcome of stellar formation and evolution. Overall, more than half of Sun-like stars are in binary and higher-order hierarchical systems, and the higher the mass of the system, the higher the multiplicity fraction \citep[e.g.,][]{2010ApJS..190....1R, 2018ApJS..235....6T, 2017ApJS..230...15M}. Systems where one or more of the individual components happen to have the fortuitous geometrical orientation to produce eclipses present excellent targets for validation and confirmation of the underlying multiplicity, as well as ideal laboratories for testing and benchmarking stellar evolution models \citep[e.g.][]{1991A&ARv...3...91A, 2010A&ARv..18...67T,2023MNRAS.522...90K, 2015ASPC..496...55O,2013MNRAS.435..943P,2018MNRAS.476.4234F,2021MNRAS.502.4479H,2019MNRAS.486.4781F,2019MNRAS.483.4060L,2022PhRvD.106d3014T,2022ApJ...926..195V,2021MNRAS.507.5832K,2021MNRAS.507..560S,kolar2025,zasche2025}. Of these, perhaps the most intriguing targets are those where the outer periods are short enough that the dynamical interaction between the components are not only detectable but allow {\it direct confirmation} of the systems' architectures \citep[e.g.][]{2022Galax..10....9B,2016MNRAS.455.4136B,2022MNRAS.515.3773B,2022MNRAS.510.1352B,2017MNRAS.467.2160R,2023MNRAS.521..558R,2022MNRAS.513.4341R,2021AJ....161..162P,2021ApJ...917...93K}. Speaking of dynamical interactions, these systems are invaluable testbeds for investigations of long-term orbital stability and secular evolution, including Von Zeipel-Lidov-Kozai cycles \citep[e.g.][]{zeipel1910,kozai1962, lidov1962,2017MNRAS.470.1657H}. The physical and orbital properties of such systems offer important new insights into how they form (e.g., initial disk fragmentation, later capture, or a combination) and evolve (e.g., similar to isolated single stars, or through complex multiple common envelope phases), and even constrain the presence of potential planets \citep[e.g.][]{1994ARA&A..32..465M,2015Natur.518..213P,2016Natur.538..483T,Tokovinin2021,Whitworth2001,2023ApJS..264...45F,2022MNRAS.517.2111P,2022MNRAS.516.1406S,2021ApJ...917...93K,2016ApJ...832..183K}. 

From an observational perspective, compact 2+2 quadruple systems present unique challenges and opportunities. For example, the smaller the outer orbit is, the more direct imaging or astrometric observations would struggle to resolve it \citep[e.g.][]{2018ApJS..235....6T, Tokovinin2021,2025AAS...24610303M}. Another difficulty is that spectral lines from four unresolved stars can lead to significant blending, making it difficult to measure individual radial velocities \citep[][]{Czekala2017,2020MNRAS.494..178P,Powell2025}. With that said, when successful, such analyses can provide a wealth of information about the stellar properties, including effective temperatures, surface gravities, and elemental abundances for all four components. When both binary components of these systems produce eclipses, the composite lightcurve can exhibit complex patterns where overlapping (blended) eclipses create intricate photometric signatures that can be difficult to disentangle. This complexity often requires sophisticated modeling techniques and long-term monitoring to accurately determine the system parameters \citep[e.g.,][]{2021arXiv210508614K,2022Galax..10....9B, 2023MNRAS.522...90K,Powell2025}. However, it also presents an opportunity: the extremely rich information content in these lightcurves can provide precise measurements of stellar masses, radii, and orbital parameters for all four stars simultaneously. 

All-sky photometric surveys provide excellent discovery platforms for such systems, and indeed the number of known eclipsing 2+2 quadruple candidates has dramatically increased in the past several years alone. Thanks to observations from ASAS-SN, Kepler, OGLE, TESS, and ZTF, hundreds of such candidates have already been detected \citep[e.g.][]{2017PASP..129j4502K,2018AJ....156..241H,2022MNRAS.517.2190R,2011AJ....141...83P,2022ApJS..258...16P,2018A&A...619A..97D,2022MNRAS.509..246H,2022arXiv221100929M,2016AcA....66..405S, 2022ApJS..259...66K,2024MNRAS.527.3995K,2024A&A...682A.164V}, dramatically expanding their family portrait and, in essence, transitioning the field from `cherry picking' to `cherry harvesting'. It is important to note that these candidates have undergone various levels of vetting and validation, and most are yet to be confirmed as genuine quadruple systems. Such confirmation requires substantial efforts, typically based on extensive follow-up observations and analysis. These can include long-term photometric and spectroscopic monitoring to obtain new eclipse times and measure the radial velocities of the two component EBs, utilizing archival data, direct imaging to resolve the individual components, and comprehensive photodynamical modeling \citep[e.g.,][]{2021ApJ...917...93K, 2023MNRAS.522...90K, 2023MNRAS.524.4220P, Powell2025}.

Here we present the latest addition of \Nquads targets to our TESS/GSFC/VSG (TGV) catalog of eclipsing stellar multiples with a 2+2 hierarchical configuration. This brings the total number of TGV quadruples to 250, and represents a meticulously curated set of uniformly-vetted, -validated, and -characterized systems where the two binary components are confirmed to originate from within $\sim0.1-0.2$ TESS pixels ($\sim2-4$ arcsec) of the target. In turn, this potentially makes the systems compact enough to detect dynamical interactions on observable timescales. For each system, we measure the ephemerides of both EBs, the respective eclipse depths and durations (including those of the secondary eclipses, if present), outline interesting systems and potential issues, include Gaia astrometric information, and provide relevant comments. 

This paper is organized as follows. In Section \ref{sec:detection} we highlight the detection, vetting, and validation of the quadruple candidates. Section \ref{sec:catalog} presents the details of the catalog and we summarize our results in Section \ref{sec:summary}.

\section{Detection, Validation, and Vetting}
\label{sec:detection}

\begin{figure*}
    \centering
    \includegraphics[width=0.99\linewidth]{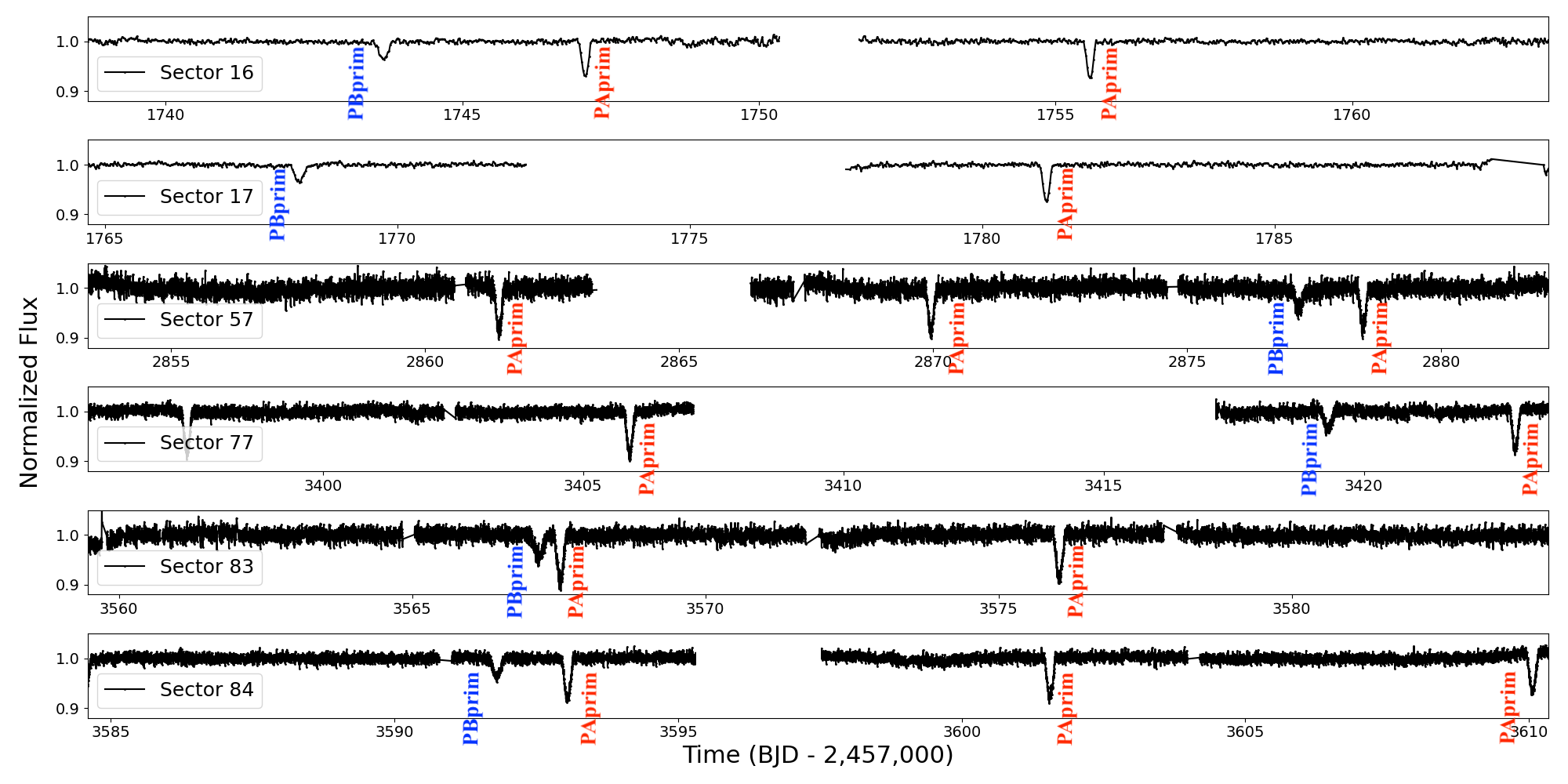}
    \caption{An illustrative example of an eclipsing quadruple candidate detected in TESS FFI \textsc{eleanor} data. The panels show six sectors of data for TIC 64832327, exhibiting two sets of primary eclipses with two distinct periods, labeled as PAprim (13 red events) and PBprim (6 blue events).}
    \label{fig:64832327_lc}
\end{figure*}

All targets presented here produce two sets of eclipses following two distinct periods, and are assumed to be unresolved at the time of writing\footnote{Targets where Gaia DR3 measurements \citep{Gaia2021} suggest that one EB may be co-moving with a resolved nearby EB -- thus potentially indicating a (very) wide quadruple system -- are beyond the scope of this work, and excluded from our TGV catalog.}. An example is shown in Figure \ref{fig:64832327_lc} for TIC 64832327 where the two component EBs produced prominent primary eclipses (and no discernible secondary eclipses). The analysis of the systems presented here followed the methodology of \citep[][K22 and K24 hereafter]{2022ApJS..259...66K,2024MNRAS.527.3995K}. For completeness, we outline the various steps below.

\subsection{Detection}

For the detection of the systems presented here, we relied on an established collaborative effort between advanced computational techniques and expert human visual inspection. The underlying process leverages the natural synergy between state-of-the-art machine learning algorithms and trained citizen scientists, enabling efficient exploration of the vast datasets generated by TESS. Briefly, this is a two-step process that generally goes as follows.

First, we identified the target stars as EB candidates through the application of a trained neural network to TESS Full Frame Image {\sc eleanor} data for all stars brighter than T = 15 mag \cite{2021AJ....161..162P, 2022RNAAS...6..111P}. Notably, the network was trained to detect light curves exhibiting one or more eclipse-like features -- rather than a strict sequence of consecutive eclipses -- focusing solely on morphological characteristics without imposing periodicity requirements \citep{2025ApJS..279...50K}. This process effectively reduced the number of unique targets necessitating subsequent manual review by two orders of magnitude, i.e., from millions of light curves per sector to tens of thousands. 

Most of the targets discussed in this catalog were identified by members of the Visual Survey Group (VSG, \citep{2022PASP..134g4401K}), comprising highly skilled volunteers conducting visual inspections of pre-selected light curves. These experts employ specialized software tools, including LcTools and custom-designed programs \citep{2019arXiv191008034S,2021arXiv210310285S}, to perform on-demand and in-depth analysis of the photometric data. Their extensive experience enables rapid and accurate identification of salient features, with typical assessment times on the order of seconds per light curve (K22, K24). The remaining targets were identified through the Eclipsing Binary Patrol project (EBP) \citep{2025ApJS..279...50K}, a Zooniverse-hosted citizen science initiative aimed at vetting and validation of the EB candidates detected by the neural network of \citep{2025ApJS..279...50K}\footnote{\url{https://www.zooniverse.org/projects/vbkostov/eclipsing-binary-patrol}}. As part of the EBP workflow, the volunteers inspect the presented lightcurve and have the option to flag interesting features such as additional eclipses. 

It is worth noting that the contributions of volunteer astronomers to professional astronomical research are substantial, resulting in numerous peer-reviewed publications and significant discoveries of a large number of transiting planets and multiple star systems \citep[e.g.,][and references therein]{2022PASP..134g4401K}. As an example, members of the VSG have visually inspected millions of lightcurves from Kepler, K2, and TESS, and have identified numerous unique objects such as triply eclipsing triples, quadruple star systems, tidally-tilted pulsators, and exocomets, underscoring the critical importance of citizen science \citep[e.g.,][]{2020NatAs...4..684H,2021AJ....162..299P,2022PASP..134g4401K,2022ApJS..263...14C,2023MNRAS.524.4220P,2024ApJ...975..121J,2025arXiv250721255H}. To ensure their analysis is as thorough as possible, the volunteers also examine publicly available archives and databases such as Simbad, Gaia, ZTF, etc., as well as alternative TESS data processing pipelines such as QLP \citep{2020RNAAS...4..204H} and SPOC \citep{2016SPIE.9913E..3EJ} for select targets. This multifaceted approach maximizes the fidelity of the potential candidates and increases the probability of them being genuine quadruple systems. 

\subsection{Validation and Vetting}

Given the relatively large pixel size of TESS (21 arcsec), it comes as no surprise that the vast majority of candidates for stellar multiples detected by the citizen scientists turn out to be false positives due to the presence of distinct, spatially resolved eclipsing binaries in close proximity to the target star. Indeed, TESS lightcurves are frequently affected by such occurrences, as well as various other astrophysical sources of contamination (variable stars, Solar System objects, etc.), and non-astrophysical artifacts \citep[e.g.,][]{2022MNRAS.513..102C,2023MNRAS.521.3749M,2025ApJS..279...50K}. Notably, the issue of contamination is neither unique to TESS, nor uncommon. Instead, it presents a general challenge across different photometric surveys and datasets, and requires careful scrutiny. As an example, \cite{2024A&A...688A..41Z} report that about one in four of the eclipsing quadruple candidates \cite{2023A&A...674A.170A} detected in OGLE data are in fact false positives due to unrelated nearby EBs. 

Consequently, each of the quadruple candidates identified by the citizen scientists must undergo a series of rigorous vetting and validation procedures. Below we elaborate on our multi-tiered methodology for addressing the contamination challenge, presenting examples, discussing known limitations and relevant considerations. For more details, we refer the reader to K22, K24, and \citep{2025ApJS..279...50K}. 

In order of decreasing angular separation between the target and nearby contaminating stars, our three `lines of defense' in validating potential quadruple candidates are as follows: 

\begin{itemize}
\item Pixel-by-pixel analysis: This is a common procedure based on manual inspection of the TESS image. Specifically, we employ Lightkurve's interactive features \citep{lightkurve} by first locating the target star's pixel position and selecting an appropriate aperture to minimize contamination from resolved sources. Next, we examine the adjacent pixels individually for the presence of nearby unrelated EBs that may mimic a second set of eclipses detected in the target's lightcurve. Finally, we rule out adjacent pixels and verify that both sets of eclipses originate from the vicinity of the target. An example of this vetting is shown in Figure \ref{fig:302537339} for the case of TIC 302537339. Candidates passing this initial inspection proceed to the next level of scrutiny -- a more detailed photocenter vetting.

\begin{figure*}
    \centering
    \includegraphics[width=0.995\linewidth]{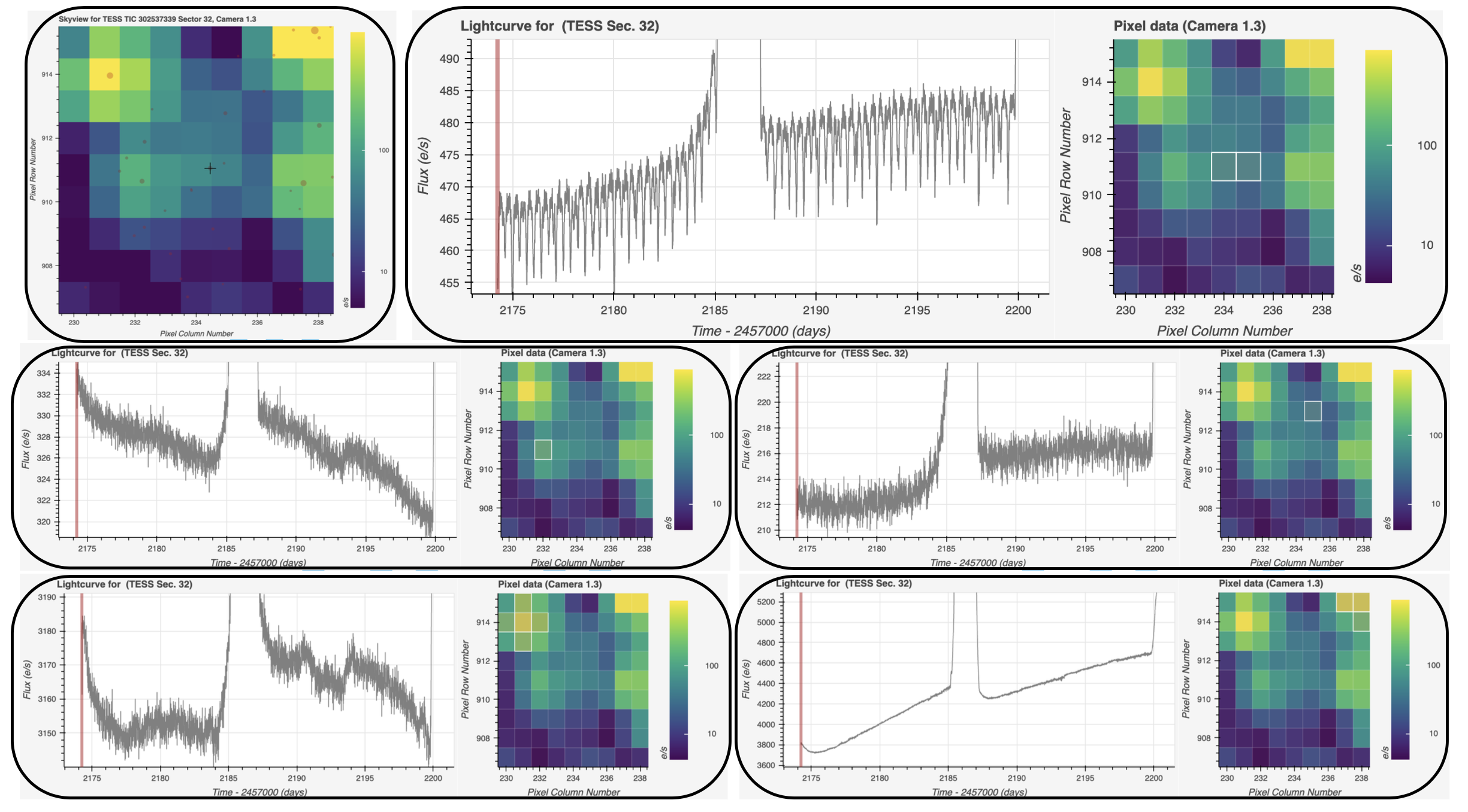}
    \caption{Pixel-by-pixel analysis of the quadruple candidate TIC 302537339 for Sector 32. First row, left panel: 9$\times$9 TESS pixels difference image centered on the target. There are several nearby field stars bright enough to contaminate the target's lightcurve and mimic a quadruple candidate. First row, right two panels: lightcurve (left) from the two pixels centered on the target (right, white contours). Second and third rows: same as above but for the nearby field stars. None of these show features resembling the eclipses seen in the target's lightcurve, confirming the signals originate from the vicinity of TIC 302537339. For clarity, the individual panels are outlined with black contours.}
    \label{fig:302537339}
\end{figure*}

\item Photocenter analysis: Another commonly-used vetting method employs a difference imaging technique to analyze the center-of-light motion during detected features of interest such as transits or eclipses. This process involves creating difference images by subtracting the in-eclipse images (how TESS sees the surrounding pixels during an eclipse) from the out-of-eclipse images (how TESS sees the surrounding pixels before and after the same eclipse) for each detected eclipse and available sector. The center-of-light of these difference images represents the pixel position of the eclipse source. To measure the pixel position of said center-of-light, we fit each difference image with a Point Spread Function (PSF) and the TESS Pixel Response Function (PRF), and adopt the average of the two measurements. Finally, the average measured photocenter pixel position from all detected eclipses is compared to the catalog pixel position of the target (according to the TIC, \citep{2019AJ....158..138S}). If the difference between the two positions is not statistically significant, the corresponding eclipses are considered to be on-target. An example of this is shown in Figure \ref{fig:470730552_cent} for TIC 470730552. Alternatively, the eclipses are marked as off-target and the candidate is flagged as a false positive. Overall, based on our experience with TESS data, and depending on the particular target, the photocenter positions can often be reliably measured to within $\sim0.1-0.2$ pixels (corresponding to $\sim2-4$ arcsec) of the target \citep[e.g.,][]{2022ApJS..259...66K,2024MNRAS.527.3995K,2025ApJS..279...50K}.

\begin{figure*}
    \centering
    \includegraphics[width=0.8\textwidth]{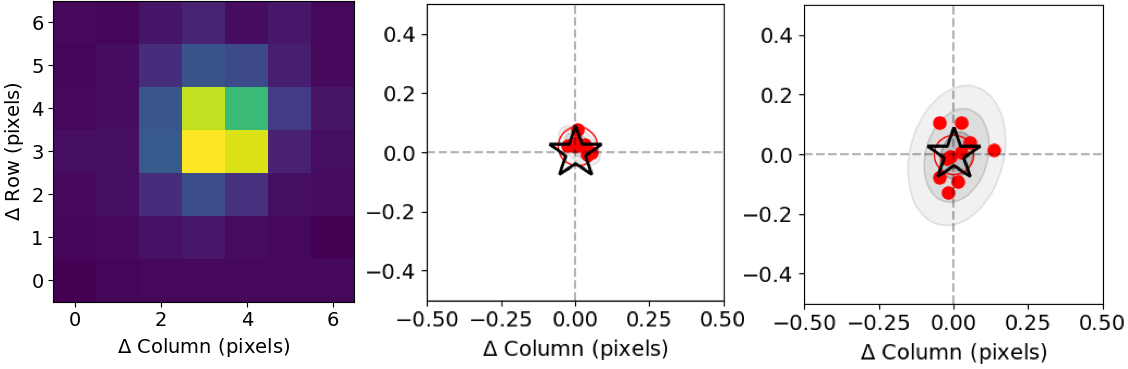}
    \caption{Left panel: 7$\times$7 TESS pixels difference image centered on TIC 470730552 for Sector 58. Middle and right panels: Measured photocenters for PA (middle) and PB (left panel). The small red dots, the large red circle, and the black star represent the per-eclipse photocenters, the average photocenter, and the pixel position of the target, respectively. Note that the scale of the left panel is 10 times larger than that of the middle and left panels.}
    \label{fig:470730552_cent}
\end{figure*}

We note that the difference images should ideally look like the left panel in Figure \ref{fig:470730552_cent}, i.e., a single, bright, well-defined pixelated ``spot'' centered near the target's position and superimposed on an otherwise dark background. In practice, the difference images can often be affected by systematics that distort them, sometimes even to the point of making them unsuitable for photocenter measurements; these systematics are target-specific and generally vary from one eclipse to another. An example of this is shown in Figure \ref{fig:2158899} for three eclipses of TIC 2158899 in Sector 44. Thus, to minimize the impact of such distortions on the photocenter measurements and ensure their reliability, we visually inspect the difference image for each eclipse. Those that do not pass scrutiny are removed from the analysis\footnote{Partially or fully blended eclipses also render the difference imaging technique inapplicable, and are thus removed from the photocenter measurements as well. This situation can arise in eclipsing quadruple systems with similar periods and ephemerides, unlike in transiting multiplanet systems.}

\begin{figure}[h]
    \centering
    \includegraphics[width=0.995\linewidth]{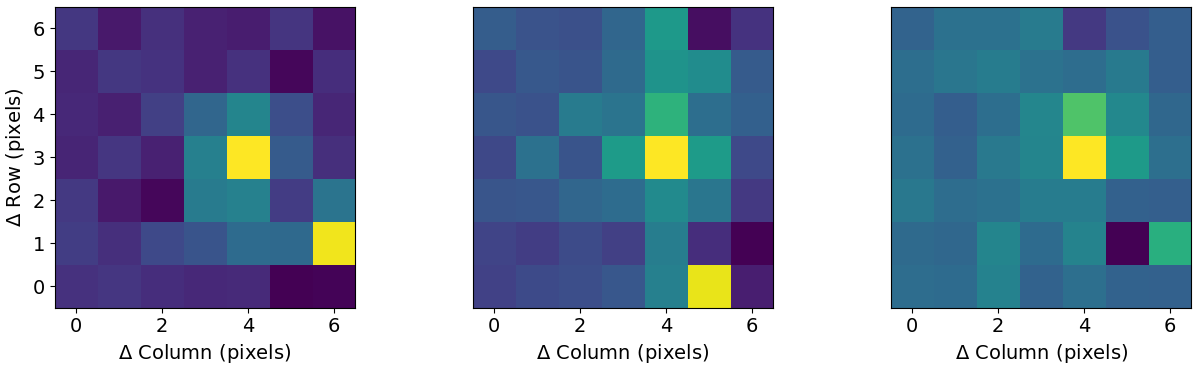}
    \caption{Same as left panel of Figure \ref{fig:470730552_cent} but for three eclipses of TIC 2158899 from Sector 44. All three difference images are affected by systematics, making them less than ideal for photocenter measurements.}
    \label{fig:2158899}
\end{figure}

\item Magnitude difference (${\rm \Delta_{mag}}$) as a function of measured eclipse depth (${\rm \delta_{ecl}}$): Finally, there are instances where a resolved field EB is sufficiently bright to produce the detected `extra' eclipses as false positives, yet is too close to the target star to pinpointing their origin. This occurs when said EB is within a magnitude difference with respect to the target star:

\begin{equation}
{\rm \Delta_{mag} = 2.5\times~log_{10}\left(\frac{1 - 2\times~\delta_{ecl}}{2\times~\delta_{ecl}}\right)}
\label{eq:delta_mag}
\end{equation}

{\noindent If the projected separation between the two is smaller than $\sim0.1-0.2$ pixels, the true source of the extra eclipses cannot be definitively determined based on TESS photocenter measurements, the candidate is marked as unclear and excluded from the catalog.} An example of this is shown in Figure \ref{fig:8698910_sv} for the case of TIC 8698910, producing two distinct sets of eclipses. However, there is a resolved star 1.83 arcsec away, TIC 668542259, with ${\rm \Delta_{mag} \approx 1.1}$ which makes pinpointing the origin of the two sets of eclipses highly challenging.
 
\begin{figure}
    \centering
    \includegraphics[width=0.995\linewidth]{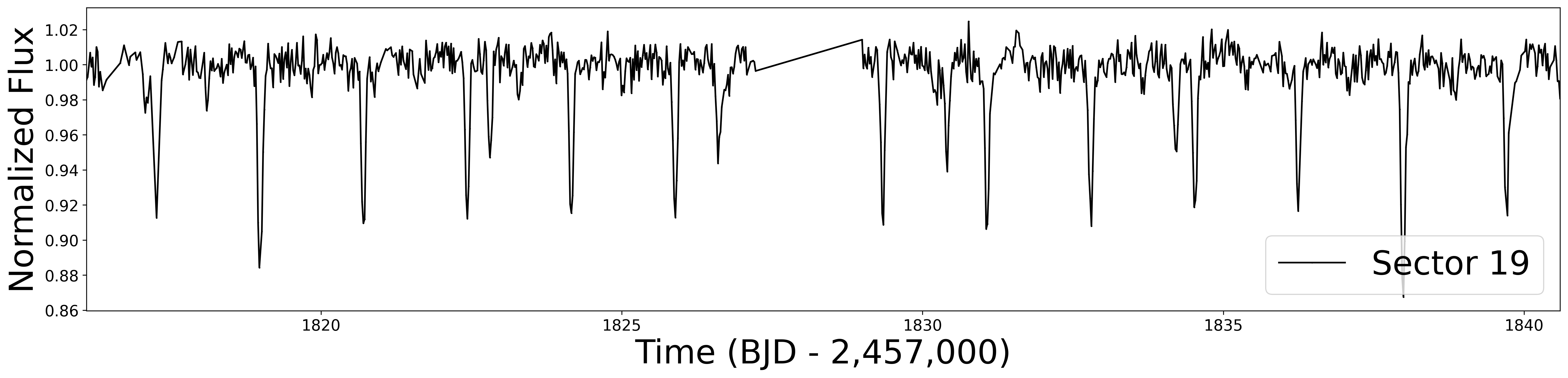}
    \includegraphics[width=0.995\linewidth]{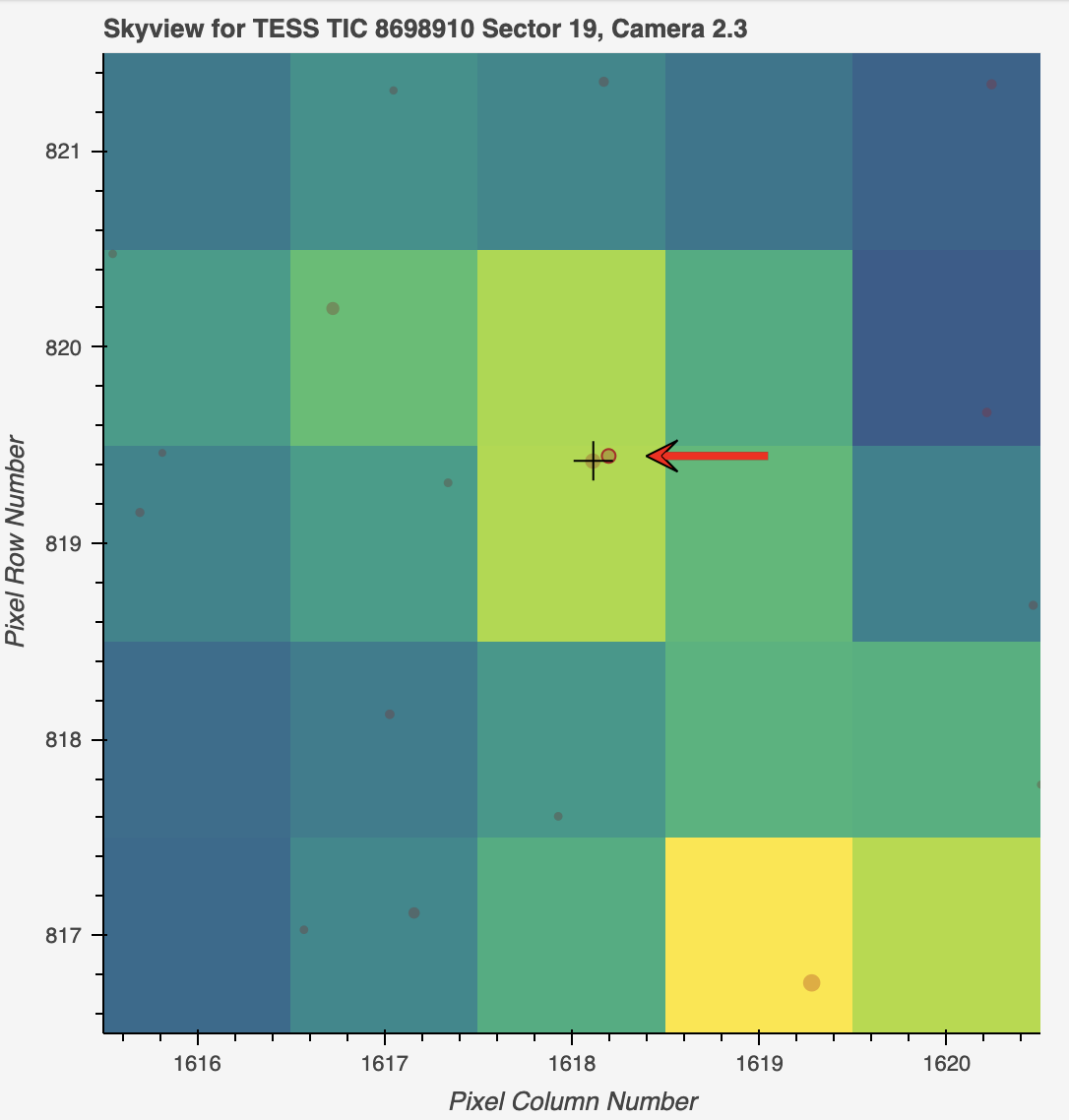}
    \caption{Upper panel: TESS FFI \textsc{eleanor} lightcurve for TIC 8698910 exhibiting two sets of eclipses. Lower panel: 5$\times$5 pixel Skyview image centered on the target (black cross-hair symbol). There is a resolved field star separated by 1.83 arcsec and $\approx$ 1.1 TESS magnitude fainter (marked with a red arrow), making photocenter vetting of the detected eclipses difficult.}
    \label{fig:8698910_sv}
\end{figure}

\end{itemize}

Altogether, this comprehensive vetting process ensures the reliability and consistency of the TGV catalog, minimizing false positives and providing a robust dataset for further research and analysis.

\subsection{Ephemeris Determination and Eclipse Characterization}

For all but three of the \Nquads candidates, we derived initial estimates of the orbital periods and eclipse times for both sets of detected eclipses utilizing the Box-Least Squares algorithm (BLS, \cite{2002A&A...391..369K}), applied to all available TESS FFI data at the time of writing. As part of this process, we removed data outliers and eclipses dominated by systematics, normalized the lightcurve on a sector-by-sector basis, stitched all sectors together, and, if necessary, detrended the data utilizing a low-order Savitsky-Golay filter. For the remaining targets, TIC 97642729, 391461666, and 79908874, the second EB does not produce enough eclipses in the available data to uniquely determine the orbital period. 

Following the preliminary BLS analysis, we improved the ephemeris measurements and measured eclipse depths and durations by implementing custom models that fit each individual eclipse with a trapezoid, Gaussian, a generalized hyperbolic secant (see Eqn. 2, K22), and a generalized Gaussian function (see Eqn. 1, K22). Based on our experience, the latter two provide comparable fits to the data, and both are significantly better than either the trapezoid or the Gaussian model. For simplicity, the measurements reported in this catalog are based on the generalized Gaussian fits.

\begin{figure}[h]
    \centering
    \includegraphics[width=0.75\linewidth]{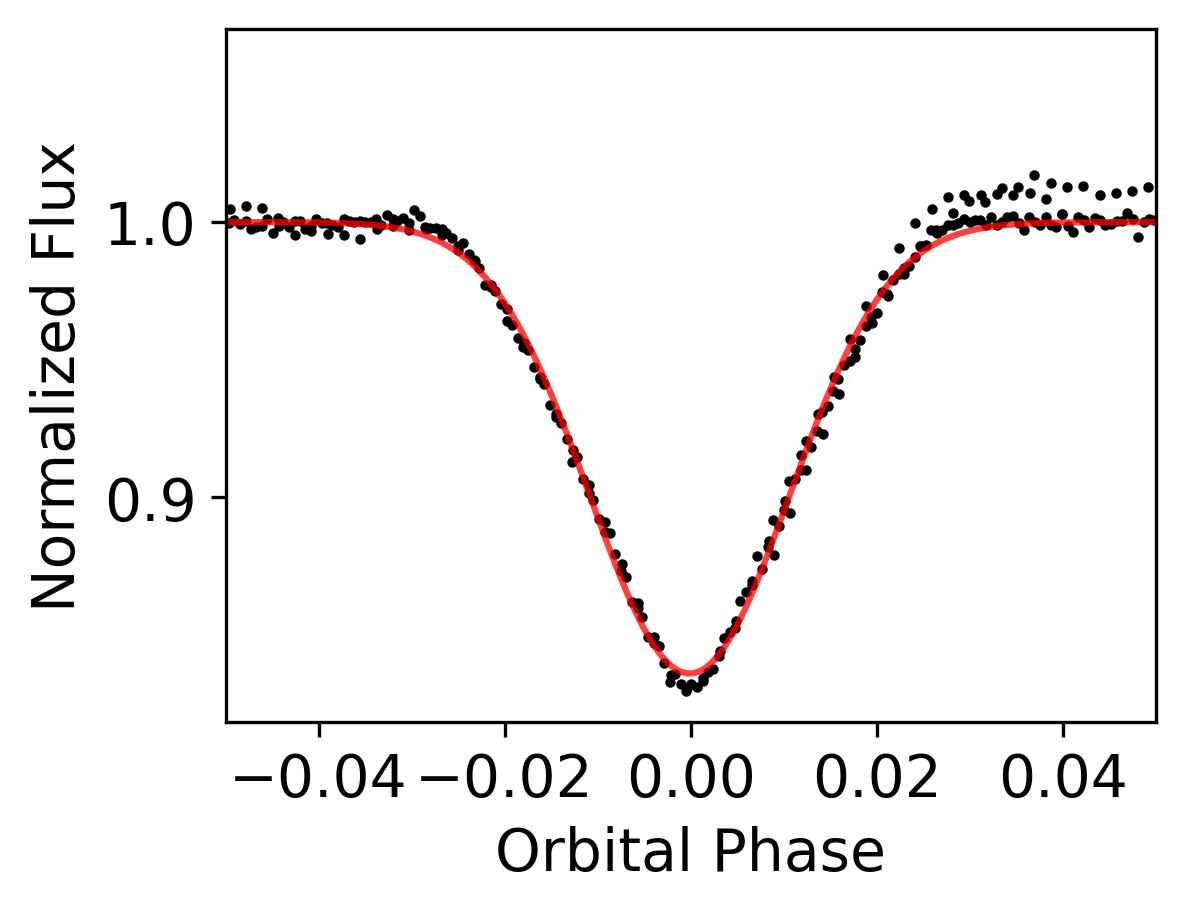}
    \includegraphics[width=0.75\linewidth]{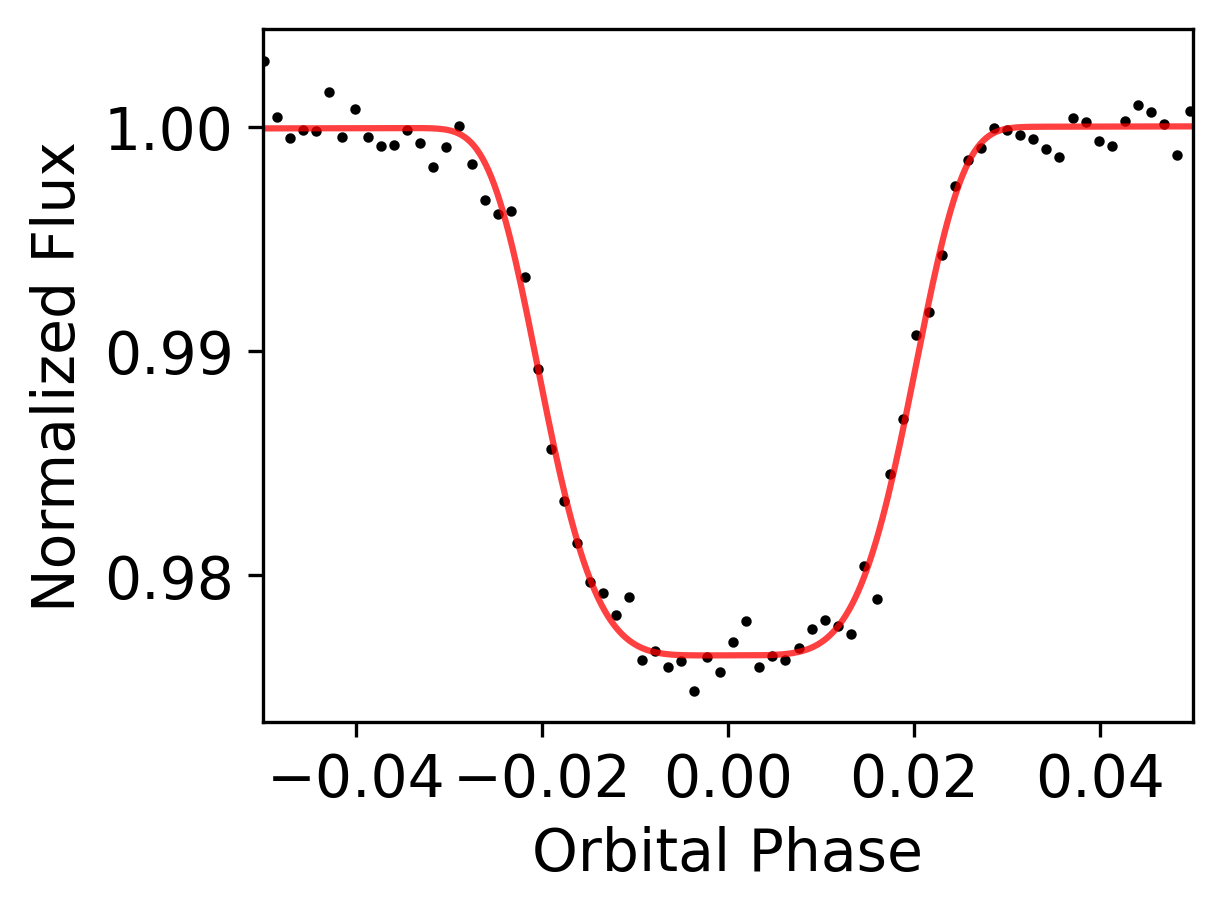}
    \includegraphics[width=0.75\linewidth]{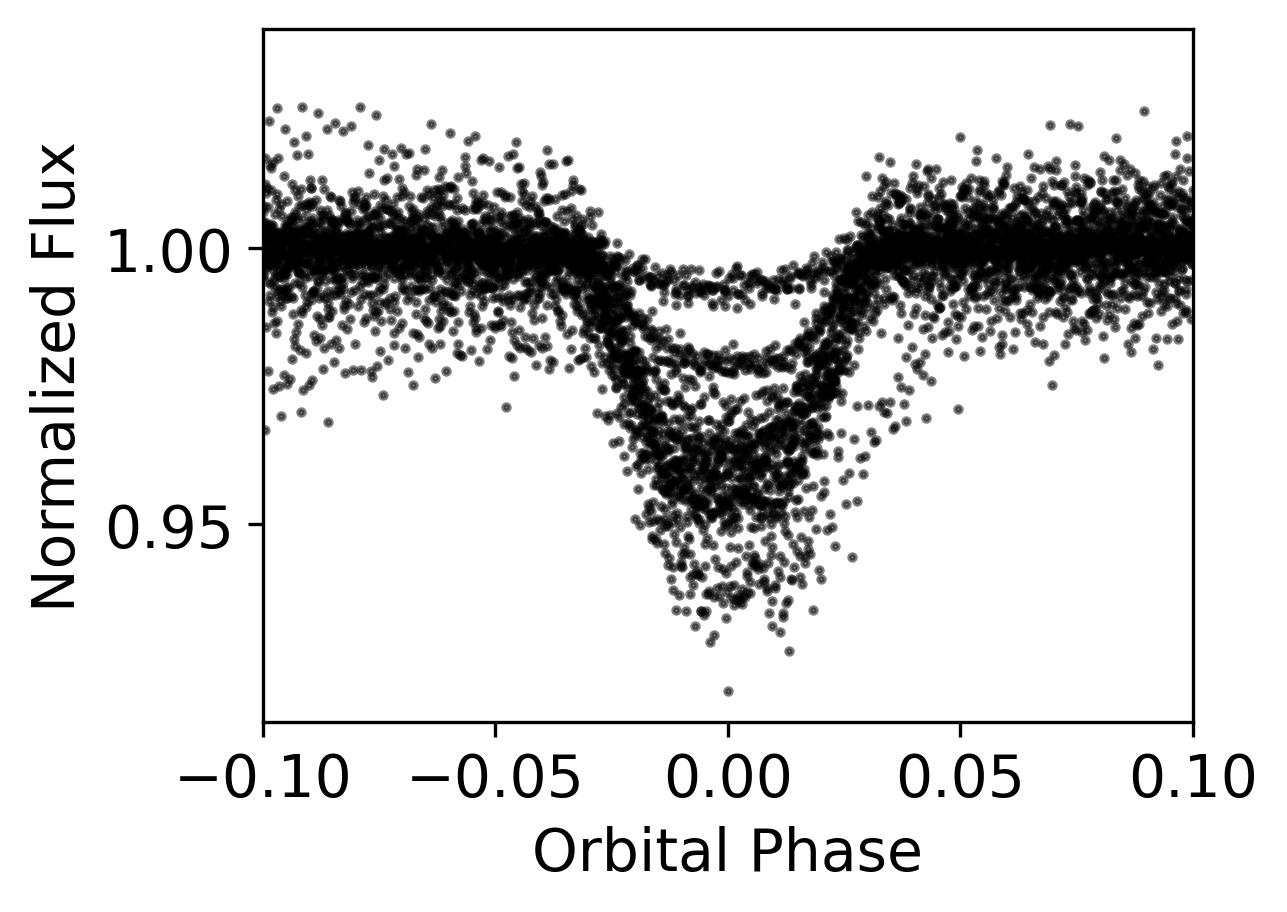}
    \caption{Upper and middle panels: Generalized Gaussian model fits (red) for the primary eclipses of the PA $\approx$ 3.96 days component (upper) and PB $\approx$ 4.94 days (middle) component of the quadruple candidate TIC 20938739. The black dots represent the TESS FFI data for Sector 51. The two eclipses have notably different shapes -- one being more V-shaped and the other more U-shaped -- yet the model fits both. Lower panel: phase-folded primary eclipses of the PA $\approx$ 3.08 days component of the quadruple candidate TIC 97642729 showing TESS FFI data for all available Sectors (6, 33, 43, 44, 45, 72, and 87). Due to systematic effects, the apparent eclipse depths vary dramatically between different sectors.}
    \label{fig:GG_example_1}
\end{figure}

\begin{figure}[h]
    \centering
    \includegraphics[width=0.995\linewidth]{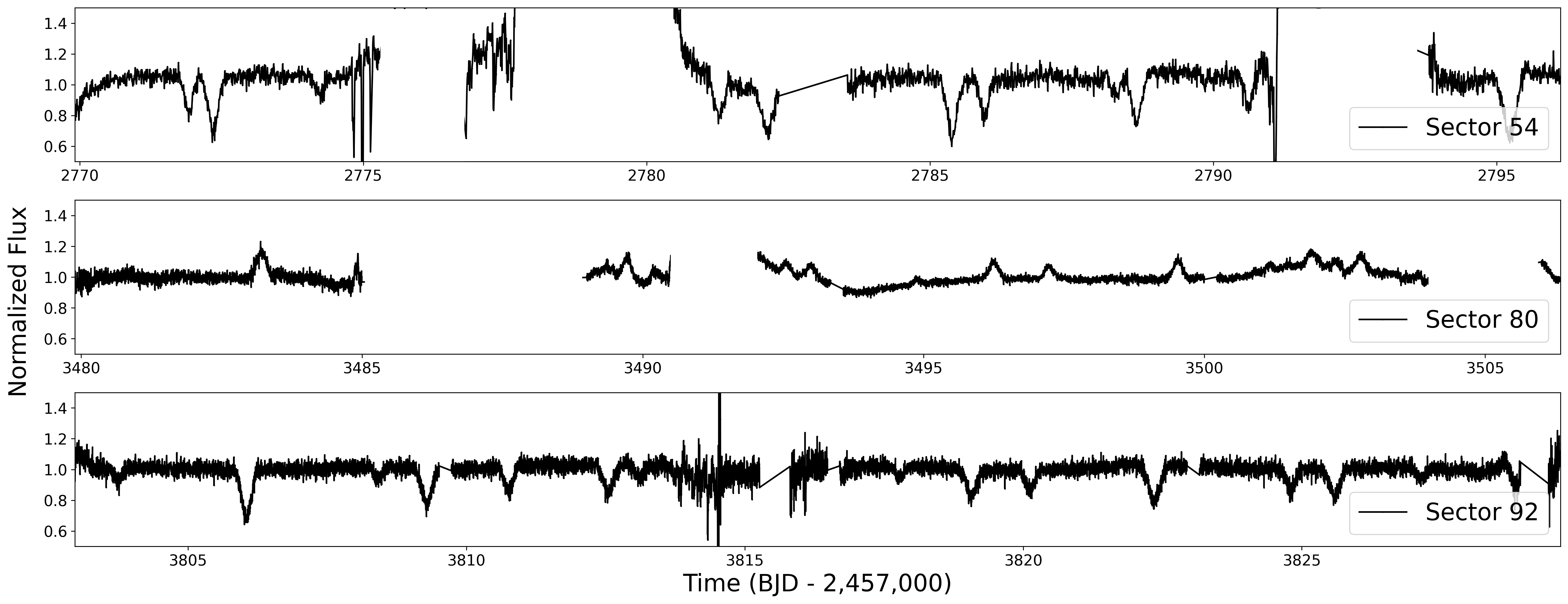}
    \caption{TIC 99629496: an example of systematics-dominated \textsc{eleanor} lightcurve where the eclipses are upside down in one sector of TESS data.}
    \label{fig:tic_99629496}
\end{figure}

\begin{figure*}
    \centering
    \includegraphics[width=0.48\textwidth]{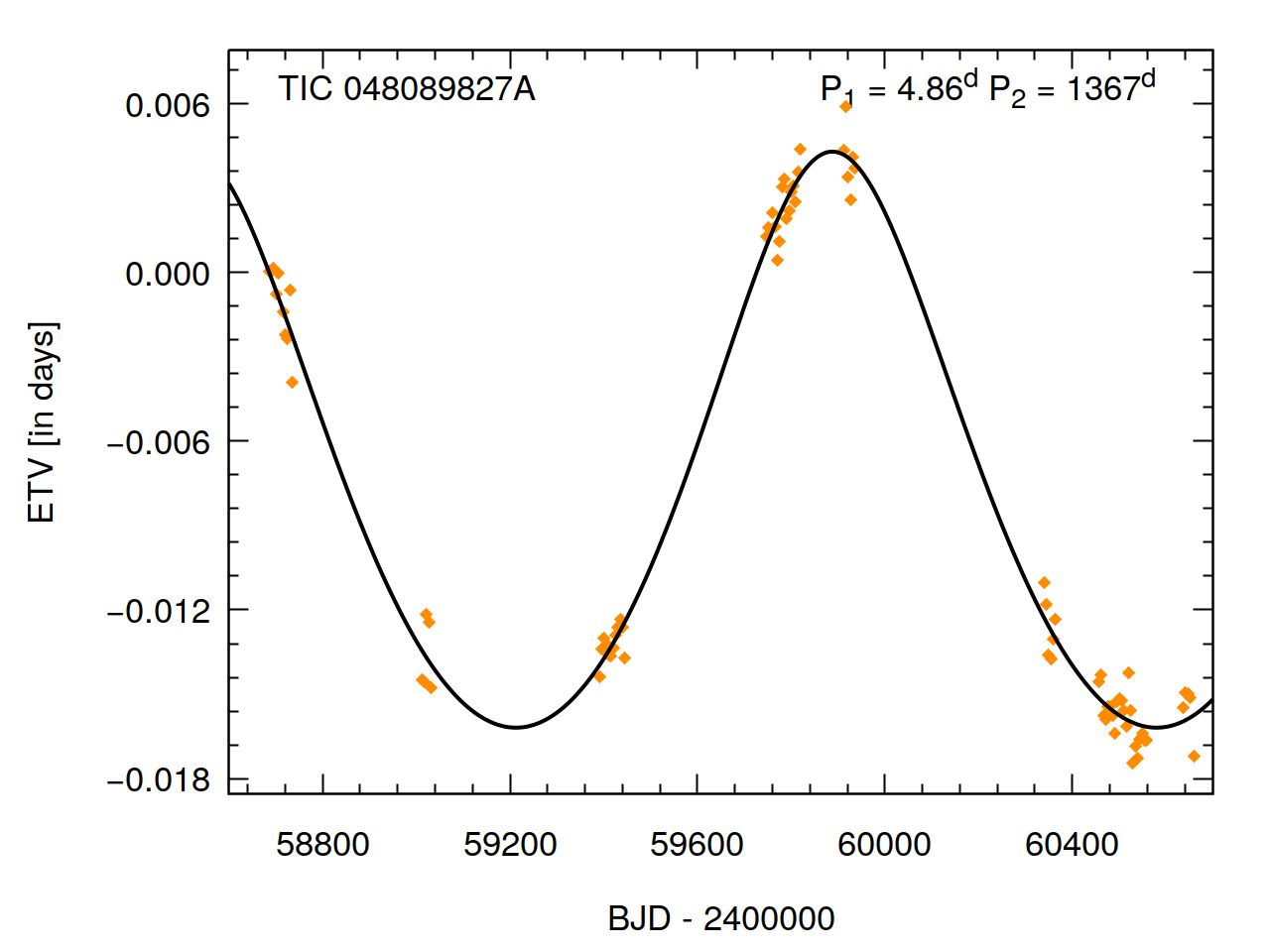}
    \includegraphics[width=0.48\textwidth]{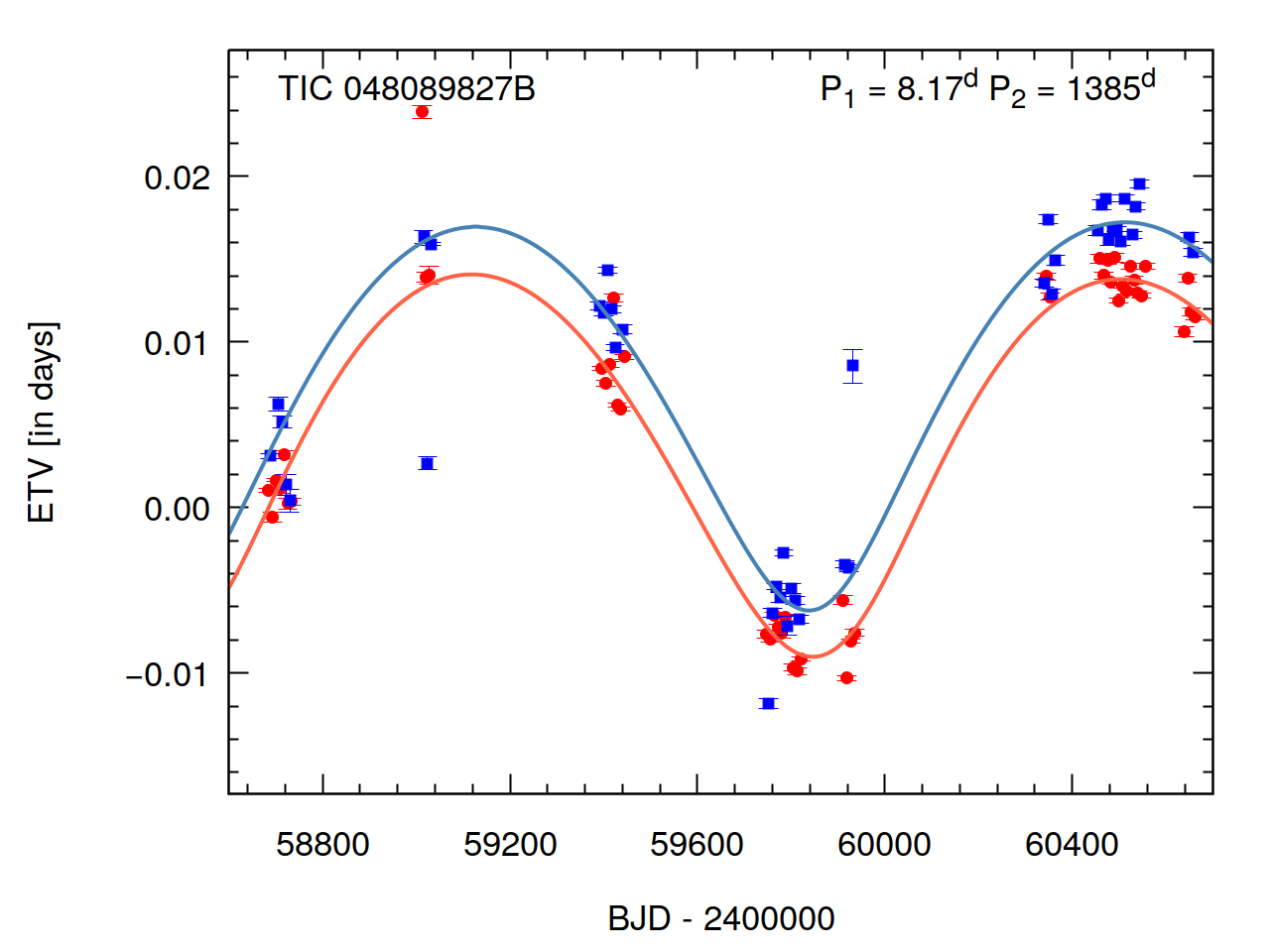}
    \caption{Measured anti-correlated ETVs for the PA $\approx$ 4.86-days component (left panel, primary eclipses) and PB $\approx$ 8.17 days components (right panel, primary (red) and secondary (blue) eclipses) for TIC 48089827. The two EBs produce dramatic, anti-correlated ETVs that confirm the target as a genuine quadruple system. The solid curves represent the best-fit models indicating that the ETVs are dominated by the light travel time effect, and suggesting an outer period of about 1,400 days.}
    \label{fig:TIC_48089827_etvs}
\end{figure*}

We note that while this approach does not provide any information on the astrophysical effects responsible for eclipses, it nevertheless represents a fast and flexible framework that comes with three key advantages:

\begin{itemize}

    \item Eclipse depths and duration: Eclipses come in a wide variety of shapes, depending on the a-priori unknown physical and orbital parameters of the system. The two generalized functions easily accommodate this variety by reliably modeling the eclipse shapes observed in our target sample. This is highlighted in Figure \ref{fig:GG_example_1} for the two components of the quadruple candidate TIC 20938739, where the primary eclipses for one of the EBs have a pronounced V-shape with a relatively sharp minimum, while those of the other have a much flatter bottom and resemble the letter U. 
    \item Eclipse depth differences between different sectors: Sometimes, eclipse depths appear to vary from one sector to another (see Figure \ref{fig:GG_example_1}). In the vast majority of cases, such depth variations are not astrophysical but instead caused by systematic effects that affect the lightcurve on a sector-by-sector basis. This is typically due to changes in the relative orientation and respective overlap between the sector-specific aperture used to extract the target's lightcurve and nearby field stars. Consequently, the effects of blending, contamination, and background subtraction can vary not only from one sector to another but also within a single sector, particularly across observation gaps caused by data downlink periods \citep[e.g.,][]{2021ApJS..257...53L, 2023Natur.623..932L, 2024AJ....167....1W, 2025ApJ...988L...4H, 2025ApJS..279...50K}. An example of one of the more extreme cases where systematics dominate the lightcurve is shown in Fig. \ref{fig:tic_99629496} for TIC 99629496, where the eclipses in Sector 80 are upside-down. As above, both the generalized hyperbolic secant and the generalized Gaussian automatically circumvent this problem by providing independent depth measurements on an eclipse-by-eclipse basis. For simplicity, the eclipse depths reported in this catalog are the median values.
    \item Eclipse timing-variations (ETVs): Many EBs exhibit deviations from strict periodicity caused by dynamical interactions with additional bodies in the system \citep[e.g.,][]{2015ASPC..496...55O,2016MNRAS.455.4136B,2025A&A...695A.209B}. Detecting such variations provides critical information about the underlying architecture of the system, and can even confirm its higher-order multiplicity. The signal is often small and easy to miss, especially when the resulting `smear' in the lightcurve folded on the BLS period is difficult to notice during a visual inspection. The generalized functions mentioned above naturally account for this issue as they provide the time of minimum light for each individual eclipse, thus relaxing the requirement for strict periodicity between consecutive eclipses. An example of measured ETVs is shown in Figure \ref{fig:TIC_48089827_etvs} for the case of the quadruple candidate TIC 48089827.
\end{itemize}

\section{The Catalog}
\label{sec:catalog}

The results presented here expand the TGV catalog with \Nquads new eclipsing quadruple star candidates, bringing the total number that we have published in K22, K24, and this work to 250. Each catalog entry provides detailed information about the target, including TIC and Gaia DR3 identifiers, sky position, TESS magnitude, measured ephemerides, primary and, where applicable, secondary eclipse depths and duration, and secondary phases. The orbital periods of the two component EBs are labeled as PA and PB, for simplicity ordered such that PA $<$ PB. For completeness, we also incorporate Gaia DR3 distance estimates, composite effective temperature, and key astrometric measurements. Supplementary notes are provided where relevant, highlighting potential issues or interesting features. Finally, to ensure catalog consistency we have assigned each target a distinct TGV identifier, starting with TGV-199. To facilitate data analysis and further investigations, we have compiled the catalog's content into a machine-readable format, with the structure and organization highlighted in Table \ref{tbl:main_table}.

\begin{figure*}
    \centering
    \includegraphics[width=0.39\textwidth]{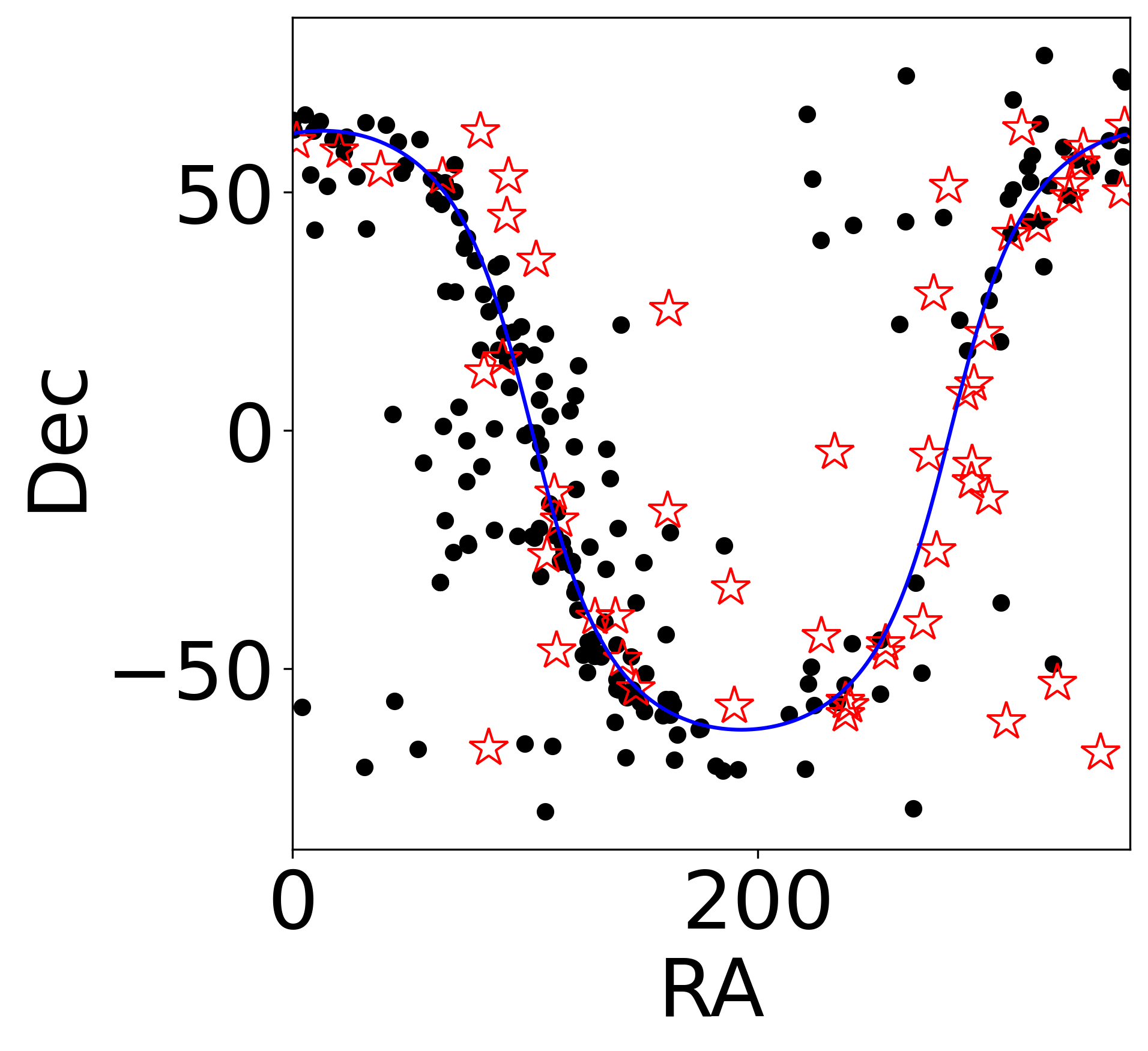}
    \includegraphics[width=0.39\textwidth]{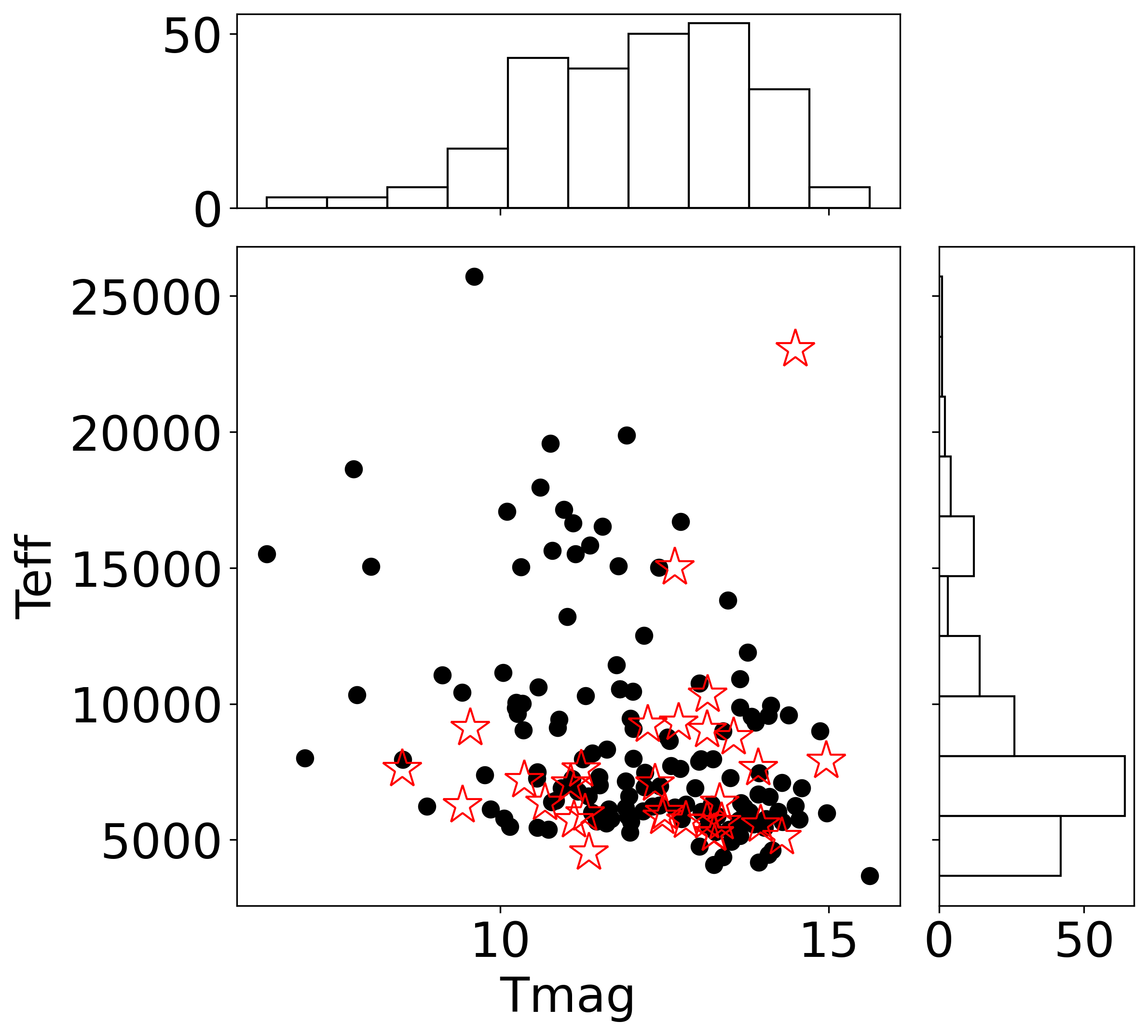}
    \includegraphics[width=0.8\textwidth]{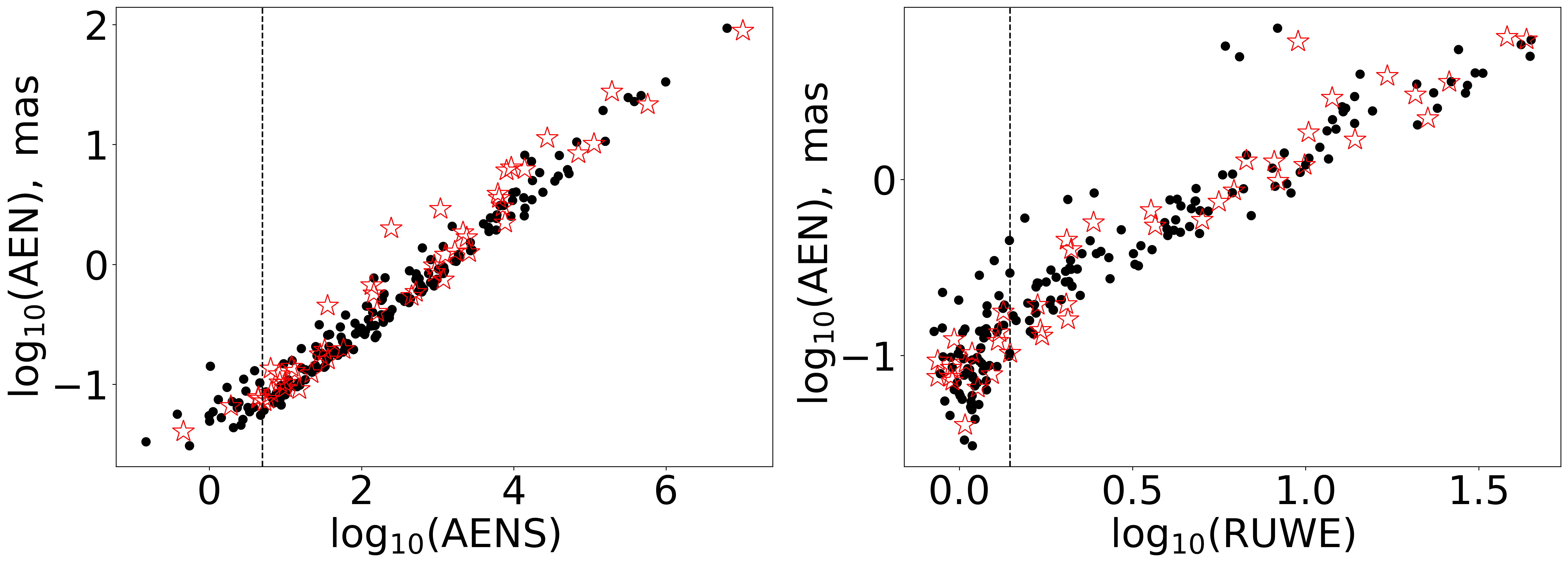}
    \caption{Upper left panel: Sky position in RA and Dec for the \Nquads quadruple candidates presented here (red stars). The black dots represent the other 198 targets in the TGV catalog, and the blue line represents the Galactic plane. Upper right: Corresponding Gaia DR3 composite effective temperature as a function of TESS magnitude. Lower panels: Gaia DR3 Astrometric Excess Noise (AEN), Astrometric Excess Noise Sig (AENS), and Renormalized Unit Weight Error (RUWE); AEN and AENS are in units of milliarcsec. Note the axes are in log10 base. The dashed vertical line in the right panel represents RUWE = 1.4, a potential indicator for unresolved companions.}
    \label{fig:basic_param}
\end{figure*}

Figure \ref{fig:basic_param} illustrates that most of the observed targets are close to the Galactic plane. This distribution aligns with the broader pattern observed in the TESS EB population (see Figure 9, \cite{2025ApJS..279...50K}) from which we are extracting these candidates. The \Nquads targets presented here encompass a range of TESS magnitudes, spanning from T = 8.5 mag (TIC 391461666) at the brightest end to T = 15 mag (TIC 359247237) for the faintest target. The average TESS magnitude is approximately 12.3, while the median is 12.7 mag. Thirty out of the \Nquads targets have effective temperatures provided by Gaia DR3. These range from 4,500 K for TIC 165052445 to the notable outlier of 23,000 K for TIC 277316707, with mean/median values of 7,700/6,700 K, respectively.

To assess the likelihood of the \Nquads candidates being genuine gravitationally-bound quadruples, we investigated several Gaia DR3 indicators commonly associated with potentially unresolved multiplicity: \verb|astrometric_excess_noise| (AEN), \verb|astrometric_excess_noise_sig| (AENS), renormalized unit weight error (RUWE) systems, and \verb|non_single_star| (NSS)\citep[e.g.][]{Belokurov2020,Penoyre2020,Stassun2021,Gandhi2022,2023MNRAS.523.2641R}. Overall, there are AEN/AENS/RUWE measurements for 49/49/46 targets, respectively. As highlighted in Figure \ref{fig:basic_param}, the vast majority of AENS values are enormous, with 47/45/28 targets showing AENS higher than 3/5/100, respectively. Approximately 40\% of the candidates demonstrate AEN values greater than 1 mas, with the top five targets having AEN in excess of 10 mas and AENS surpassing 70,000. The most dramatic case is TIC 139995365 -- a well-isolated target with hardly any contamination from nearby source -- where AEN and AENS are $\approx89$ mas and $\sim10$ million, respectively. 

In terms of RUWE, the values are greater than 1.4 -- a typically-considered threshold for the presence of unresolved companions \citep[e.g.,][]{Stassun2021} -- for 28 of the \Nquads targets. Additionally, four targets have non-zero NSS values: TIC 258507555 (NSS = 1), TIC 412074304 (NSS = 2), TIC 430752710 (NSS = 2), and TIC 466310009 (NSS = 2). The corresponding AEN, AENS, and RUWE are in the range of 0.1-1.3 mas, 22-2585, and 1-8.1, respectively; TIC 258507555 and TIC 466310009 are discussed in further detail below. These considerations suggest that a potentially significant fraction of the \Nquads quadruple candidates presented here may indeed be genuine quadruple systems of two EBs orbiting the common center of mass, and observed in motion thanks to Gaia's precise astrometric measurements.

\subsection{Period, Eclipse Depth and Duration Distributions}

The orbital periods of the individual components in a 2+2 quadruple star are one of the fundamental properties describing the system. These act as critical tracers of its formation history and key factors for predicting its evolution: for example, theoretical models of resonant capture in 2+2 quadruples indicate that while period ratios of 1:1 should be rare due to the inefficiency of the mechanism, period ratios of 3:2 and 2:1 are expected to be relatively common \citep{2018MNRAS.475.5215B,2020MNRAS.493.5583T}. Although determining such ratios is generally challenging, eclipsing quadruple systems represent ideal targets to do so, and indeed several recent studies have leveraged various photometric datasets to directly measure periods for hundreds of these systems (e.g., Zasche et al. 2019, 2022, 2024, 2025; K22, K24). 

\begin{figure*}
    \centering
    \includegraphics[width=0.9\textwidth]{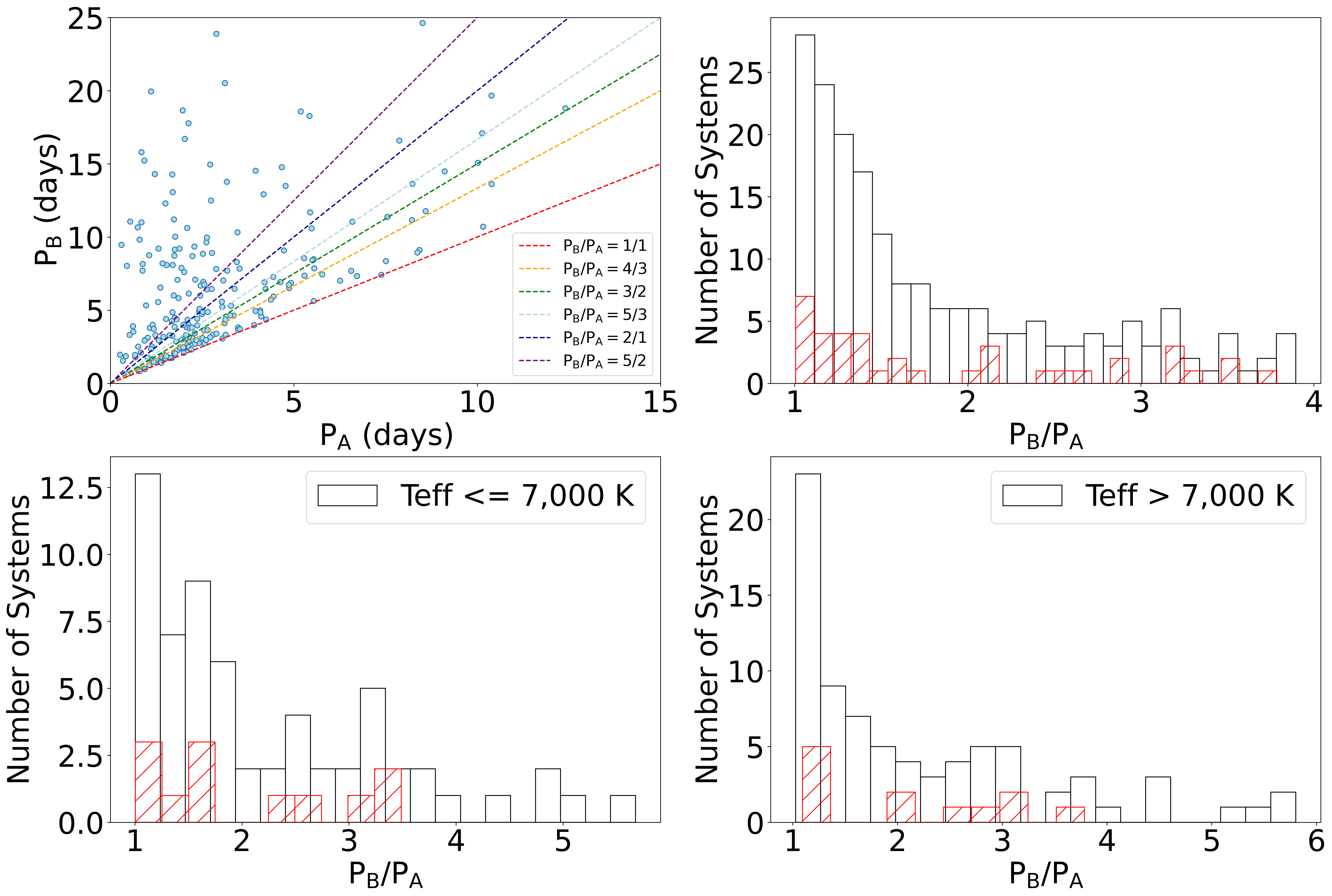}
    \caption{Upper left panel: Measured periods of the A and B binaries of all 250 quadruple candidates in the TGV catalog, compared to several integer ratios of PB/PA. Upper right panel: Corresponding period ratios PB/PA, such that PA $<$ PB. The black histogram represents all 250 targets in the TGV catalog, and the red histogram -- the \Nquads presented here. Lower panels: Same as upper right panel but separated for targets with Teff $< 7,000$ K (left) and Teff $> 7,000$ K (right). The panels are meant for comparison with Figure 16 from Zasche et al. (2023).}
    \label{fig:periods_}
\end{figure*}

\begin{figure*}
    \centering
    \includegraphics[width=0.9\textwidth]{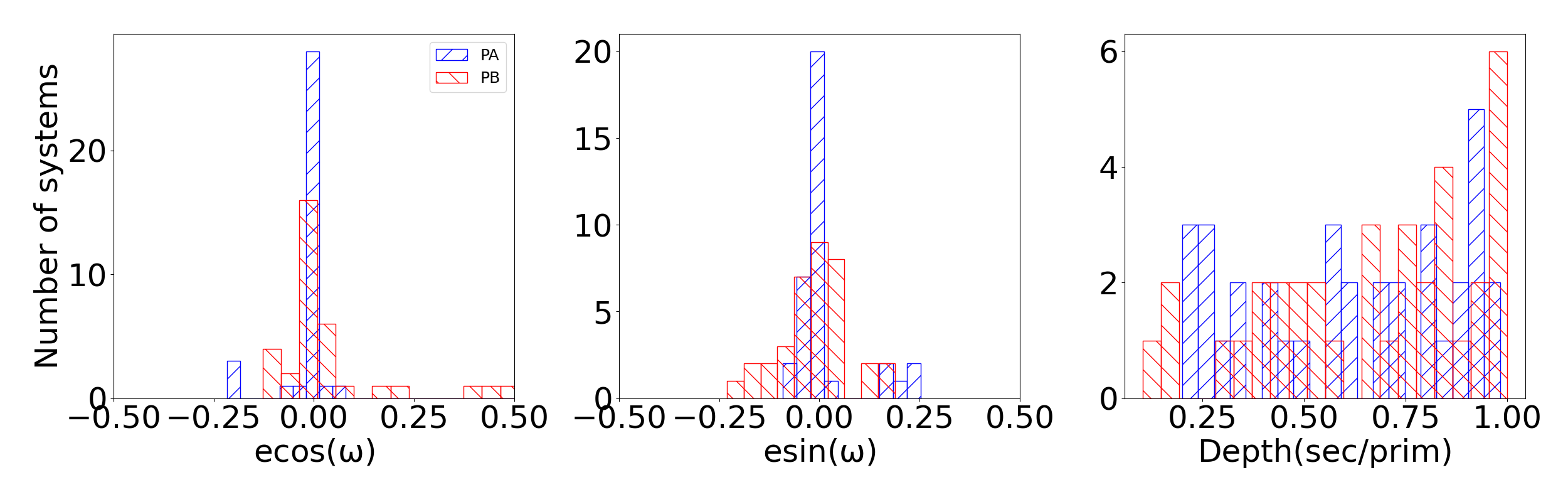}
    \includegraphics[width=0.9\textwidth]{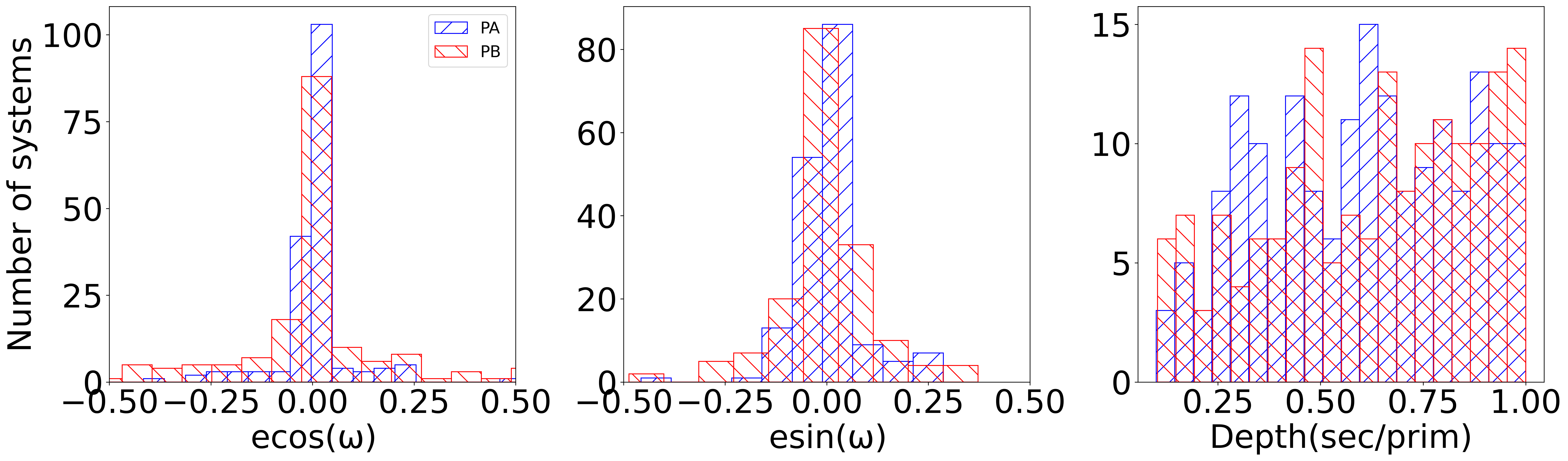}
    \caption{Upper left and middle panels: measured ${e \cos{\omega}}$ and ${e \sin{\omega}}$ for the \Nquads quadruple candidates presented here; Upper right panel: measured secondary-to-primary eclipse depths. The blue/red colors in the lower panels correspond to binary A/B, respectively. Lower panels: same as upper panels, but for all 250 TGV quadruple candidates.}
    \label{fig:periods_ALL}
\end{figure*}

In Figure \ref{fig:periods_} we show the measured periods PA and PB for each of the \Nquads systems presented here. Overall, these findings are in line with the results of K22 and K24. Specifically, granted the number of targets is relatively small, we find no exact 1:1, 3:2, and 2:1 period ratios. Only four systems are within 1\% of low-order integer ratios: TIC 229652559 (PB/PA = 1.005), TIC 122124665 (PB/PA = 1.51), TIC 20938739 (PB/PA = 1.249), and TIC 122994468 (PB/PA = 1.24); TIC 430752710 is within 5\% of 2:1 (PB/PA = 2.05). For the systems that produce significant secondary eclipses, we measure the corresponding depths and duration, and estimate the respective ${\rm e\cos(\omega)}$ and ${\rm e\sin(\omega)}$. These are illustrated in the upper panels of Figure \ref{fig:periods_ALL}, showing that most of the components EBs have nearly circular orbits, and the secondary versus primary depth ratios are mostly evenly distributed between 0 and 1. For completeness, in Figure \ref{fig:basic_param}, \ref{fig:periods_}, and \ref{fig:periods_ALL} we also show the basic parameters and orbital properties of all 250 TGV quadruple candidates. 

\subsection{Discussion}

The chance geometric orientation required for binary stars to produce eclipses make these relatively rare. Indeed, of the hundreds of millions of stellar binaries spread across the entire sky (some of which have been monitored for centuries), only a relatively small number -- several million -- have the appropriate orbital configuration to be confirmed as eclipsing thanks to observations from ASAS-SN, ATLAS, CoRoT, Gaia, Kepler, OGLE, TESS, ZTF \citep[e.g.,][and references therein]{2023A&A...674A..16M,2023MNRAS.523.2641R,2025ApJS..279...50K}. A tiny fraction of these -- less than a thousand at the time of writing -- have been found to be, in fact, eclipsing triple- and higher-order candidates for stellar multiples \citep[e.g.,][and references therein]{2022Galax..10....9B}. It is worth pointing out that this roughly speaking one in a hundred chance to see, e.g., tertiary eclipses in 2+1 triples or eclipses from both EBs in 2+2 quadruples is only a lower limit. The probability of detecting such systems is highly dependent on their overall architecture -- which is practically unknown a-priori -- and thus the fraction of EBs in non-eclipsing higher-order multiples is likely much higher \citep[e.g.,][]{Tokovinin2021}.

It is essential to clarify that the 250 targets listed in our full TGV catalog do not signify an exhaustive inventory of eclipsing 2+2 quadruples identified in TESS data. Rather, they represent our best effort to compile a sample of candidates that pass thorough scrutiny against false positives as possible. Naturally, it is certainly possible there are yet more candidates hidden in the data. However, based on our experience we would be surprised if these still-undiscovered stellar gems number in the hundreds as members of the VSG have already conducted visual inspections of millions of TESS EB lightcurves and identified thousands of potential candidates. Most of these, however, did not pass our vetting and validation criteria. Thus, while a comprehensive study of completeness and reliability is beyond the scope of this work, to facilitate future investigations we provide in Table \ref{tab:fps} a representative sample of $\sim500$ candidates for eclipsing quadruples that failed two common tests: (i) photocenter motion during eclipse; and (ii) resolved field star that cannot be definitively ruled out as an unrelated EB based on TESS data alone. 

\begin{table}[h]
    \centering
    \begin{tabular}{c|c}
        TIC & False Positive \\
        \hline
        2768366 & CO \\
        2775663 & CO \\
        2844449 & CO \\
        4254645 & FSCP \\
        5049897 & CO \\
        5092088 & CO \\
        5109750 & CO \\
        8698910 & FSCP \\
        10072325 & CO \\
        11469030 & CO \\
        11793277 & CO \\
        \hline
    \end{tabular}
    \caption{Example false positives mimicking two sets of on-target eclipses in TESS lightcurves. The full table is available as an on-line supplement. "CO" refers to "Centroid Offset", i.e., photocenter measurements show that one EB (or both) is off-target; "FSCP" refers to "Field Star in Central Pixel", i.e., there is a resolved field star that is bright enough to produce one EB (or both) as contamination, but is too close to the target star ($\lesssim 0.1-0.2$ pixels) for the photocenter measurements to pinpoint the origin of the eclipses.}
    \label{tab:fps}
\end{table}

\subsubsection{Interesting Systems}
\label{sec:individual_systems}

Below we list several systems exhibiting interesting features in addition to the two sets of detected eclipses. 

\begin{description}
\item[$\bullet$ TIC 48089827 (TGV-204)] TESS observed TIC 48089827 in 15 sectors (14, 15, 26, 40, 41, 53, 54, 55, 59, 75, 79, 80, 81, 82, and 86), providing an excellent baseline for long-term ETV measurements. The target produced two sets of eclipses with PA $\approx$ 4.09 days and PB $\approx$ 4.86 days, the latter exhibiting primary and secondary eclipses. Because the two periods are rather similar, the corresponding eclipses are often partially or fully blended (see Figure \ref{fig:48089827_lc}). As demonstrated in Figure \ref{fig:TIC_48089827_etvs}, the two EBs show dramatic, anti-correlated ETVs -- the signature effect of light travel time \citep[e.g.][]{1973A&AS...12....1F, 2016MNRAS.455.4136B}. This confirms TIC 48089827 as a gravitationally-bound eclipsing quadruple star. For illustrative purposes, we fit a simple model to the measured ETVs, and estimated an outer period of about 1,400 days. This is the longest confirmed outer period in the TGV catalog. 

\begin{figure*}
    \centering
    \includegraphics[width=0.8\textwidth]{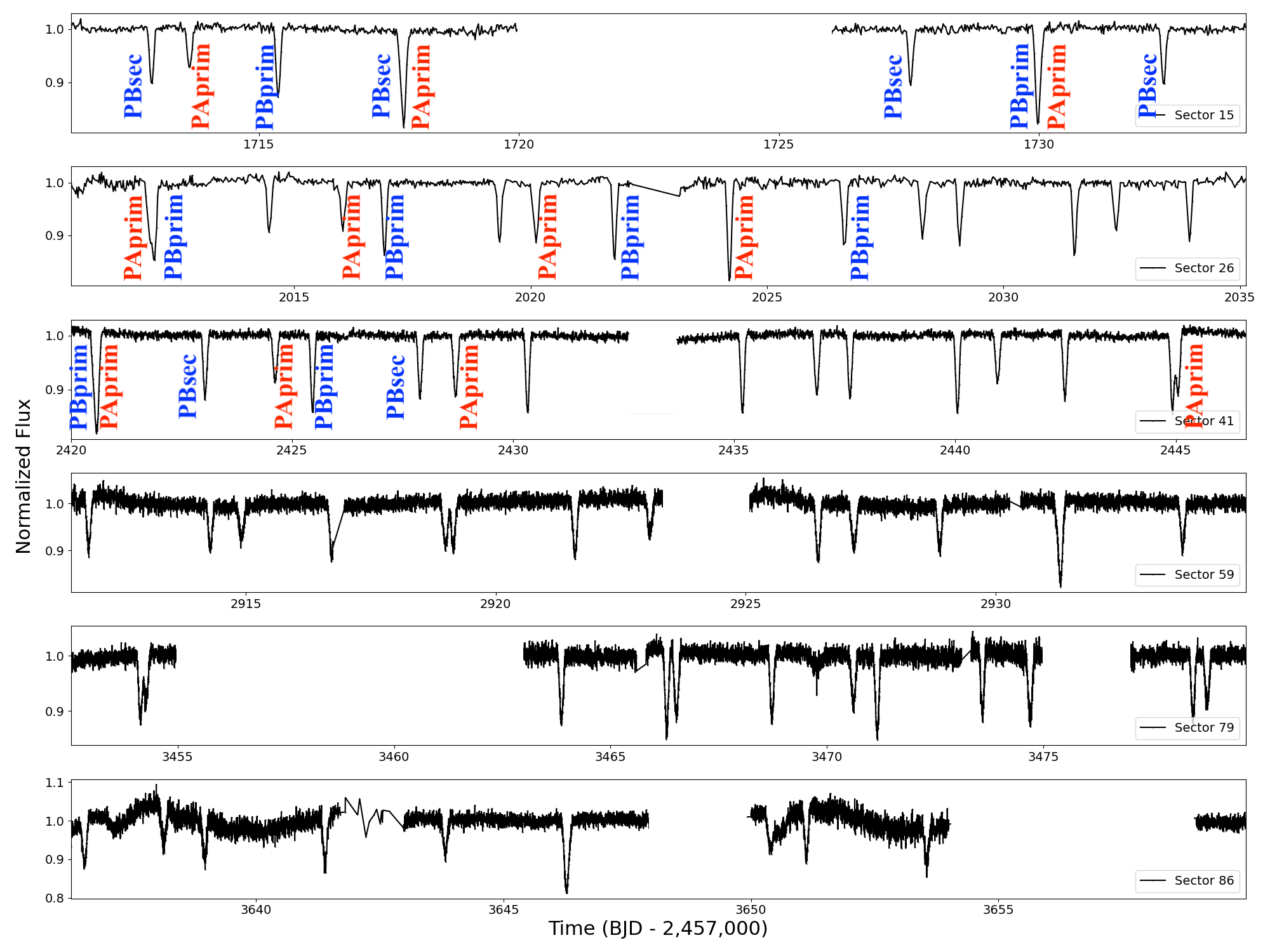}
    \caption{Same as Figure \ref{fig:64832327_lc} but illustrating 6 sectors of TESS data (out of 15 total) for TIC 48089827.}
    \label{fig:48089827_lc}
\end{figure*}

\item[$\bullet$ TIC 289822938 and TIC 352830705] TESS observed TIC 289822938 (TGV-233) in Sectors 9, 36, 62, 89, where the target produced two sets of eclipses with PA $\approx$ 0.92 days and PB $\approx$ 15.24 days. As illustrated in Figure \ref{fig:289822938}, the corresponding ETVs show clear anti-correlated modulations between the two primary eclipses. Like the case of TIC 48089827, this practically confirms the physically bound quadruple nature of TIC 289822938. 

TIC 352830705 (TGV-238) was observed in 13 sectors and produced two EBs with PA $\approx$ 3.44 days and PB $\approx$ 8.29 days. Like TIC 48089827 and TIC 289822938, the two EBs exhibit prominent anti-correlated ETVs consistent with a light travel time effect around the common center of mass, confirming the target as a genuine 2+2 quadruple system.

The ETV coverage for both of these targets is rather sparse, making robust measurements of the outer period difficult. For illustrative purposes, we fit a double sine model to TIC 352830705, and estimated a potential outer period of about 1,700 days.

\begin{figure*}
    \centering
    \includegraphics[width=0.8\textwidth]{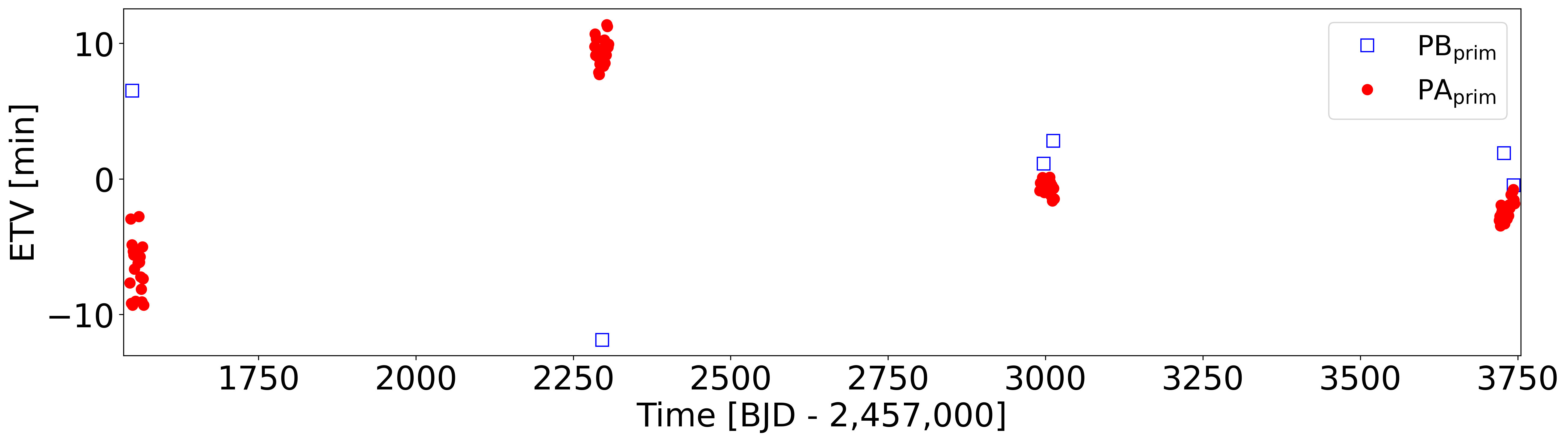}
    \includegraphics[width=0.8\textwidth]{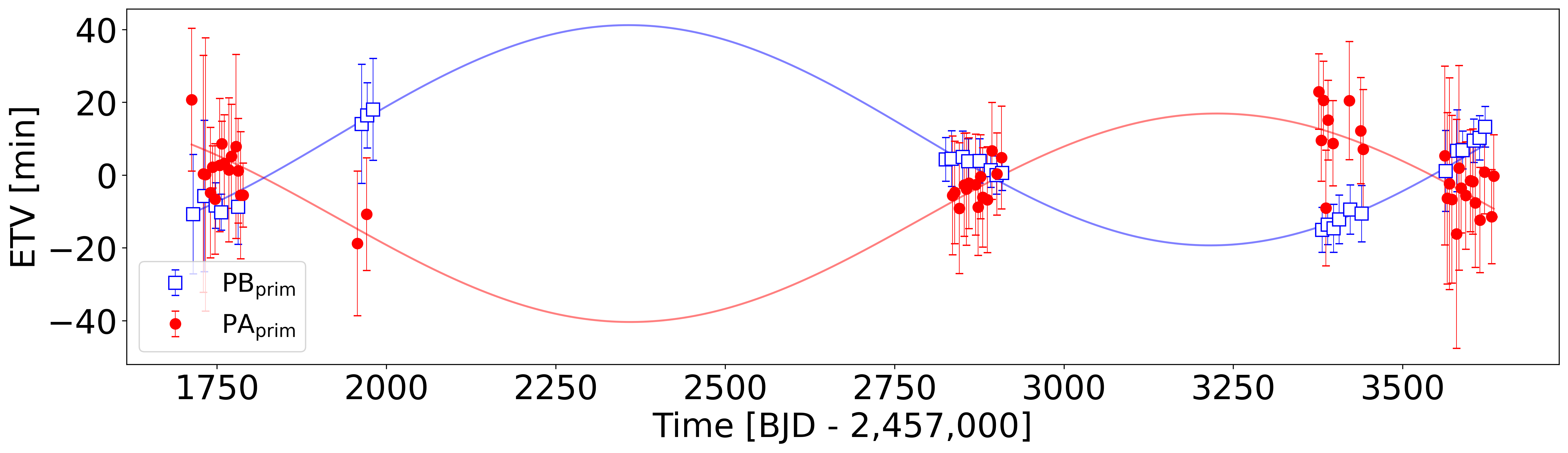}
    \caption{Same as Figure \ref{fig:TIC_48089827_etvs} but for TIC 289822938 (upper panel) and TIC 352830705 (lower panel). Both targets produce prominent anti-correlated ETVs, although the sparse coverage makes estimating the outer period difficult. For illustrative purposes, the lower panel shows a double sine fit to TIC 352830705 with a potential outer period of about 1,700 days.}
    \label{fig:289822938}
\end{figure*}
\item[$\bullet$ TIC 466310009 and TIC 258507555] 

    TESS observed TIC 466310009 (TGV-249) in Sectors 13, 27, 67, and 94, and produced two sets of eclipses with periods PA $\approx$ 7.87 days (deep) and PB $\approx$ 16.6 days (much shallower). As seen from Figure \ref{fig:466310009}, PA shows non-linear primary and secondary ETVs; the PB eclipses are too few for meaningful ETV measurements. Upcoming TESS observations in Sectors 101, 102, 103, and 104 will help further constrain the architecture of the system.

\begin{figure*}
    \centering
    \includegraphics[width=0.8\textwidth]{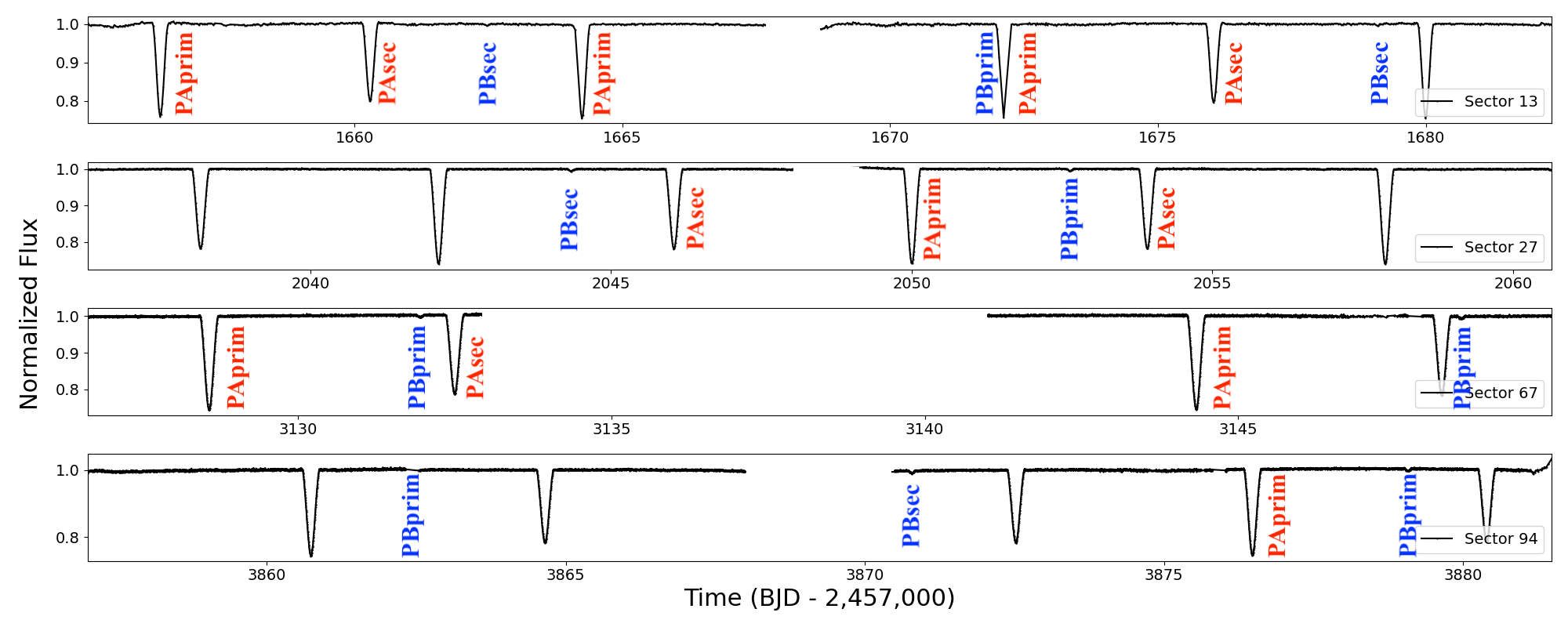}
    \includegraphics[width=0.8\textwidth]{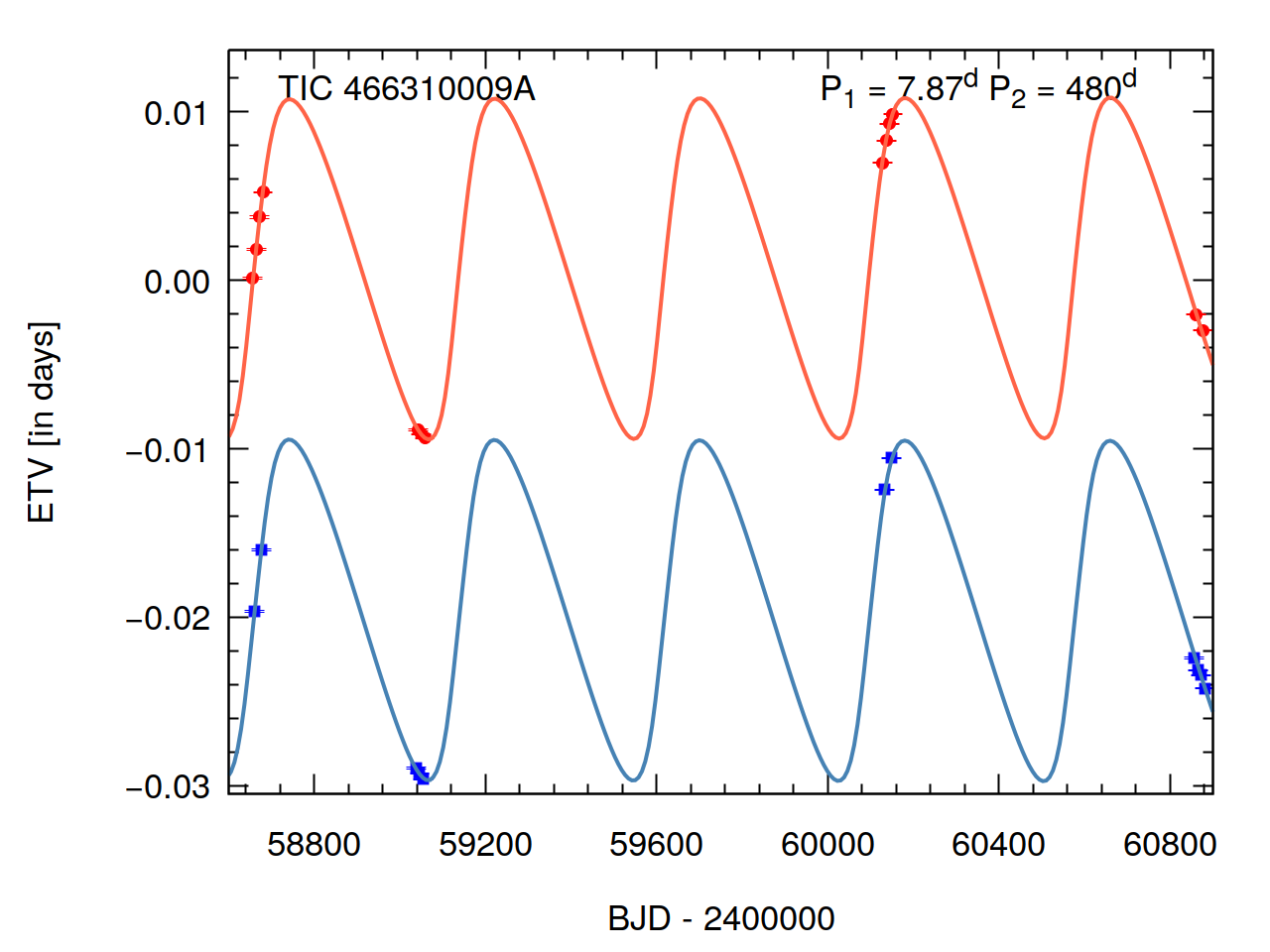}
    \caption{Upper panels: same as Figure \ref{fig:48089827_lc} but for TIC 466310009. The lower panel shows a preliminary analytic light travel-time effect and dynamical effect fit to the primary (red) and secondary (blue) eclipses of PA. The model indicates an outer period of about 480 days, consistent with Gaia. 
    }
    \label{fig:466310009}
\end{figure*}

    As illustrated in Figure \ref{fig:258507555}, TIC 258507555 (TGV-229) was observed in Sectors 19, 60, and 73, and produced two sets of eclipses with periods PA $\approx$ 0.87 days (shallower) and PB $\approx$ 7.7 days (much deeper). The target was close to the detector edge in Sector 19 and the data are rather poor; a substantial portion of Sector 73 is strongly affected by systematics as well. Overall, there are no indications for significant ETVs on PA (see Figure \ref{fig:258507555}, lower panel); PB has too few eclipses for meaningful measurements.

\begin{figure*}
    \centering
    \includegraphics[width=0.8\textwidth]{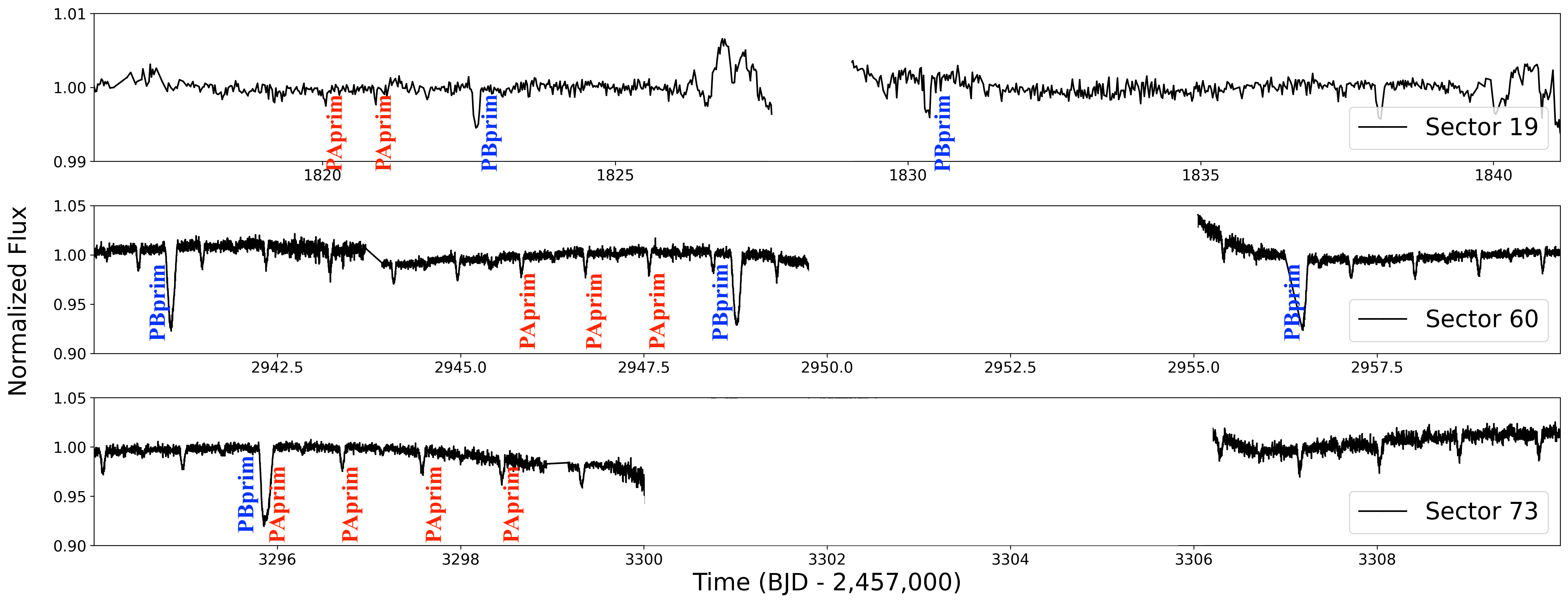}
    \includegraphics[width=0.8\textwidth]{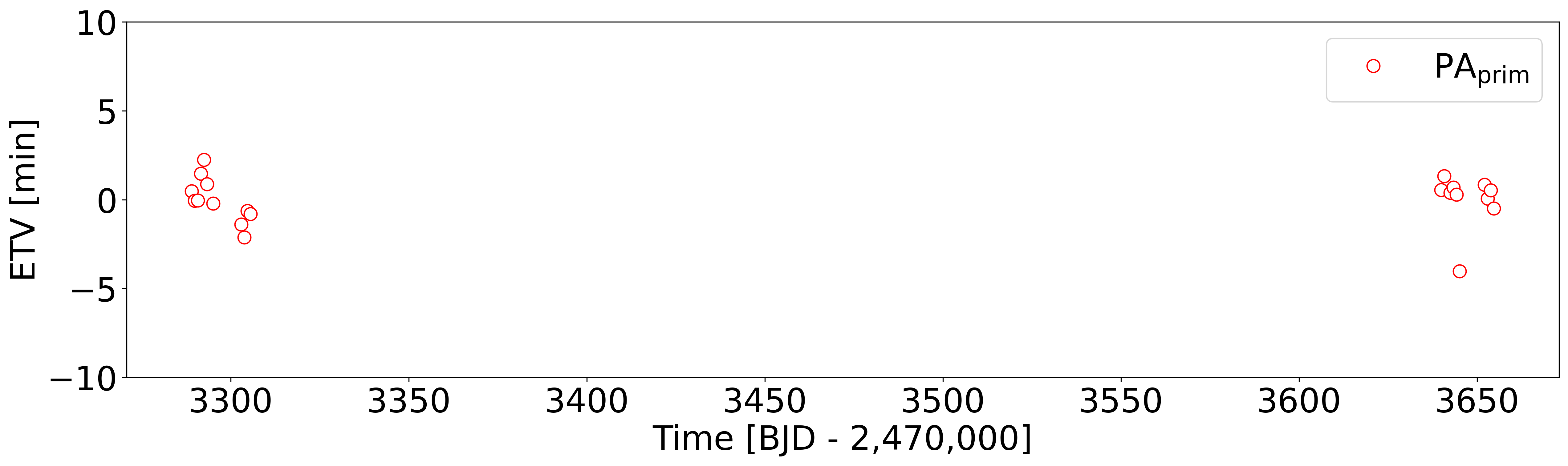}
    \caption{Upper panels: same as Figure \ref{fig:466310009} but for TIC 258507555. Large sections of the lightcurve are dominated by systematics, especially in Sector 19 when the target is close to the detector edge. Lower panel: same as Figure \ref{fig:TIC_48089827_etvs} but for TIC 258507555. There are no indications for significant ETVs on PA. Here, Gaia measured an outer period ${\rm P_{out} \approx277}$ days and eccentricity ${\rm e_{out} \approx0.1}$. The ETVs in TESS are too sparse to estimate the orbital period.}
    \label{fig:258507555}
\end{figure*}

Interestingly, TIC 258507555 and TIC 466310009 are not only flagged as non-single-stars by Gaia, but even have fully resolved astrometric outer orbits. This essentially confirms that both candidates are indeed gravitationally-bound quadruple systems with relatively compact outer orbits. TIC 258507555 has an outer period ${\rm P_{out} \approx277}$ days, relatively small outer eccentricity ${\rm e_{out} \approx0.1}$, and inclination of ${\rm i_{out} = 86.6\pm1.3}$ deg; the outer orbit for TIC 466310009 has a period of ${\rm P_{out} \approx474}$ days, an eccentricity of ${\rm e_{out} \approx0.5}$, and inclination of ${\rm i_{out} = 90.2\pm1.2}$ deg. While a comprehensive analysis of these two systems is beyond the scope of this work, the measured PA ETVs of TIC 466310009 are qualitatively consistent with Gaia's orbit as highlighted in Figure \ref{fig:466310009}\footnote{TIC 258507555 produced too few eclipses for meaningful ETV constraints on the outer orbit.}. As a test, we fitted a preliminary analytic model based on Roemer's delay (or light traveltime effect, LTTE) and dynamical effect (DE) \citep[e.g.,][]{2015MNRAS.448..946B, 2016MNRAS.455.4136B} to the PA primary and secondary ETVs, assuming a co-planar outer orbit. The model suggests that the DE is about three times larger than the LTTE, and indicates an outer period of about 480 days, in line with Gaia, as well as component masses of 3.5 MSun for binary A, and 1.0 MSun for binary B, respectively, albeit with considerable uncertainty.  
\item [$\bullet$ TIC 165052445 (TGV-225)] TIC 165052445 is a known EB observed by TESS in Sectors 20, 47, and 60. As illustrated in Figure \ref{fig:165052445}, the target produced two sets of eclipses with periods PA $\approx$ 1.49 days and PB $\approx$ 1.73 days, the latter identified as part of the HAT survey \citep{2011AJ....141..166H}. As mentioned in Table \ref{tbl:main_table}, both EBs are detected in ASAS-SN and ZTF data (also see discussion below), strengthening our interpretation of TIC 165052445 as a bona-fide quadruple system. Additionally, there is resolved nearby star, TIC 741891699, at a projected separation of about 3.8 arcsec, which is much too faint to produce either EB as contamination. Interestingly, TIC 165052445 and TIC 74189169 have similar enough parallax ($9.13\pm0.03$ mas vs $9.07\pm0.09$ mas) and proper motion (${\rm \mu_{RA}=61.38\pm0.03}$ mas/yr vs ${\rm \mu_{RA}=59.83\pm0.09}$ mas/yr, ${\rm \mu_{Dec}=-63.67\pm0.02}$ mas/yr vs ${\rm \mu_{Dec}=-61.46\pm0.09}$ mas/yr) that the two likely represent a wide co-moving quintuple system with a (2+2)+1 hierarchical configuration. 

\begin{figure*}
    \centering
    \includegraphics[width=0.8\textwidth]{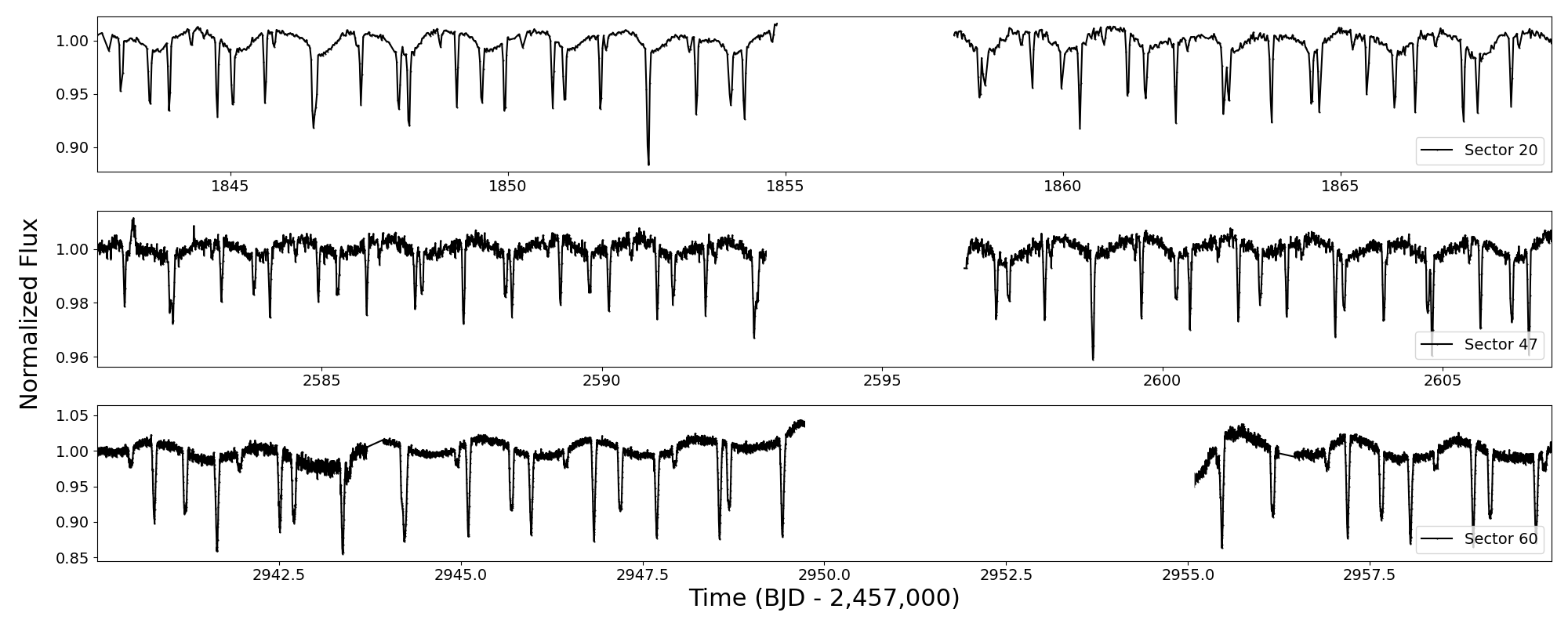}
    \caption{TESS FFI eleanor lightcurve of the 2+2 quadruple candidate TIC 165052445. A resolved nearby star (separation of about 3.8 arcsec), TIC 741891699, has similar parallax and proper motion to the target, suggesting the two may be a potentially co-moving wide quintuple. The apparent depth changes in both sets of eclipses are systematic in nature, due to contamination from TIC 165052447 ($\sim$3 pixels away and $\sim$2 magnitude brighter)}
    \label{fig:165052445}
\end{figure*}
\item[$\bullet$ TIC 277316707 (TGV-231)] \cite{IJspeert2021} reported TIC 277316707 as an eclipsing quadruple candidate with P1 = 1.93217 days and P2 = 6.965 days. We independently discovered the target, and during our analysis of the system noticed that P2 reported by \cite{IJspeert2021} is incorrect. As highlighted in Figure \ref{fig:277316707}, the correct value from TESS is P2 = 1.475034 days (corresponding to PA in this catalog). We confirmed this value with photometry from ASAS-SN, where both sets of eclipses are clearly visible in the phase-folded data. The correct P2 period is not a low-integer ratio of 6.965 ($6.965/1.475034\approx4.7$), suggesting that the latter is potentially a different signal. We searched for such a signal in the pixel-by-pixel lightcurves of a $11\times11$ TESS pixels image centered of the target but could not find evidence for it. Thus, we consider TIC 277316707 as a new quadruple candidate, and include it in the TGV catalog presented here. 

\begin{figure*}
    \centering
    \includegraphics[width=0.8\textwidth]{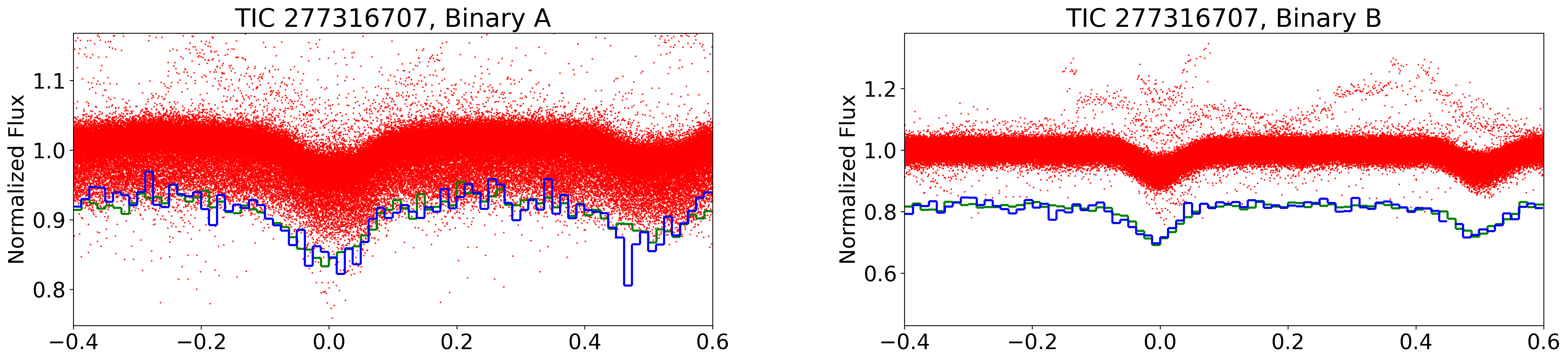}
    \includegraphics[width=0.8\textwidth]{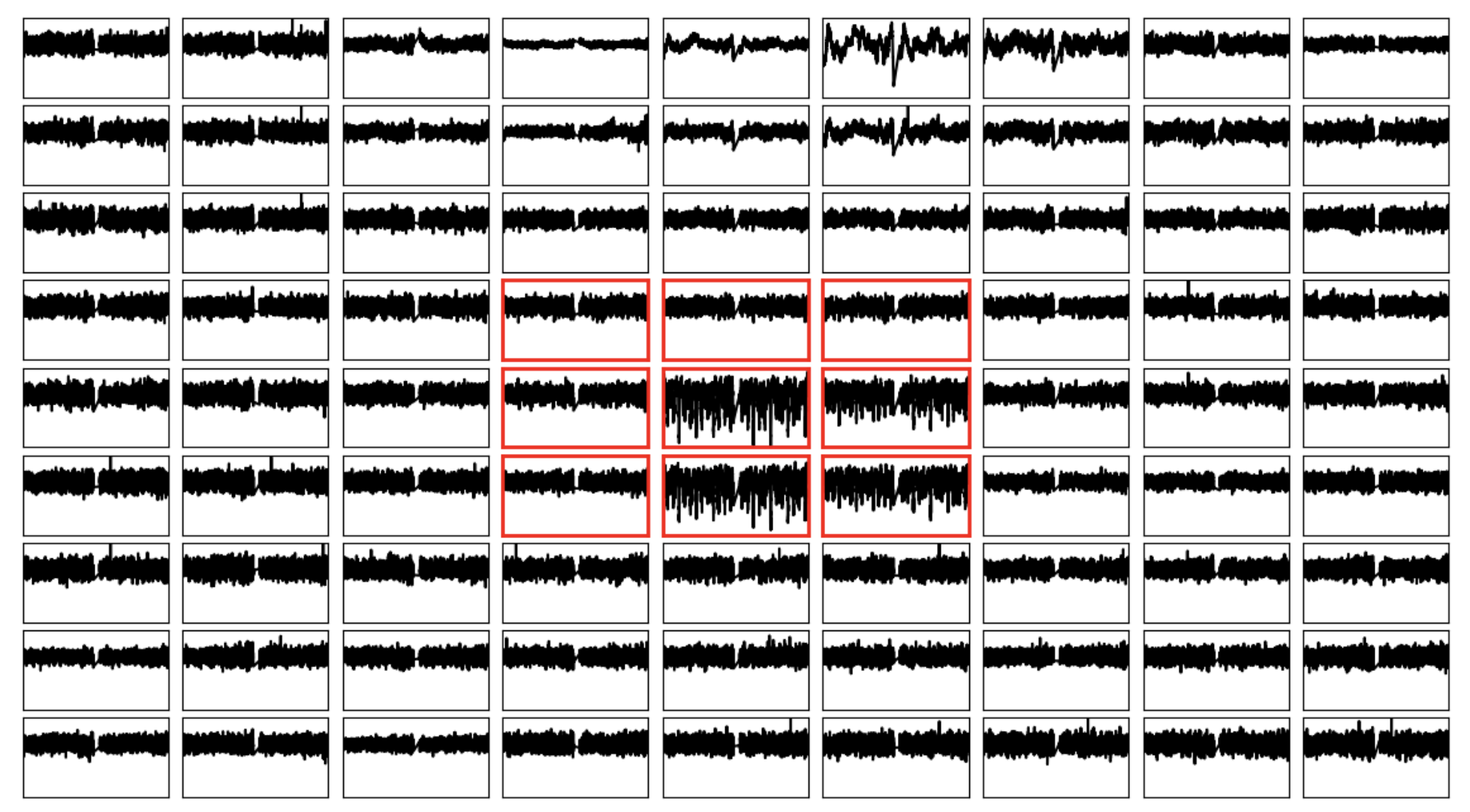}
    \caption{Upper panels: Phase-folded TESS (red) and ASAS-SN data (green and blue, for the corresponding g and V bands) for quadruple candidate TIC 277316707. We measured orbital periods for the two EB components of PA = 1.475034 days (left, compared to 6.965 days as reported by \cite{IJspeert2021}) and PB = 1.93217 days (right, consistent with \cite{IJspeert2021}). Lower panels: $11\times11$ pixel-by-pixel lightcurves centered of the target for Sector 27. The red contours represent the \textsc{eleanor} aperture used to extract the lightcurve. We do not see a 6.965 days signal in the pixels surrounding the target. }
    \label{fig:277316707}
\end{figure*}

\item[$\bullet$ Bright candidates with deep eclipses] New observations of the quadruple candidates presented here have can help confirm their nature and constrain the underlying orbital configuration. These observations can include obtaining new eclipse times, measuring radial velocities, and directly resolving the individual components through high-resolution imaging. To assist such potential investigations, we provide in Table \ref{tab:good_for_obs} a list of relatively bright systems (T $\leq 12$ mag) with eclipses deeper than 1\%.  

\end{description}

\subsection{Comparison with Archival Data}

In the context of eclipsing binary stars and transiting exoplanets it is not unreasonable to consider as `archival' observations with data that extend back in time on the order of decades (and more). Interestingly, TESS is already reaching the lower limit of this working definition, observing continuously for more than 7 years and still going strong \citep{2015JATIS...1a4003R}. And as demonstrated here and in other works, the synergy of ever-increasing baseline and high-precision photometry provides sensitivity to years-long period changes in eclipsing multiple stellar systems \citep[e.g.,][]{2025A&A...695A.209B}. To investigate even longer-term timescales, we searched the ASAS-SN SkyPatrol\footnote{\url{http://asas-sn.ifa.hawaii.edu/documentation/index.html}}, \textsc{DASCH}\footnote{\url{https://daschlab.readthedocs.io/en/latest/}}, and ZTF\footnote{\url{http://atua.caltech.edu/ZTF/Zubercal.html}} databases. 

Before discussing the results from this search, we would like to note two important considerations. Specifically, the ASAS-SN and ZTF pixel scales (8 arcsec and 1 arcsec, respectively) are much smaller than that of TESS (21 arcsec), and so is the potential contamination from nearby sources. Thus, detecting two sets of eclipses for a particular target in either of the ground-based datasets dramatically strengthens the quadruple interpretation of the system. With that said, the observational cadence of the ASAS-SN and ZTF photometry is relatively low compared to TESS, and it is not unlikely for the first two to miss some (and potentially even all) eclipses seen in the latter. Therefore, not detecting these in the archival data does not rule out the quadruple candidate. 

\begin{figure*}
    \centering
    \includegraphics[width=0.38\textwidth]{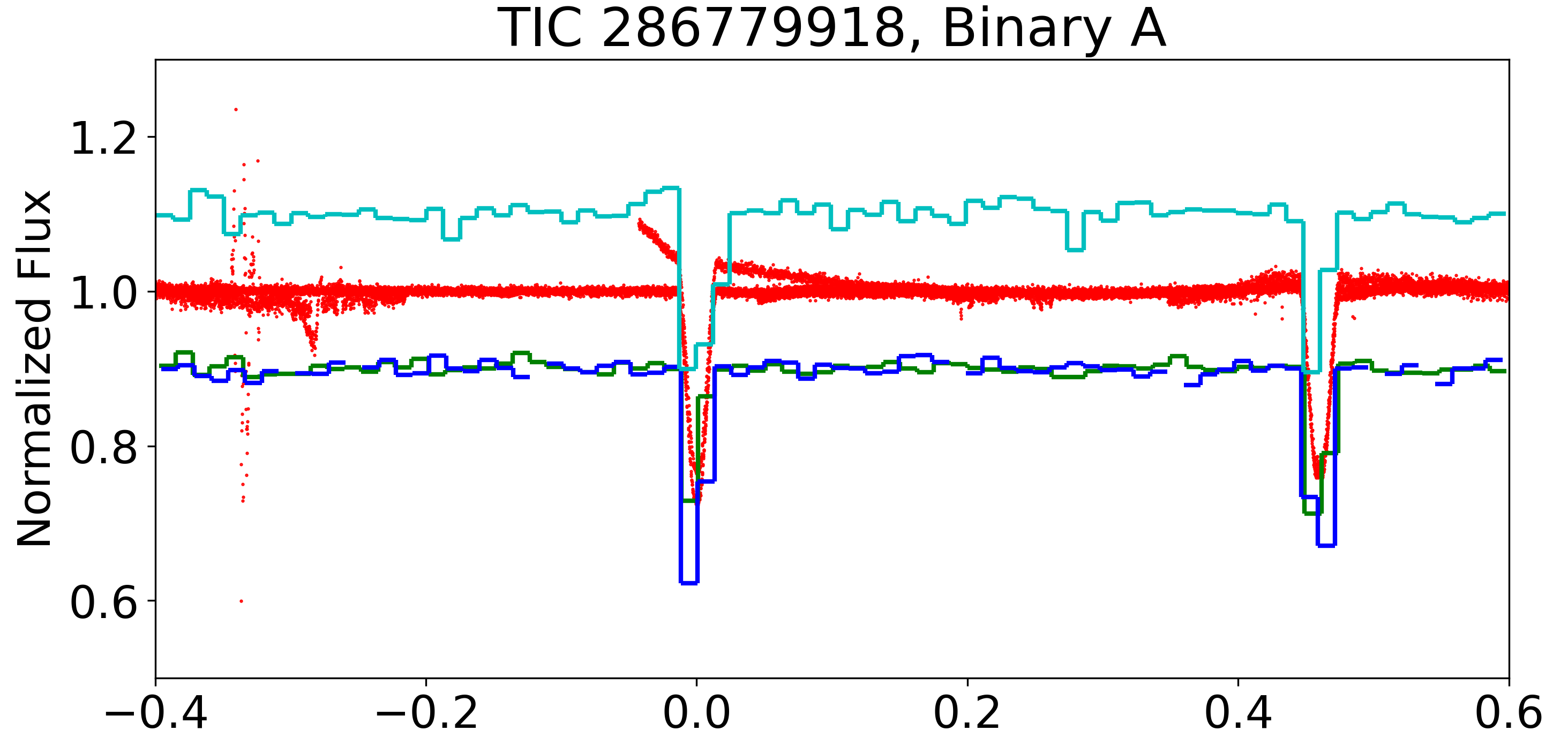}
    \includegraphics[width=0.38\textwidth]{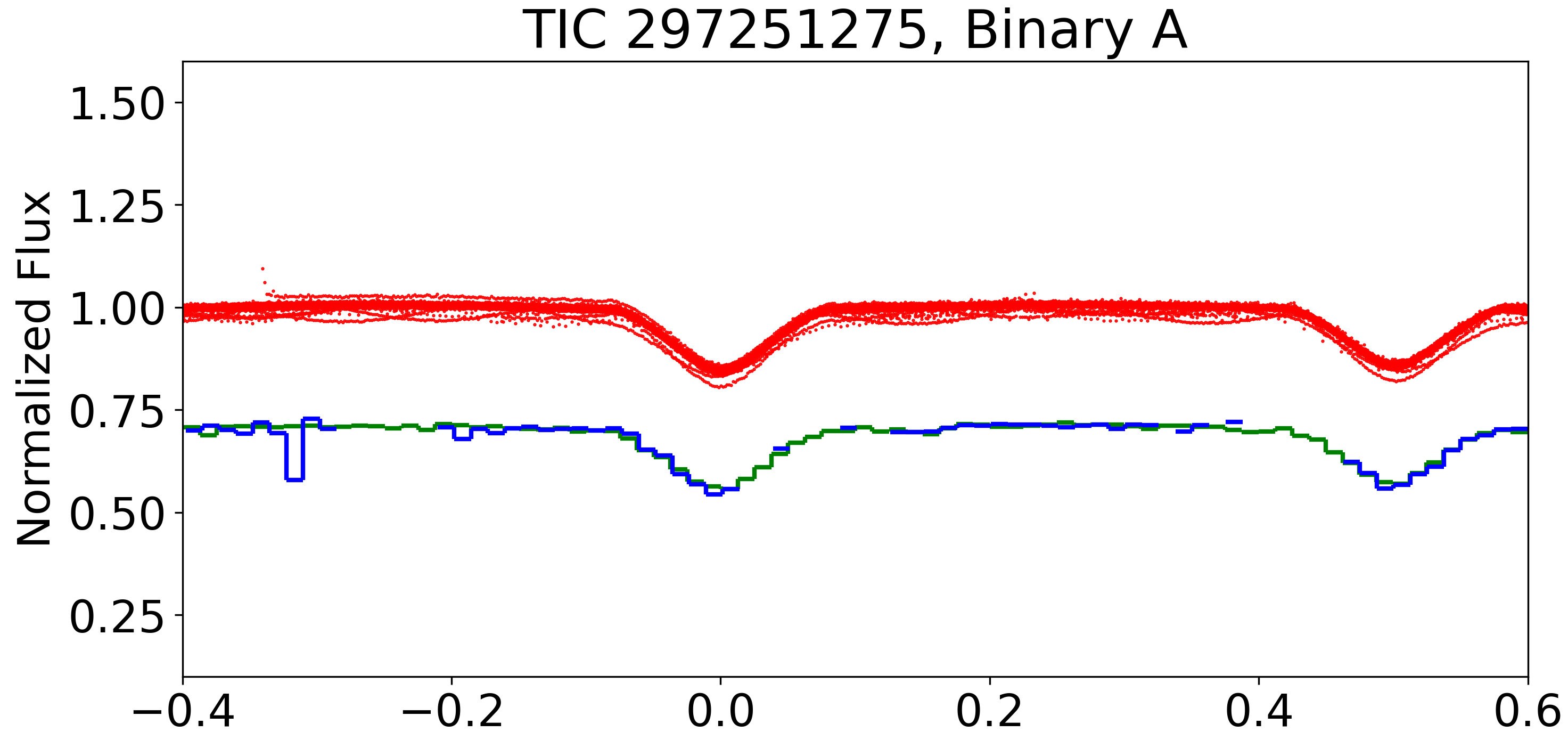}
    \includegraphics[width=0.8\textwidth]{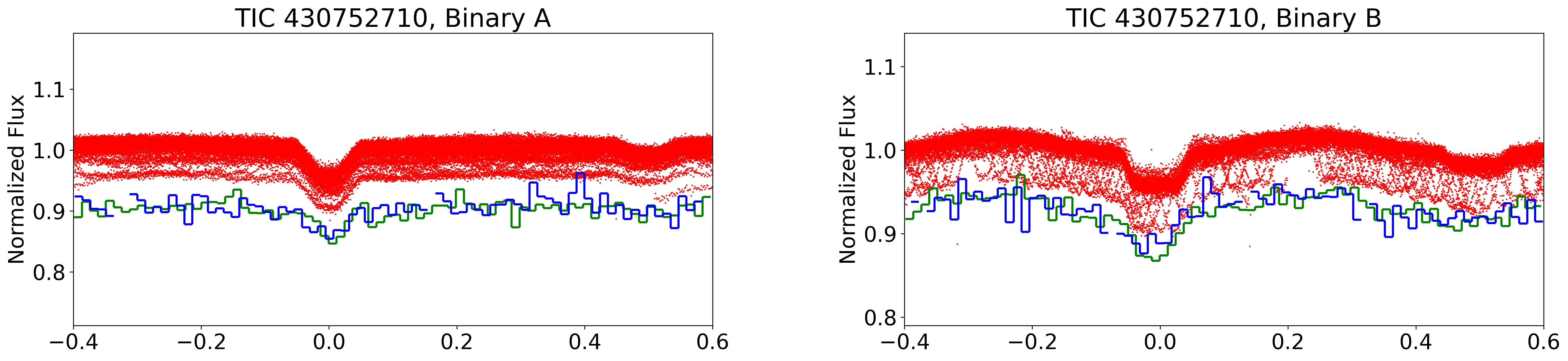}
    \includegraphics[width=0.8\textwidth]{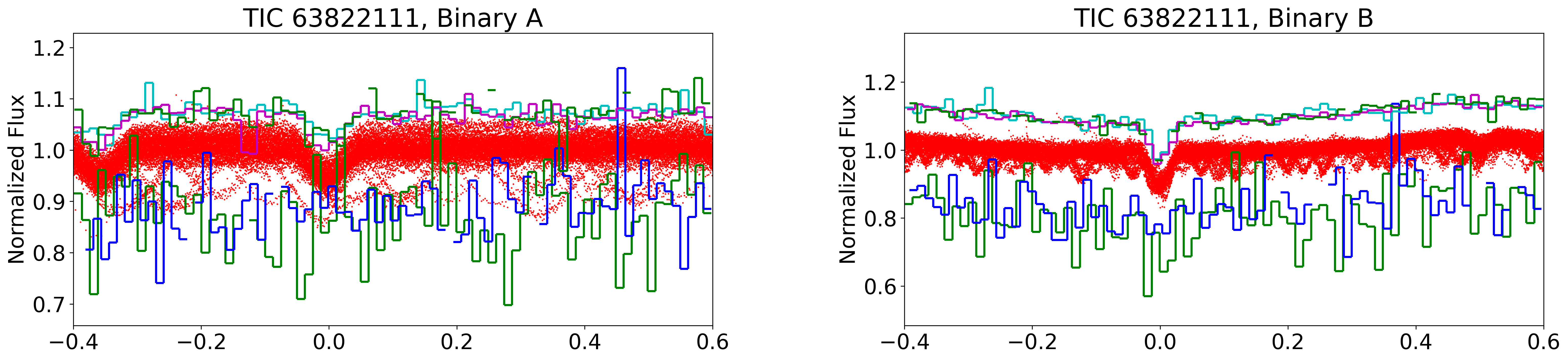}
    \includegraphics[width=0.8\textwidth]{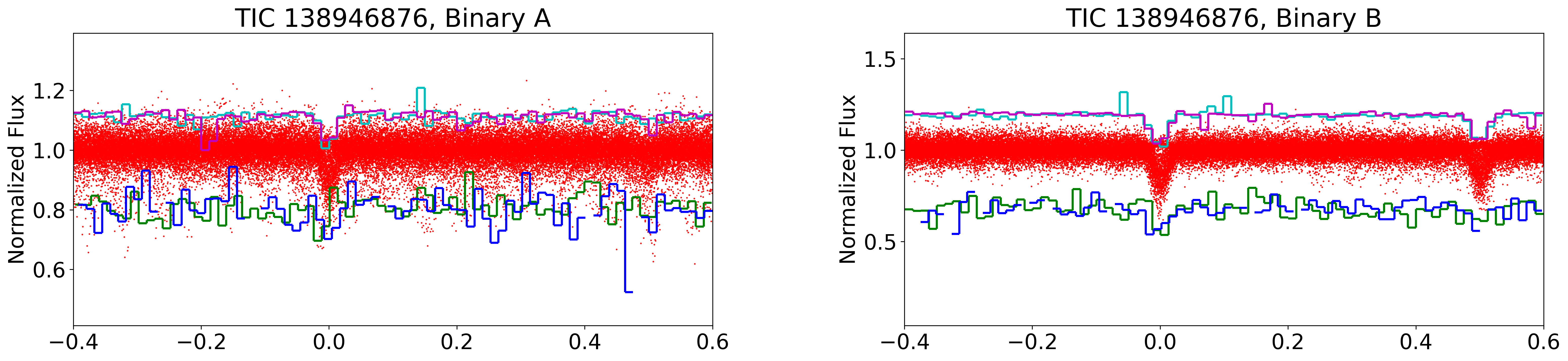}
    \caption{Comparison between TESS (red), ASAS-SN (green and blue), and ZTF (cyan, magenta, and green, for the zg, zr, and zi bands, respectively) photometry for the several quadruple candidates. The ground-based photometry is binned for clarity. First row: both primary and secondary eclipses seen for one component of the respective quadruple candidate. Second row: both primary and secondary seen for both components. Third row: Both EBs are present in ZTF but neither in ASAS-SN. Last row: Both EBs present in ZTF but only one or the other in ASAS-SN.}
    \label{fig:asas_sn}
\end{figure*}

Our cross-match query returned lightcurves for 46, 54 and 25 targets from ASAS-SN, DASCH, and ZTF, respectively. Of these, we found that at least one of the component EBs is unambiguously detected in the phase-folded ASAS-SN data in 39 out of the 46 lightcurves, and both EBs can be seen in 16 lightcurves. ZTF detects at least one of the EBs in 21 out of the 25 lightcurves, and both EBs in 14 out of the 25 lightcurves. We note that some of these eclipses are present in one of the datasets but not in the other, such that at least one EB is present in the ground-based photometry for 46 unique targets. Figure \ref{fig:asas_sn} highlights the successful recovery of (i) the primary and secondary eclipses for binary A of TIC 286779918 and TIC 297251275; (ii) both EBs of TIC 430752710; (iii) both EBs of TIC 63822111 in ZTF but neither in ASAS-SN; (iv) both EBs of TIC 138946876 in ZTF but only one of them in ASAS-SN. Overall, we find consistency between the periods in all data sets to within the statistical uncertainties.

The eclipses of 6 targets are quite clear in DASCH data: TIC 79062805 (binary B), TIC 139995365 (binary B), TIC 286779918 (binary A), TIC 297251275 (binary A), TIC 391461666 (binary A), and TIC 466310009 (binary A). Overall, all six EBs seem to be keeping pace between TESS and DASCH, showing no indications of dramatic apsidal motion dramatic or eclipse depth variations on century-long timescales. The latter is in line with the nearly edge-on outer orbit of TIC 466310009, where despite the short outer period, high eccentricity, and strong dynamical interactions between the two components, binary A continues to produce eclipses for decades. 

\begin{figure*}
    \centering
    \includegraphics[width=0.38\textwidth]{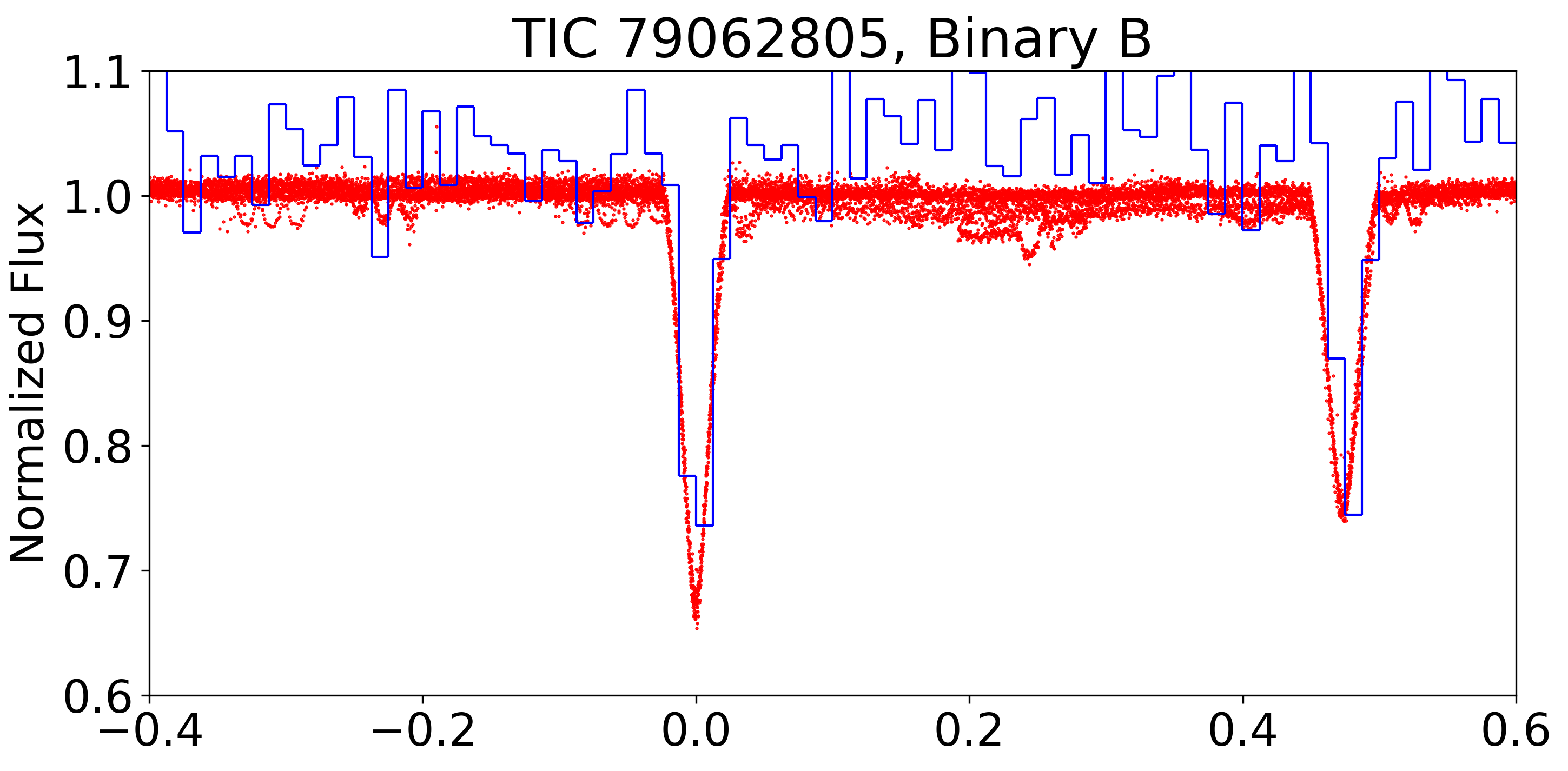}
    \includegraphics[width=0.38\textwidth]{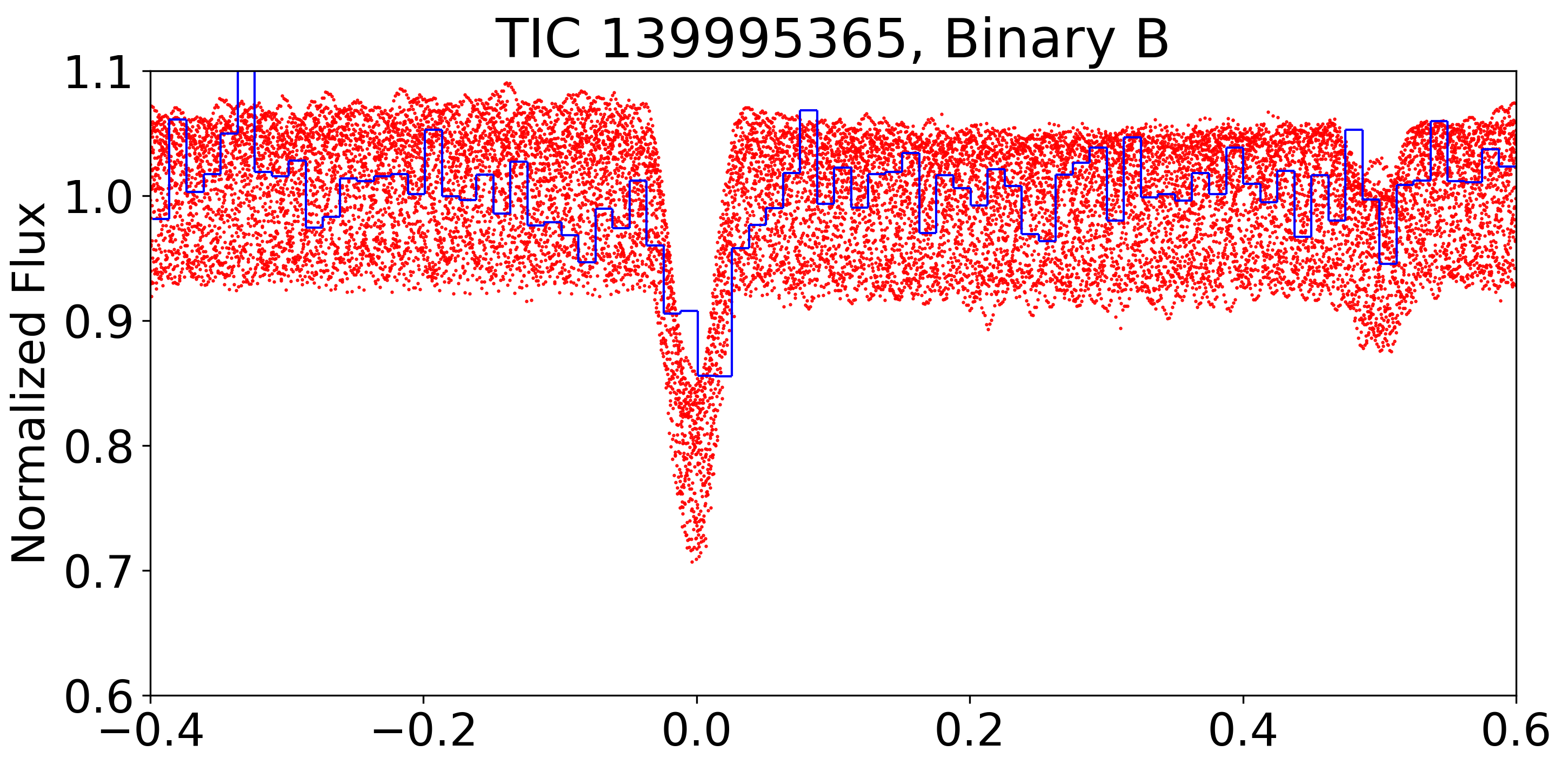}
    \includegraphics[width=0.38\textwidth]{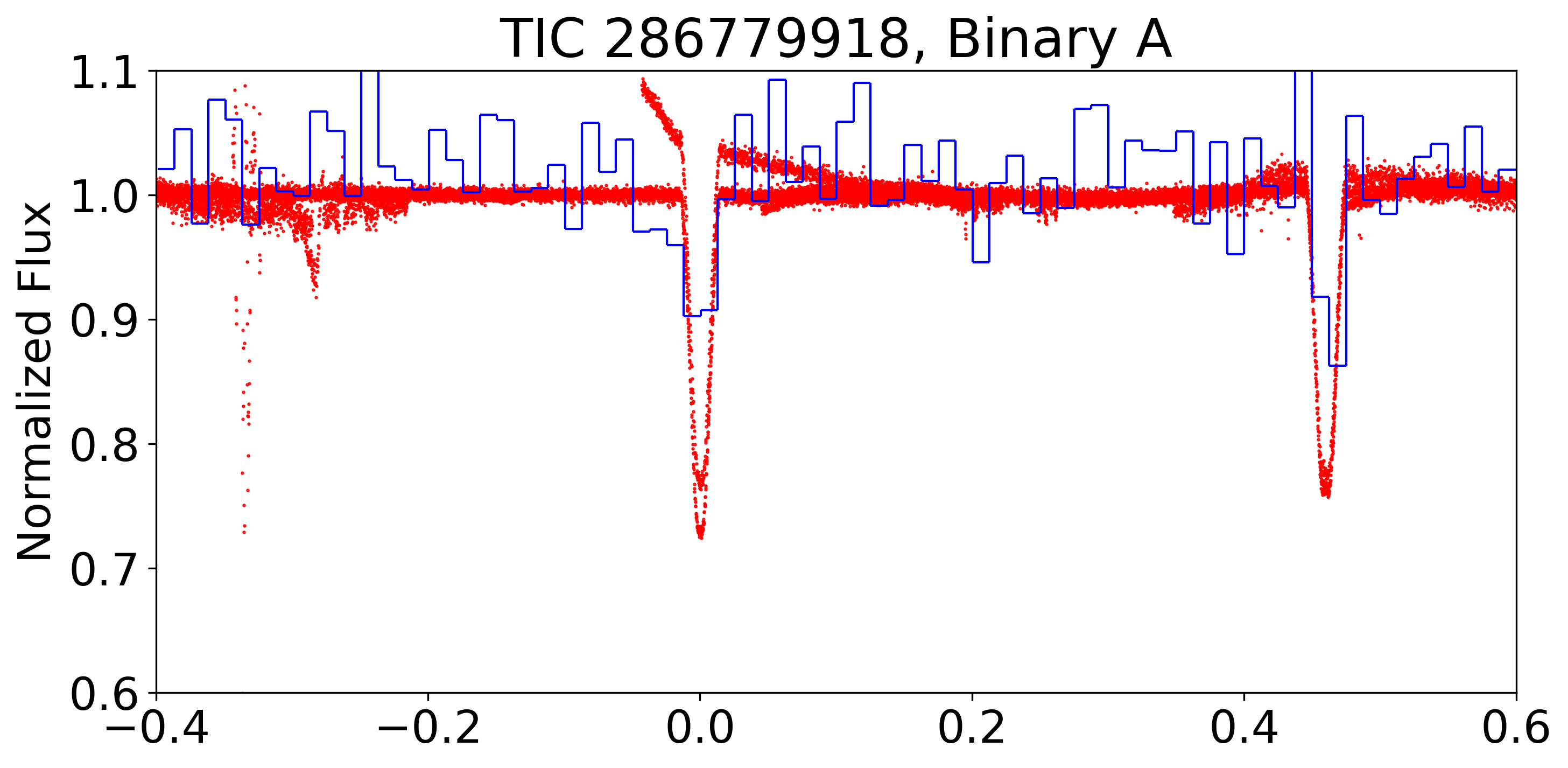}
    \includegraphics[width=0.38\textwidth]{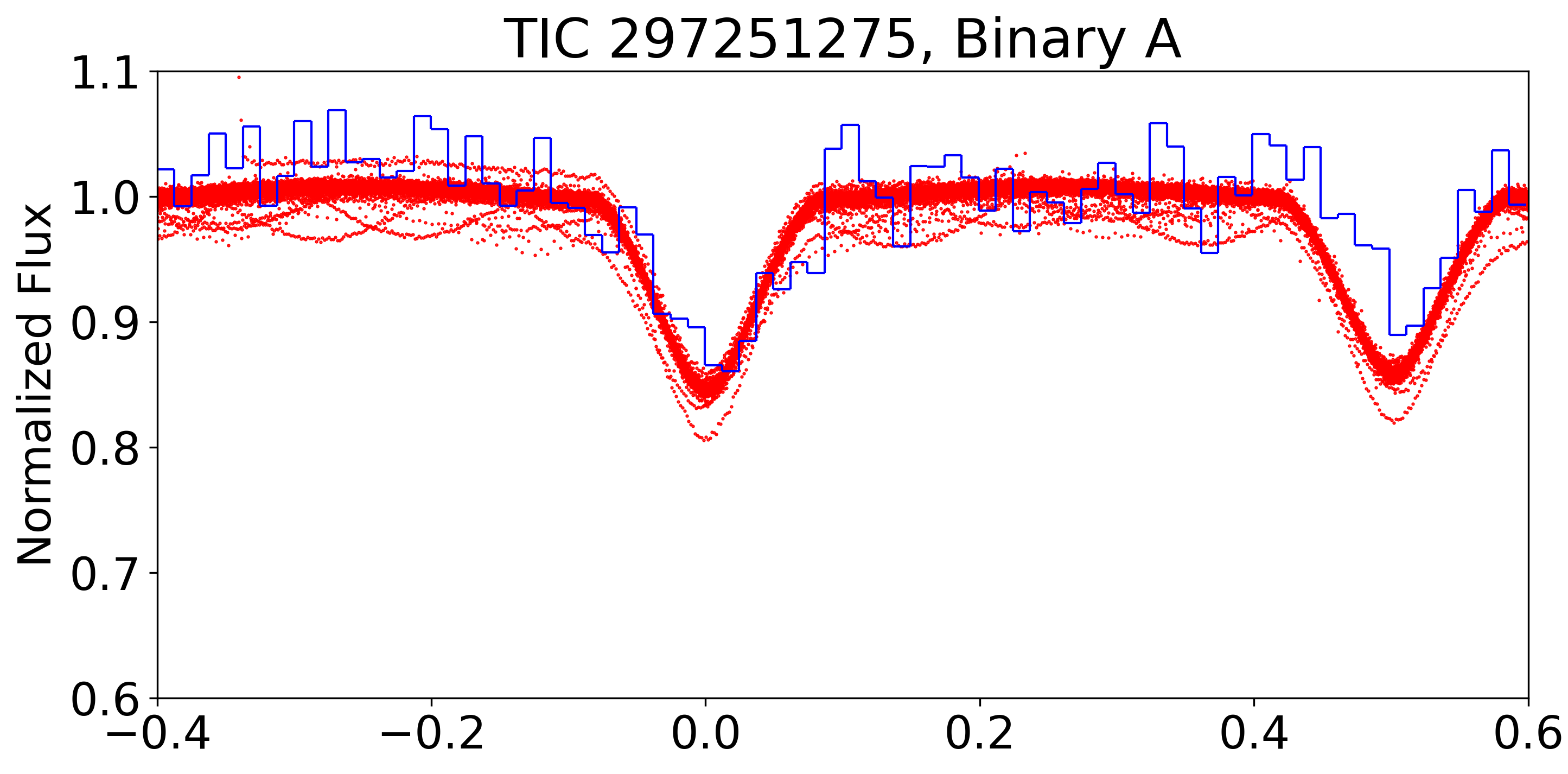}
    \includegraphics[width=0.38\textwidth]{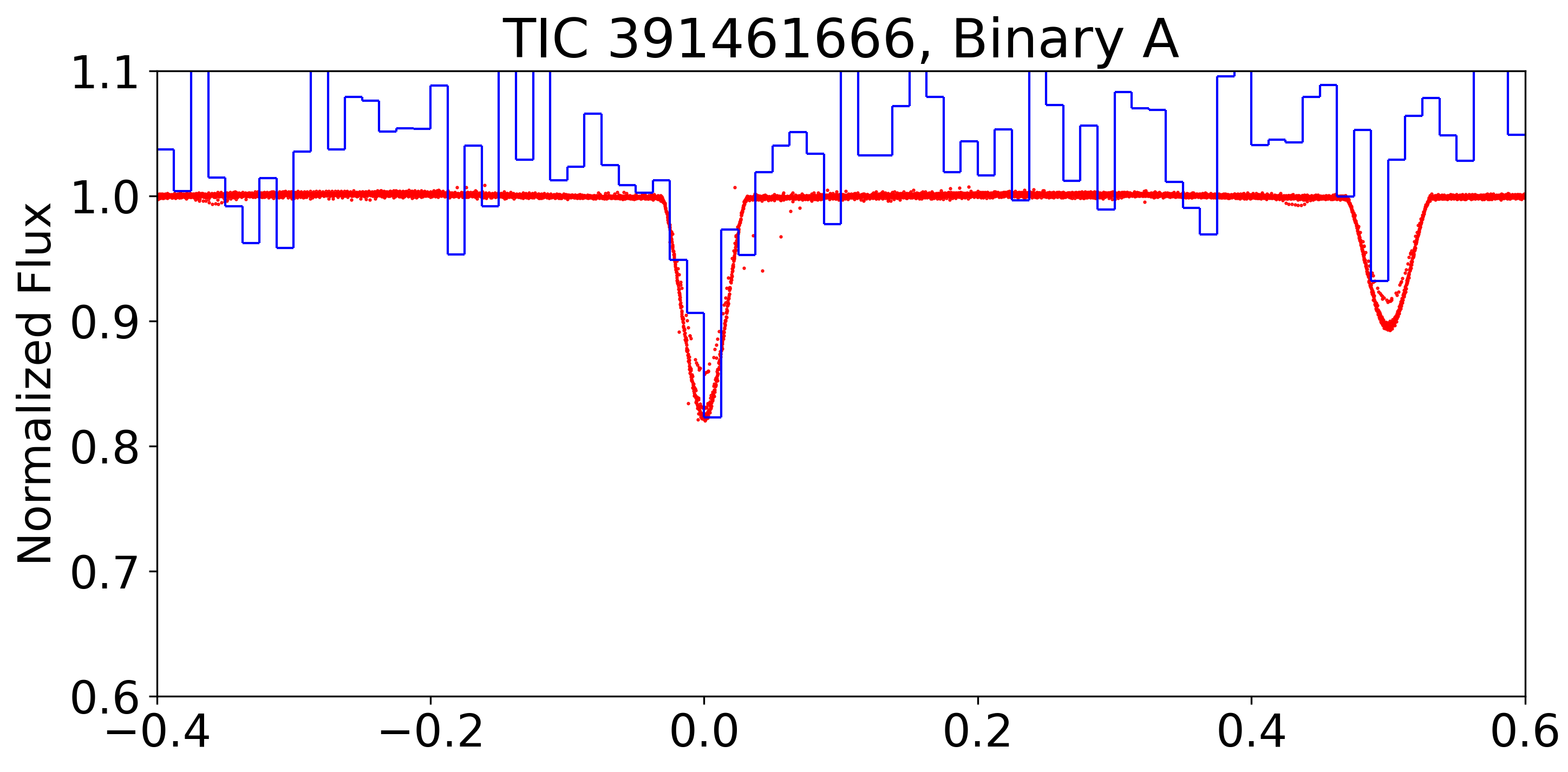}
    \includegraphics[width=0.38\textwidth]{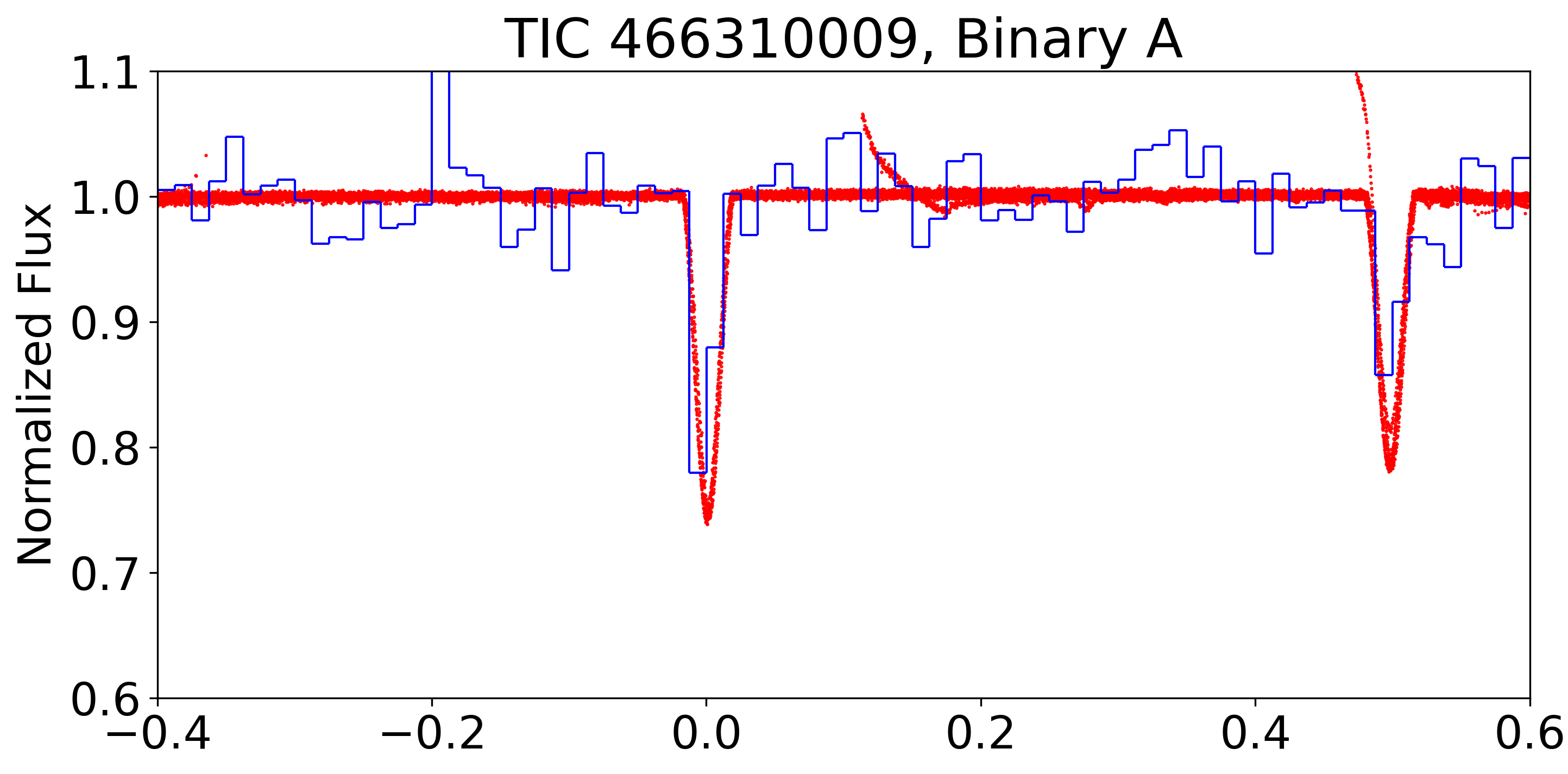}
    \caption{Same as Figure \ref{fig:asas_sn} but showing a comparison between TESS (red) and DASCH (blue) for the 6 targets where the TESS eclipses are seen in DASCH. There are no indications of dramatic depth changes or apsidal motion of century-long timescales.}
    \label{fig:placeholder}
\end{figure*}

Finally, for completeness we also cross-matched our candidates against APOGEE spectroscopic binaries listed in \cite{2021AJ....162..184K}. None of our targets appears in the catalog as either SB3 or SB4.

\section{Summary}
\label{sec:summary}

We have presented the second expansion of our TESS/GSFC/VSG (TGV) catalog, adding \Nquads new candidates for eclipsing quadruple star systems. The results from this work bring the total TGV tally to 250, and represent our efforts to provide to the community a curated set of uniformly-vetted-, -validated, and -characterized hierarchical 2+2 systems discovered in TESS FFI data by sophisticated machine learning and highly-trained volunteers. All targets passed extensive scrutiny, including photocenter measurements, ephemerides determination, and cross-match with available ground-based photometry. Each candidate exhibits two distinct sets of eclipses, all originating within $\sim0.1-0.2$ pixels ($\sim2-4$ arcsec) of the target; there are no resolved stars within said separation that are bright enough to be the source of either set. For each target, the catalog provides identifying information, measured ephemerides, eclipse depths and durations, as well as relevant notes and comments. Several targets display pronounced eclipse timing variations (ETVs), indicating dynamical interactions between the two eclipsing binary components; the ETVs for one system complete a $\sim 1,400$-days outer orbit during TESS observations. Many of the candidates exhibit notable astrometric excess noise and high renormalized unit weight error in Gaia measurements, suggesting detectable astrometric motion in a wider quadruple configuration. Two targets, TIC 258507555 and TIC 466310009, have complete orbital solutions from Gaia for the outer orbits. The former has an outer period of about 278 days, placing it among the top 5 most compact quadruples, and the latter is nearly edge-on with an outer inclination of $\approx90.2$ degrees.

\clearpage
\acknowledgments
This paper includes data collected by the \emph{TESS} mission, which are publicly available from the Mikulski Archive for Space Telescopes (MAST). Funding for the \emph{TESS} mission is provided by NASA's Science Mission directorate.

Resources supporting this work were provided by the NASA High-End Computing (HEC) Program through the NASA Center for Climate Simulation (NCCS) at Goddard Space Flight Center.  

This research has made use of the Exoplanet Follow-up Observation Program website, which is operated by the California Institute of Technology, under contract with the National Aeronautics and Space Administration under the Exoplanet Exploration Program. 

V.\,B.\,K., J.\,O.\, and W.\,W. are grateful for financial support from NSF grant AST-2206814. V.\,B.\,K. acknowledges support from NASA grant 80NSSC23K0270. TB acknowledges the financial support of the Hungarian National Research, Development and Innovation Office -- NKFIH Grant K-147131. TB acknowledges the funding from the HUN-REN Hungarian Research Network. GH thanks the Polish National Center for Science (NCN) for financial support through grant 2021/43/B/ST9/02972

This work has made use of data from the European Space Agency (ESA) mission {\it Gaia} (\url{https://www.cosmos.esa.int/gaia}), processed by the {\it Gaia} Data Processing and Analysis Consortium (DPAC, \url{https://www.cosmos.esa.int/web/gaia/dpac/consortium}). Funding for the DPAC has been provided by national institutions, in particular the institutions participating in the {\it Gaia} Multilateral Agreement.

\facilities{
\emph{Gaia},
MAST,
TESS,
ASAS-SN,
TRES,
ZTF
}

\software{
{\tt Astropy} \citep{astropy:2013, astropy:2018, astropy:2022}, 
{\tt Eleanor} \citep{eleanor},
{\tt IPython} \citep{ipython},
{\tt Keras} \citep{keras},
{\tt LcTools} \citep{2019arXiv191008034S,2021arXiv210310285S},
{\tt Lightcurvefactory} \citep{2013MNRAS.428.1656B,2017MNRAS.467.2160R,2018MNRAS.478.5135B},
{\tt Lightkurve} \citep{lightkurve},
{\tt Matplotlib} \citep{matplotlib},
{\tt Mpi4py} \citep{mpi4py2008},
{\tt NumPy} \citep{numpy}, 
{\tt Pandas} \citep{pandas},
{\tt Scikit-learn} \citep{scikit-learn},
{\tt SciPy} \citep{scipy},
{\tt Tensorflow} \citep{tensorflow},
{\tt Tess-point} \citep{tess-point}
}

\clearpage

\bibliography{refs}{}
\bibliographystyle{aasjournal}

\newpage

\newpage
\clearpage

{\onecolumngrid
\begin{longtable}{lllllllllllllll}

\caption{Parameters of the \Nquads quadruple candidates presented here, including ephemerides, eclipse depths and durations, and additional comments. The full table is available online.}\\

\hline
\hline
TIC ID & RA & Dec & Bin & Period & T$_{\rm 0,~p}$ & T$_{\rm 0,~s}$ & ${\rm \phi_{s}}$ & Dep$_{p}$ & Dep$_{p}$(spr) & Dep$_{s}$ & Dep$_{s}$(spr) & Dur$_{p}$ & Dur$_{s}$ \\
- & degrees & degrees & - & d & d & d & & ppt & ppt & ppt & ppt & hr & hr\\
\hline
\endhead

1216203 & 273.4655 & -5.2411 & A &  2.3247 & 3506.1242 & 3502.7167 & 0.47 & 126 &  & 67 &  & 5.8 & 5.3 \\
& & & B & 2.1994 & 3502.0225 & 3481.2407 & 0.45 & 69 &  & 30 &  & 3.1 & 5.2 \\
\multicolumn{14}{l}{Additional information: TGV-199, Gaia DR3 4173434534889039616, Tmag: 10.35, Teff:, Dist:} \\
\multicolumn{14}{l}{Comments: Only one sector available at the time of writing; PB in ASAS-SN } \\
\hline
3121385 & 138.6631 & -39.0524 & A &  17.4925 & 3011.4576 & 3005.6961 & 0.33 & 122 & 117-126 & 78 & 75-89 & 4.6 & 4.7 \\
& & & B & 20.1544 & 2996.8688 & 3004.4404 & 0.62 & 146 & 141-149 & 133 & 125-148 & 6.1 & 4.4 \\
\multicolumn{14}{l}{Additional information: TGV-200, Gaia DR3 5430510714397083520, Tmag: 13.12, Teff:, Dist: 440 pc} \\
\multicolumn{14}{l}{Comments: Apsidal motion on both PA and PB; PA and PB in ASAS-SN } \\
\hline
20938739 & 233.0364 & -4.5114 & A &  3.9561 & 2694.0852 & 2700.0188 & 0.50 & 171 &  & 158 &  & 4.4 & 4.5 \\
& & & B & 4.9446 & 2700.7489 &  &  & 22 &  &  &  & 5.5 &  \\
\multicolumn{14}{l}{Additional information: TGV-201, Gaia DR3 4401400571324534016, Tmag: 9.44, Teff:, Dist: 345 pc} \\
\multicolumn{14}{l}{Comments: From VSG and PHT\footnote{\url{https://www.zooniverse.org/projects/nora-dot-eisner/planet-hunters-tess/talk/2112/2512585?comment=4121105&page=1}} and EBP\footnote{\url{https://www.zooniverse.org/projects/vbkostov/eclipsing-binary-patrol/talk/6324/3452938?comment=5820654&page=1}}; Only one sector available at the time of writing;}\\
\multicolumn{14}{l}{Comments: PB may be slightly off; There may be ${\rm B_{sec}}$ but hard to tell} \\
\hline
26993740 & 356.2475 & 50.0971 & A &  1.8842 & 1766.4777 & 2854.5747 & 0.52 & 100 & 6-120 & 20 &  & 2.0 & 3.3 \\
& & & B & 9.2238 & 2854.965 & 2859.4045 & 0.52 & 54 & 44-142 & 17 &  & 3.0 & 2.5 \\
\multicolumn{14}{l}{Additional information: TGV-202, Gaia DR3 1943564205724787840, Tmag: 11.71, Teff:, Dist: 188 pc} \\
\multicolumn{14}{l}{Comments: ${\rm T_{0}(B_{sec})}$ may be slightly off; PA in ASAS-SN } \\
\hline
27258509 & 292.9677 & 9.8500 & A &  2.1253 & 2771.5387 &  &  & 47 & 43-55 &  &  & 2.7 &  \\
& & & B & 17.7779 & 2770.1229 & 3526.9428 & 0.43 & 90 & 84-95 & 38 & 35-44 & 3.9 & 4.9 \\
\multicolumn{14}{l}{Additional information: TGV-203, Gaia DR3 4302472704155454592, Tmag: 13.09, Teff:, Dist: 1026 pc} \\
\multicolumn{14}{l}{Comments: PA in ASAS-SN } \\
\hline
48089827 & 282.0706 & 51.2979 & A &  4.8637 & 3660.8489 & 3658.4146 & 0.50 & 140 & 101-154 & 115 & 89-131 & 3.8 & 3.6 \\
& & & B & 8.1718 & 3638.1317 & 3650.4113 & 0.50 & 97 & 88-121 & 72 & 57-93 & 4.3 & 4.2 \\
\multicolumn{14}{l}{Additional information: TGV-204, Gaia DR3 2144698898581885184, Tmag: 13.53, Teff:, Dist: 1383 pc} \\
\multicolumn{14}{l}{Comments: Dramatic ETVs with Pout ~ 1,400 d; PA and PB in ZTF } \\
\hline
48677841 & 299.3129 & -14.1831 & A &  2.0582 & 3825.5941 & 2770.7586 & 0.50 & 73 &  & 23 &  & 4.2 & 3.9 \\
& & & B & 3.3182 & 3824.1794 &  &  & 23 &  &  &  & 3.5 &  \\
\multicolumn{14}{l}{Additional information: TGV-205, Gaia DR3 6878567390734910208, Tmag: 10.68, Teff: 6356 K, Dist: 551 pc} \\
\multicolumn{14}{l}{Comments: Only one sector available at the time of writing; PA in ASAS-SN } \\
\hline
\multicolumn{14}{c}{{TGV-N = TESS/GSFC/VSG quadruple candidate -N; }}\\
\multicolumn{14}{c}{{${\rm T_{0,~p/s}}$ = Time of primary/secondary (in BJD-2,457,000); ${\rm Dep_{p/s}}$ = Depth of primary/secondary; ${\rm Dur_{p/s}}$ = Duration of prim/sec}}\\
\multicolumn{14}{c}{{${\rm \phi_{s}}$ = Secondary phase; Teff = Composite effective temperature; ppt = parts-per-thousand}}\\
\multicolumn{14}{c}{{Dep$_{p/s}$(spr) = Spread of non-blended primary/secondary eclipse depths across available sectors}}\\

\label{tbl:main_table}
\end{longtable}
}

\clearpage

\begin{table}
\begin{tabular}[t]{cc}   
\begin{tabular}[t]{c|c|c|c} 
\hline
\hline
TIC & Tmag & ${\rm Depth_A (\%)}$ & ${\rm Depth_B (\%)}$ \\
\hline
391461666 & 8.52 & 17 & 4 \\
99875938 & 9.43 & 13 & 3 \\
20938739 & 9.44 & 17 & 2 \\
52877118 & 9.55 & 18 & 16 \\
79908874 & 10.15 & 3 & 2 \\
297251275 & 10.18 & 15 & 3 \\
1216203 & 10.35 & 7 & 12 \\
130946041 & 10.40 & 11 & 0.6 \\
139995365 & 10.42 & 11 & 26 \\
79062805 & 10.46 & 2 & 32 \\
466310009 & 10.47 & 25 & 0.6 \\
48677841 & 10.68 & 7 & 2 \\
430752710 & 11.08 & 5 & 3 \\
258507555 & 11.13 & 2 & 6 \\
412074304 & 11.24 & 1 & 3 \\
286779918 & 11.30 & 27 & 0.5 \\
165052445 & 11.35 & 4 & 7 \\
26993740 & 11.71 & 10 & 5 \\
\hline
\end{tabular} &  
\end{tabular}

\caption{Eclipsing quadruple star candidates brighter than Tmag = 12, and primary eclipses deeper than 1\% for at least one of the component EBs.}
\label{tab:good_for_obs}
\end{table}

\appendix
\setcounter{figure}{0}
\renewcommand{\thefigure}{A\arabic{figure}}

\section{Quadruple Candidates With Sparse Eclipses}

Some of the targets presented here were flagged by members of our team as worthy of further investigations quite early in the TESS mission. At the time, the number of available sectors was small and so was the number of interesting features in the corresponding lightcurves. Several EBs produced a single extra transit-like event, quickly confirmed to be on-target through photocenter measurements, that could be interpreted as potentially due to a circumbinary planet. Naturally, this was quite an exciting possibility as such planets are few and far between. As TESS continued observing, however, the events repeated in a strictly-periodic pattern indicating they are in fact produced by a second EB with an orbital period (much) longer than a TESS sector. 

\begin{figure*}
    \includegraphics[width=0.99\textwidth]{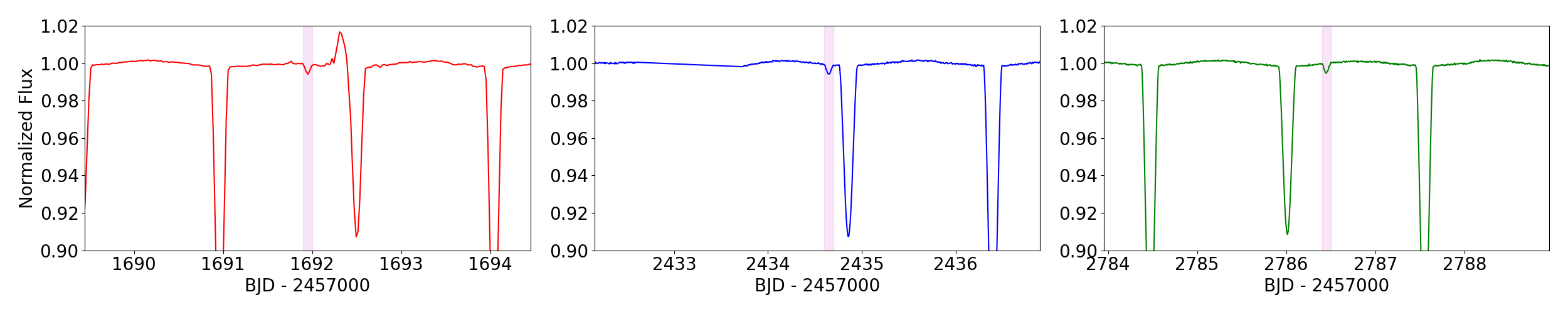}
    \caption{TESS FITSH lightcurve of TIC 391461666 for Sectors 14 (left), 41 (middle), and 54 (right), highlighting the three transit/eclipse-like event. The period of the second EB, PB = 351.8/N days, where N is an integer value, cannot be uniquely determined at the time of writing due to the large separation between the three eclipses it produces.}
    \label{fig:391461666_lc}
\end{figure*}

\begin{figure*}
    \centering
    \includegraphics[width=0.49\linewidth]{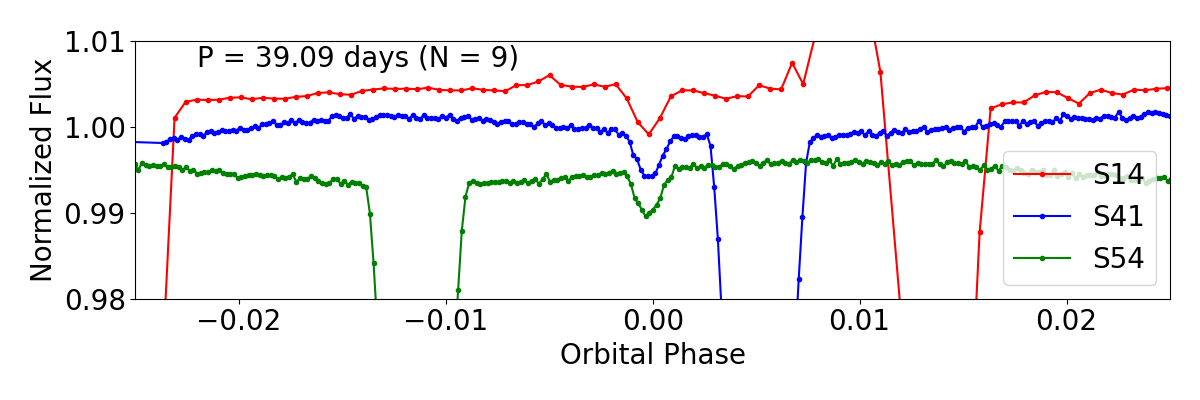}
    \includegraphics[width=0.49\linewidth]{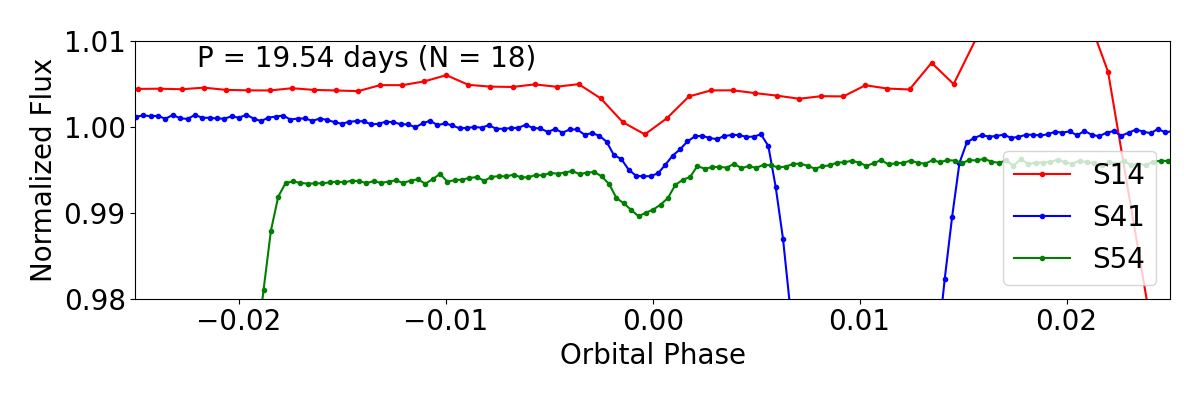}
    \includegraphics[width=0.49\linewidth]{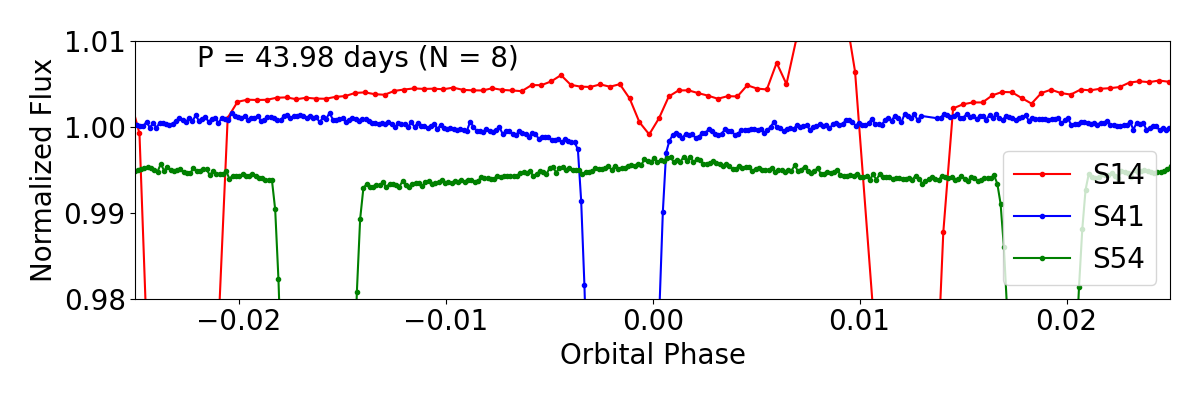}
    \includegraphics[width=0.49\linewidth]{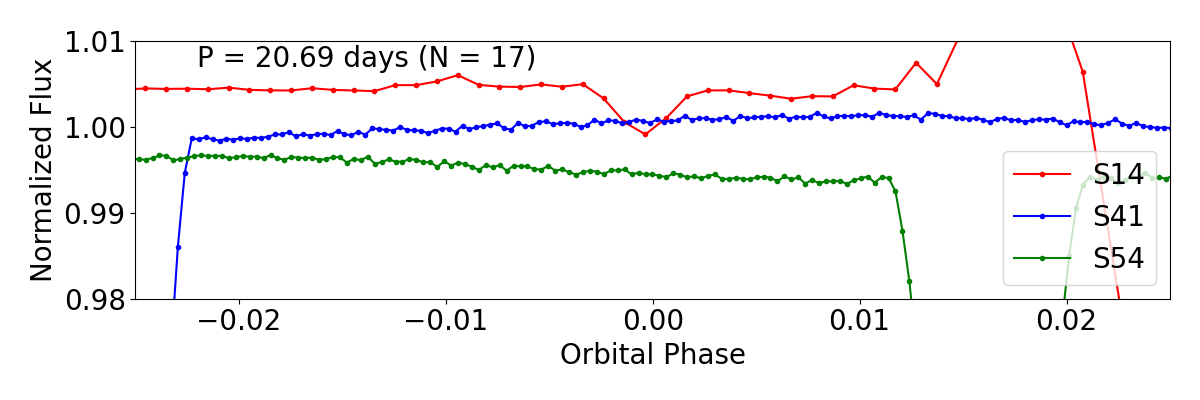}
    \caption{Phase-folded lightcurve for TIC 391461666. Upper panels: fold on two possible periods of the second EB, PB = 351.8/N days: PB = 39.09 days, (i.e., N = 9, upper left), and PB = 19.54 days (i.e., N = 18, upper right). Both provide a convincing case for binary B and thus a potential true period. Lower panels: same as upper panes, but for two values of N for which PB = 351.8/N days can be ruled out from the available data: N = 8 (lower left) and N = 17 (lower right).}
    \label{fig:391461666_good}
\end{figure*}

One of the more illustrative examples of this situation is TIC 391461666 (TGV-241), where the lightcurve shows an EB with PA $\approx$ 3.08 days and one extra transit/eclipse-like event in each of the three sectors the target was observed in (Sectors 14, 41, and 54). The three events are similar in depth, duration, and overall shape, and are separated by hundreds of days (see Figure \ref{fig:391461666_lc}). Specifically, the first and the second are about 742.7 days apart, while the second and the third are separated by about 351.8 days. Assuming the three events are the same eclipse, the gaps between them are too large to uniquely determine the period of the EB, with multiple integer ratios of P(N) = 351.8/N days producing viable phase-folded lightcurves. The two most promising options seem to be for P(9) = 39.09 days (i.e., N = 9), and P(18) = 19.54 days (i.e., N = 18), both producing a convincing phase-folded lightcurve (see upper panels, Figure \ref{fig:391461666_good}). While all other values of N between 7 and 20 can be ruled out from the available data (see lower panels, Figure \ref{fig:391461666_good}), smaller values of N = 2-6 would also work. It is also possible that the three events represent two primary and one secondary eclipse. Alas, TESS will not observe TIC 391461666 again and we cannot confirm PB at the time of writing. Nevertheless, the target is likely an eclipsing 2+2 quadruple and is thus included in the catalog presented here. 

\begin{figure*}
    \centering
    \includegraphics[width=0.99\linewidth]{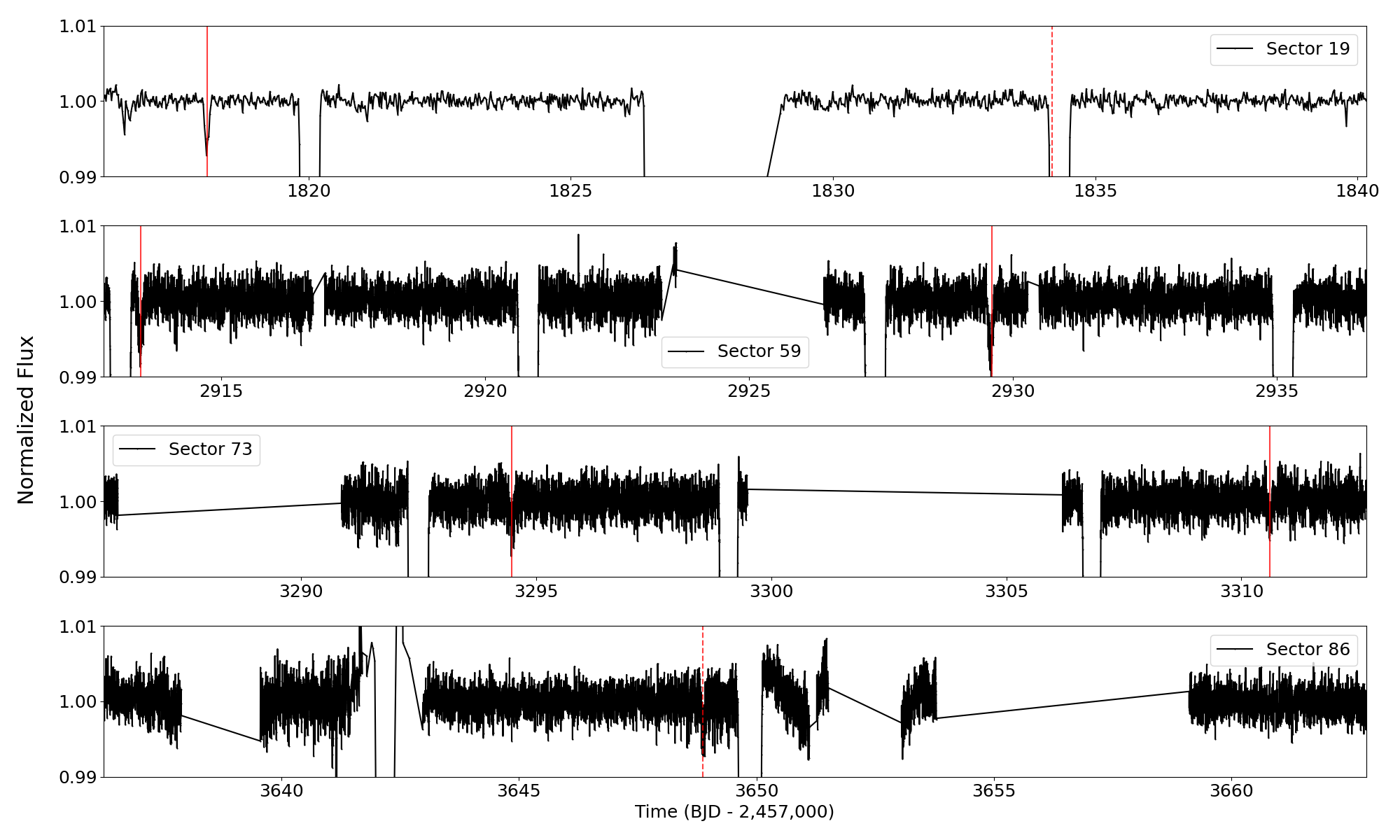}
    \caption{TESS-Gaia lightcurve of TIC 286779918 highlighting the six extra eclipse-like events (vertical red lines).}
    \label{fig:286779918}
\end{figure*}

\begin{figure*}
    \centering
    \includegraphics[width=0.49\linewidth]{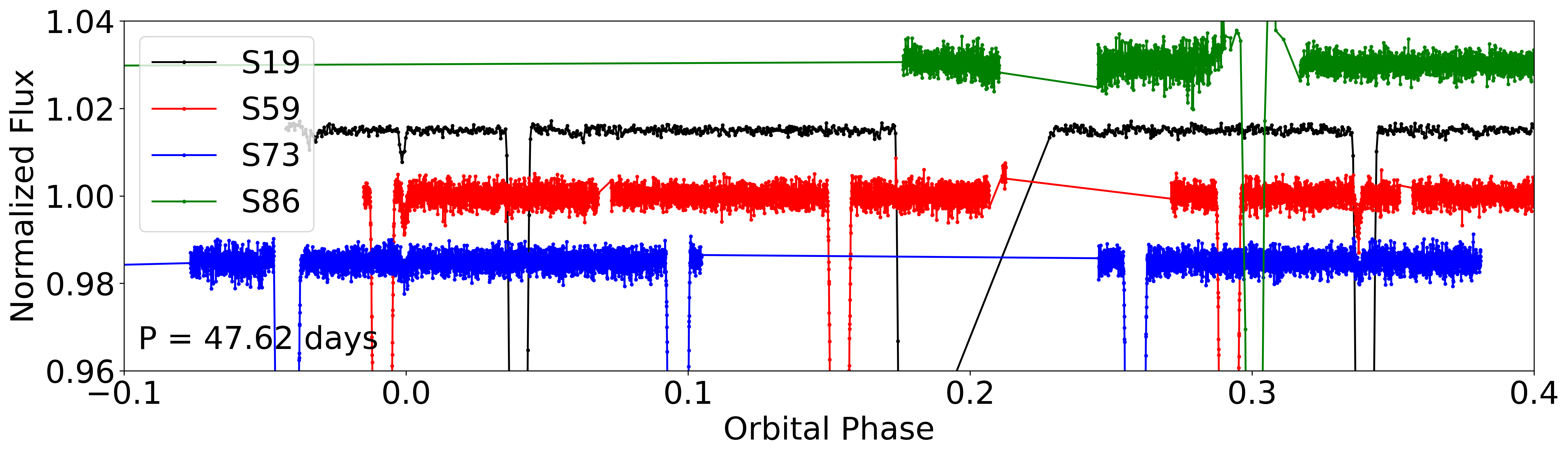}
    \includegraphics[width=0.49\linewidth]{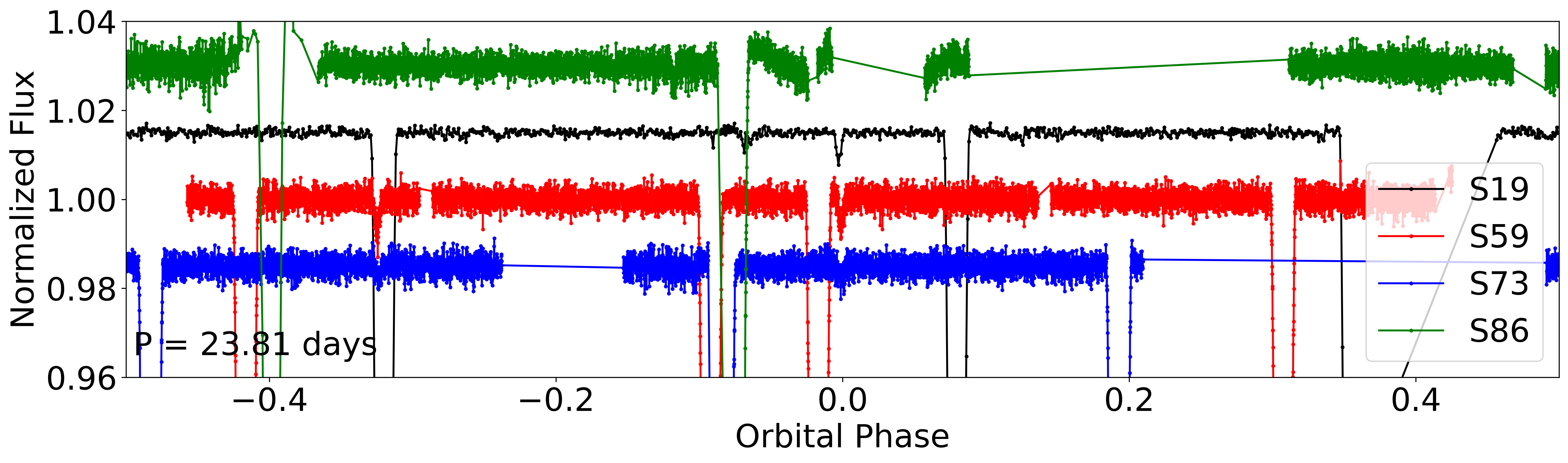}
    \includegraphics[width=0.49\linewidth]{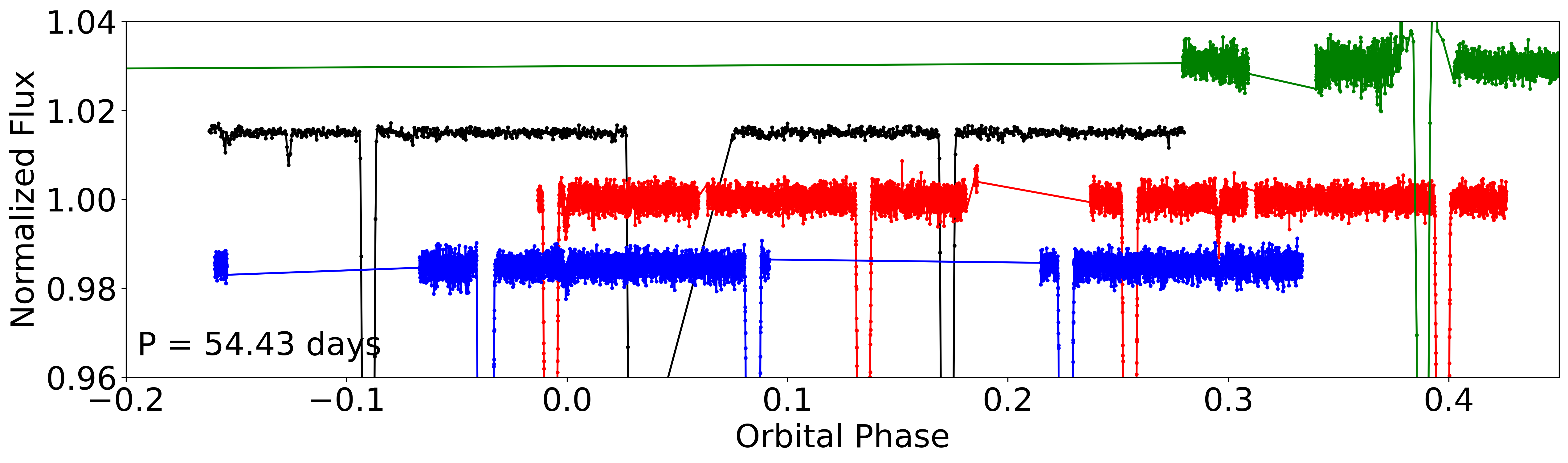}
    \includegraphics[width=0.49\linewidth]{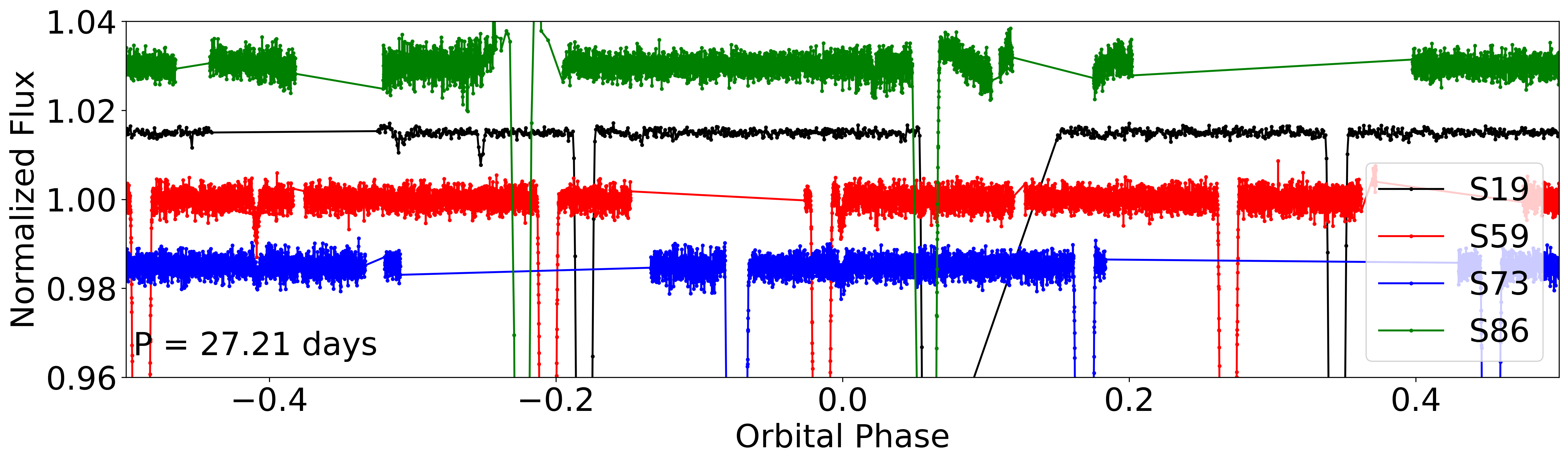}
    \caption{Same as Figure \ref{fig:391461666_good} but for TIC 286779918. Upper panels: two test periods that fold three of the extra events: PB $\approx 47.6$ days (N = 8, upper left) and PB $\approx 23.8$ days (N = 16, upper right). However, neither periods works for all six events. Lower panels: two test periods that fold no more than two of the extra events (centered near zero phase): PB $\approx 54.4$ days (N = 7, lower left) and PB $\approx 27.2$ days (N = 14, lower right). Note that the x-axes are different between the different panels, to highlight the phase of potential secondary events.}
    \label{fig:286779918_ok_fold}
\end{figure*}

Another example is TIC 286779918 (TGV-232), which produced six additional eclipse-like events (in addition to the much deeper EB, see Figure \ref{fig:286779918}). The two pairs of extra events in Sectors 59 and 73 have practically the same separation between the events of $\sim$16.125 days, suggesting that the pairs are potentially related. However, a period of $\sim$16.125 days does not produce a consistent phase fold of the entire lightcurve 
Instead, assuming the two pairs have the same origin, we can speculate that they are primary and secondary eclipses separated by PB(N)${\rm \approx381/N}$ days periods between Sectors 50 and 73. Overall, three of the events fold well on a period of PB $\approx 47.6$ days (i.e., N = 8) and PB $\approx 23.81$ days (i.e., N = 16 ) (see upper panels, Figure \ref{fig:286779918_ok_fold}). Another two events fold well on said periods for Sectors 59 and 73, and potentially 19 if another event is assumed to blended with the deep PA eclipse -- but the event from Sector 86 does not. Like the case for TIC 391461666, the available data allows ruling out several values of N. Altogether, this indicates that the six events may not have a common origin, and suggests that the above assumption is likely incorrect. Thus, we cannot confirm PB from the available TESS data at the time of writing. With that said, the clear periodicity of $\sim$16.125 days strongly suggests a second EB, justifying the inclusion of TIC 286779918 in this catalog as a 2+2 eclipsing quadruple candidate.  

\begin{figure*}
  \epsscale{1.17}
  \includegraphics[width=0.49\linewidth]{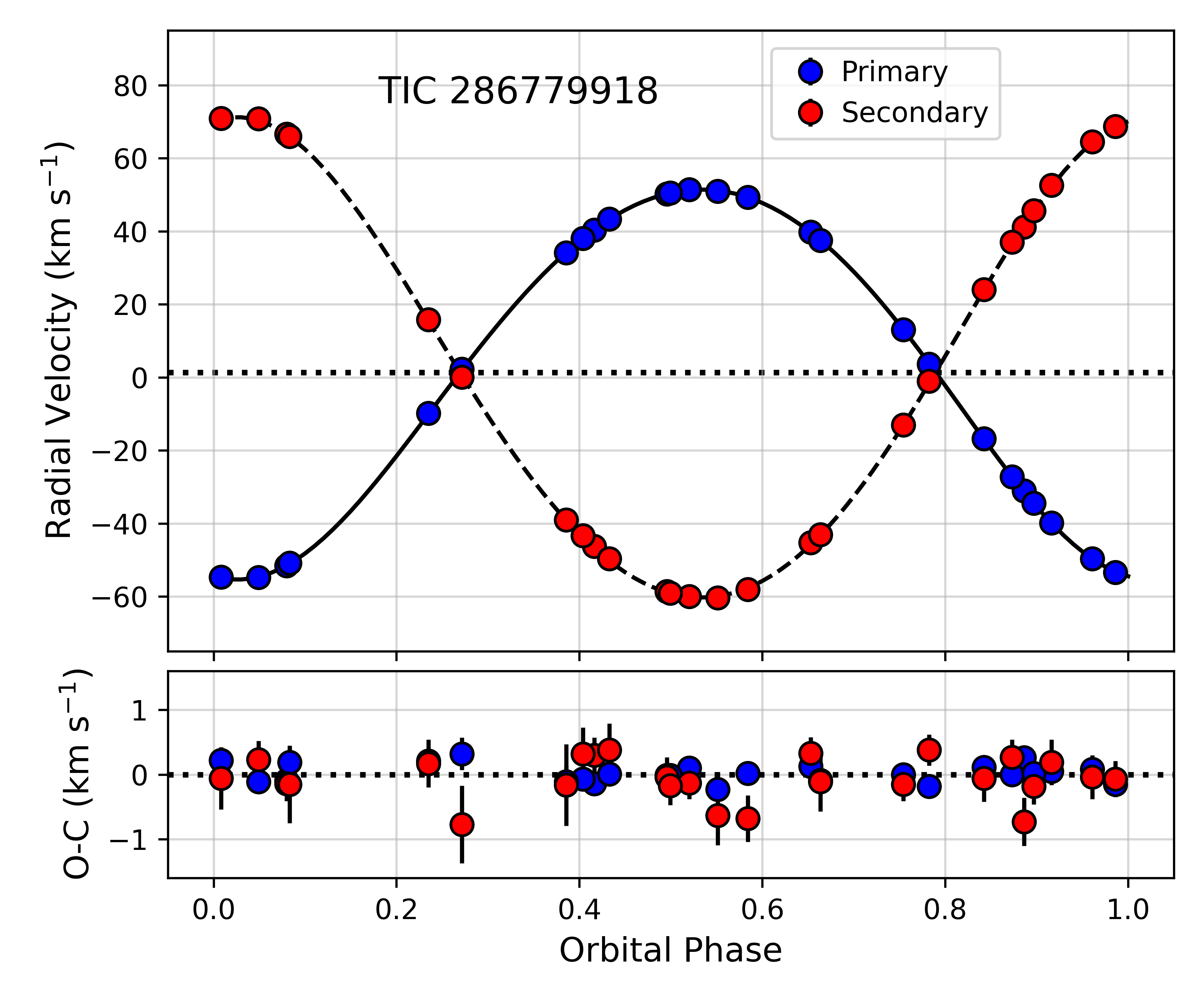}
  \figcaption{Measured radial velocities and orbit model for TIC 286779918, with residuals shown at the bottom. The dotted line in the top panel represents the center-of-mass velocity. \label{fig:tic286orbit}}
\end{figure*}

In an attempt to detect the period of binary B by different means, we observed the target spectroscopically at the Center for Astrophysics with the Tillinghast Reflector Echelle Spectrograph \citep[TRES;][]{Furesz2008, Szentgyorgyi:2007} on the 1.5m Tillinghast reflector, located at the Fred L.\ Whipple Observatory on Mount Hopkins (Arizona, USA). We gathered 26 spectra covering the wavelength range 3800--9100~\AA\ at a resolving power of $R \approx 44,\!000$, with signal-to-noise ratios of 29--75 per resolution element of 6.8~\kms. Reductions were carried out with a dedicated pipeline \citep[see][]{2012Natur.486..375B}. All spectra are double-lined and, as illustrated in Figure~\ref{fig:tic286orbit}, both PA components are clearly visible in the radial velocity measurements. The latter were extracted with the two-dimensional cross-correlation algorithm TODCOR \citep{1994ApJ...420..806Z}. The synthetic templates were taken from a library of calculated spectra based on model atmospheres by Castelli \& Kurucz \citep{2003IAUS..210P.A20C}, and a line list tuned by hand to improve the match to real stars. We adopted solar metallicity and surface gravities of $\log g = 4.5$ for both PA stars, along with best-fitting effective temperatures of 6250 and 6000~K for the primary and secondary, respectively, and rotational broadenings of 6~\kms\ for both stars. The measured velocities and corresponding uncertainties are listed in Table~\ref{tab:tic286rvs}. We determined a flux ratio between the components of $\ell_2/\ell_1 = 0.305 \pm 0.005$ at the mean wavelength of our observations ($\sim$5187~\AA). A spectroscopic orbital solution gives the elements presented in Table~\ref{tab:tic286elem}, where the symbols have their usual meaning. The orbital period derived from the spectroscopic measurements is P = 14.29592 days, fully consistent with the PA = 14.296107 days derived from the TESS photometry for binary A. The rms velocity residuals are 0.14 and 0.34~\kms\ for the primary and secondary, and the derived parameters are listed at the bottom of the table. Examination of the spectra with an extension of TODCOR to three dimensions \citep[TRICOR;][]{1995ApJ...452..863Z} revealed no additional stars down to our detection threshold of approximately 1\% of the flux of the primary. Thus, our attempt to detect PB spectroscopically failed and the orbital period of binary B remains a mystery at the time of writing. 

\setlength{\tabcolsep}{6pt}
\begin{deluxetable}{lcccc}
\tablewidth{0pc}
\tablecaption{Radial Velocities for TIC 286779918 \label{tab:tic286rvs}}
\tablehead{
\colhead{BJD} &
\colhead{$RV_1$} &
\colhead{$RV_2$} &
\colhead{$\sigma_1$} &
\colhead{$\sigma_2$}
\\
\colhead{(2,400,000+)} &
\colhead{(\kms)} &
\colhead{(\kms)} &
\colhead{(\kms)} &
\colhead{(\kms)}
}
\startdata
 60413.6375  &  \phn2.29         &  \phn0.10         &  0.25  &  0.60 \\
 60417.6382  &     50.91         &  $-$60.32\phs     &  0.19  &  0.46 \\
 60572.9649  &     40.35         &  $-$46.28\phs     &  0.11  &  0.27 \\
 60593.9793  &  $-$31.09\phs     &     41.17         &  0.15  &  0.37 \\
 60604.9379  &     39.81         &  $-$45.24\phs     &  0.10  &  0.25 \\
 60621.9420  &  $-$16.81\phs     &     24.07         &  0.15  &  0.36 \\
 60646.8408  &     49.31         &  $-$58.09\phs     &  0.15  &  0.36 \\
 60653.9306  &  $-$51.58\phs     &     66.57         &  0.11  &  0.27 \\
 60663.9729  &  \phn3.61         &  $-$\phn0.99\phs  &  0.10  &  0.24 \\
 60666.8844  &  $-$53.35\phs     &     68.74         &  0.12  &  0.28 \\
 60677.8653  &     13.10         &  $-$12.99\phs     &  0.11  &  0.26 \\
 60684.7397  &  $-$\phn9.84\phs  &     15.84         &  0.15  &  0.37 \\
 60693.8626  &  $-$27.21\phs     &     37.07         &  0.11  &  0.27 \\
 60710.6766  &  $-$54.72\phs     &     70.78         &  0.12  &  0.29 \\
 60723.7091  &  $-$49.62\phs     &     64.47         &  0.14  &  0.34 \\
 60731.7058  &     51.37         &  $-$59.98\phs     &  0.10  &  0.25 \\
 60738.6836  &  $-$54.71\phs     &     70.89         &  0.20  &  0.48 \\
 60745.6503  &     50.24         &  $-$58.63\phs     &  0.13  &  0.31 \\
 60745.7010  &     50.45         &  $-$59.02\phs     &  0.13  &  0.30 \\
 60751.6612  &  $-$39.85\phs     &     52.63         &  0.15  &  0.35 \\
 60758.6351  &     38.07         &  $-$43.36\phs     &  0.17  &  0.41 \\
 60765.6803  &  $-$34.48\phs     &     45.61         &  0.12  &  0.28 \\
 60772.6682  &     34.15         &  $-$39.06\phs     &  0.26  &  0.63 \\
 60776.6396  &     37.46         &  $-$43.06\phs     &  0.19  &  0.46 \\
 60782.6409  &  $-$50.82\phs     &     65.97         &  0.25  &  0.60 \\
 60787.6392  &     43.31         &  $-$49.65\phs     &  0.17  &  0.41 \\
\enddata
\end{deluxetable}
\setlength{\tabcolsep}{6pt}  

\setlength{\tabcolsep}{6pt}
\begin{deluxetable}{lc}
\tablewidth{0pc}
\tablecaption{Spectroscopic Orbital Parameters for TIC 286779918\ \label{tab:tic286elem}}
\tablehead{
\colhead{Parameter} &
\colhead{Value}
}
\startdata
$P$ (day)                  & $14.29592 \pm 0.00033$\phn      \\
$T_{\rm peri}$ (BJD)       & $60667.084 \pm 0.024$\phm{2222} \\
$e$                        & $0.06367 \pm 0.00072$           \\
$\omega_1$ (deg)           & $169.39 \pm 0.62$\phn\phn       \\
$K_1$ (\kms)               & $53.379 \pm 0.036$\phn          \\
$K_2$ (\kms)               & $65.742 \pm 0.085$\phn          \\
$\gamma$ (\kms)            & $+1.370 \pm 0.032$\phs          \\
$\Delta_{\rm RV}$ (\kms)   & $-0.238 \pm 0.079$\phs          \\ [0.5ex]
\hline \\ [-1.5ex]
\multicolumn{2}{c}{Derived Properties} \\ [0.5ex]
\hline \\ [-1.5ex]
$M_1 \sin^3 i$ ($M_{\odot}$)  & $1.3734 \pm 0.0039$          \\
$M_2 \sin^3 i$ ($M_{\odot}$)  & $1.1152 \pm 0.0022$          \\
$q \equiv M_2/M_1$            & $0.8119 \pm 0.0012$          \\
$a \sin i$ ($R_{\odot}$)      & $33.594 \pm 0.027$\phn       \\
\enddata
\tablecomments{The time of periastron passage, $T_{\rm peri}$, is referred to JD~2,400,000. Parameter $\Delta_{\rm RV}$ represents a small velocity zero-point offset between the primary and secondary that is likely due to template mismatch. It was taken into account to avoid biasing the minimum masses. $a \sin i$ is the total projected semimajor axis of the orbit.}
\end{deluxetable}
\setlength{\tabcolsep}{6pt}  

\section{Eclipse timing measurements}

Here we list the times of eclipse minima for both binaries in quadruple systems that exhibit eclipse timing variations. The full tables are available as machine-readable online supplement. 

\begin{table}[htbp]
    \centering
    \begin{minipage}{0.48\textwidth}
        \centering
        \caption{Measured eclipse times for TIC 48089827 binary A. Table available in full as a machine-readable online supplement.}
        \begin{tabular}{@{}lcc@{}}
            \hline
            \hline
            Time & Cycle & std. dev. \\
            (BJD - 2,457,000) & & (min) \\
            \hline
            1683.7606 & -0.5 & 0.15 \\
            1686.1924 & 0.0 & 0.08 \\
            1688.6233 & 0.5 & 0.30 \\
            1691.0559 & 1.0 & 0.23 \\
            1693.4881 & 1.5 & 0.12 \\
            1695.9194 & 2.0 & 0.17 \\
            1698.3509 & 2.5 & 0.28 \\
            1700.7826 & 3.0 & 0.31 \\
            1703.2141 & 3.5 & 0.20 \\
            1705.6469 & 4.0 & 0.17 \\
            \hline
        \end{tabular}
    \end{minipage}%
    \hfill
    \begin{minipage}{0.48\textwidth}
        \centering
        \caption{Measured eclipse times for TIC 48089827 binary B. Table available in full as a machine-readable online supplement.}
        \begin{tabular}{@{}lcc@{}}
            \hline
            \hline
            Time & Cycle & std. dev. \\
            (BJD - 2,457,000) & & (min) \\
            \hline
            1685.0630 & 0.0 & 0.21 \\
            1689.1509 & 0.5 & 0.20 \\
            1693.2331 & 1.0 & 0.43 \\
            1701.4072 & 2.0 & 0.14 \\
            1705.4976 & 2.5 & 0.64 \\
            1709.5784 & 3.0 & 0.45 \\
            1713.6683 & 3.5 & 0.51 \\
            1717.7522 & 4.0 & 0.34 \\
            1721.8363 & 4.5 & 0.94 \\
            1725.9211 & 5.0 & 0.51 \\
            \hline
        \end{tabular}
    \end{minipage}
\end{table}

\begin{table}[htbp]
    \centering
    \begin{minipage}{0.48\textwidth}
        \centering
        \caption{Measured eclipse times for TIC 64832327 binary A.}
        \begin{tabular}{@{}lcc@{}}
            \hline
            \hline
            Time & Cycle & std. dev. \\
            (BJD - 2,457,000) & & (min) \\
            \hline
            1747.0691 & 0.0 & 0.48 \\
            1755.5776 & 1.0 & 0.31 \\
            1772.5891 & 3.0 & 0.34 \\
            1781.0966 & 4.0 & 0.50 \\
            1789.6050 & 5.0 & 0.76 \\
            2861.4514 & 131.0 & 0.26 \\
            2869.9610 & 132.0 & 0.22 \\
            2878.4665 & 133.0 & 0.23 \\
            3397.3888 & 194.0 & 0.19 \\
            3405.8946 & 195.0 & 0.26 \\
            3422.9092 & 197.0 & 0.19 \\
            3567.5269 & 214.0 & 0.23 \\
            3576.0360 & 215.0 & 0.28 \\
            3593.0489 & 217.0 & 0.24 \\
            3601.5551 & 218.0 & 0.26 \\
            3610.0627 & 219.0 & 0.26 \\
            \hline
        \end{tabular}
    \end{minipage}%
    \hfill
    \begin{minipage}{0.48\textwidth}
        \centering
        \caption{Measured eclipse times for TIC 64832327 binary B.}
        \begin{tabular}{@{}lcc@{}}
            \hline
            \hline
            Time & Cycle & std. dev. \\
            (BJD - 2,457,000) & & (min) \\
            \hline
            1743.6741 & 0.0 & 1.16 \\
            1768.3148 & 1.0 & 0.56 \\
            2877.1967 & 46.0 & 0.54 \\
            3419.3081 & 68.0 & 0.62 \\
            3567.1540 & 74.0 & 0.61 \\
            3591.7944 & 75.0 & 0.60 \\
            \hline
        \end{tabular}
    \end{minipage}
\end{table}

\begin{table}[htbp]
    \centering
    \begin{minipage}{0.48\textwidth}
        \centering
        \caption{Measured eclipse times for TIC 79062805 binary A. Table available in full as a machine-readable online supplement.}
        \begin{tabular}{@{}lcc@{}}
            \hline
            \hline
            Time & Cycle & std. dev. \\
            (BJD - 2,457,000) & & (min) \\
            \hline
            1626.6611 & 0.0 & 0.25 \\
            1628.6019 & 1.0 & 6.83 \\
            1630.5342 & 2.0 & 0.56 \\
            1631.4871 & 2.5 & 5.76 \\
            1633.4341 & 3.5 & 2.90 \\
            1634.4043 & 4.0 & 0.59 \\
            1635.3645 & 4.5 & 3.14 \\
            1636.3470 & 5.0 & 1.07 \\
            1637.3336 & 5.5 & 6.57 \\
            1638.2823 & 6.0 & 1.55 \\
            \hline
        \end{tabular}
    \end{minipage}%
    \hfill
    \begin{minipage}{0.48\textwidth}
        \centering
        \caption{Measured eclipse times for TIC 79062805 binary B. Table available in full as a machine-readable online supplement.}
        \begin{tabular}{@{}lcc@{}}
            \hline
            \hline
            Time & Cycle & std. dev. \\
            (BJD - 2,457,000) & & (min) \\
            \hline
            1628.9435 & 0.0 & 0.10 \\
            1632.6812 & 0.5 & 0.09 \\
            1636.8354 & 1.0 & 0.08 \\
            1640.5729 & 1.5 & 0.07 \\
            1644.7275 & 2.0 & 0.09 \\
            1648.4645 & 2.5 & 0.07 \\
            1652.6193 & 3.0 & 0.05 \\
            2362.8961 & 93.0 & 0.04 \\
            2366.6324 & 93.5 & 0.11 \\
            2370.7882 & 94.0 & 0.05 \\
            \hline
        \end{tabular}
    \end{minipage}
\end{table}

\begin{table}[htbp]
    \centering
    \begin{minipage}{0.48\textwidth}
        \centering
        \caption{Measured eclipse times for TIC 466310009 binary A. Table available in full as a machine-readable online supplement.}
        \begin{tabular}{@{}lcc@{}}
            \hline
            \hline
            Time & Cycle & std. dev. \\
            (BJD - 2,457,000) & & (min) \\
            \hline
            1656.3752 & 0.0 & 0.07 \\
            1660.2918 & 0.5 & 0.07 \\
            1664.2496 & 1.0 & 0.07 \\
            1672.1243 & 2.0 & 0.12 \\
            1676.0409 & 2.5 & 0.06 \\
            1679.9985 & 3.0 & 0.05 \\
            2038.1734 & 48.5 & 0.06 \\
            2042.1298 & 49.0 & 0.05 \\
            2046.0458 & 49.5 & 0.06 \\
            \hline
        \end{tabular}
    \end{minipage}%
    \hfill
    \begin{minipage}{0.48\textwidth}
        \centering
        \caption{Measured eclipse times for TIC 466310009 binary B.}
        \begin{tabular}{@{}lcc@{}}
            \hline
            \hline
            Time & Cycle & std. dev. \\
            (BJD - 2,457,000) & & (min) \\
            \hline
            1654.1661 & -89.0 & 1.41 \\
            1662.4729 & -88.5 & 1.16 \\
            1670.7855 & -88.0 & 0.57 \\
            1679.0948 & -87.5 & 1.63 \\
            2044.3319 & -65.5 & 0.46 \\
            2052.6348 & -65.0 & 0.43 \\
            3131.9462 & 0.0 & 0.79 \\
            3148.5567 & 1.0 & 1.10 \\
            3870.7848 & 44.5 & 0.71 \\
            3879.0814 & 45.0 & 0.70 \\
            \hline
        \end{tabular}
    \end{minipage}
\end{table}

\begin{table}[htbp]
    \centering
    \begin{minipage}{0.48\textwidth}
        \centering
        \caption{Measured eclipse times for TIC 352830705 binary A. Table available in full as a machine-readable online supplement.}
        \begin{tabular}{@{}lcc@{}}
            \hline
            \hline
            Time & Cycle & std. dev. \\
            (BJD - 2,457,000) & & (min) \\
            \hline
            1714.9704 & 0.0 & 0.23 \\
            1723.2611 & 1.0 & 0.20 \\
            1727.4926 & 1.5 & 3.05 \\
            1731.5509 & 2.0 & 0.97 \\
            1735.7765 & 2.5 & 1.54 \\
            1739.8403 & 3.0 & 0.67 \\
            1744.0689 & 3.5 & 8.35 \\
            1748.1306 & 4.0 & 0.36 \\
            1752.3591 & 4.5 & 3.84 \\
            1756.4190 & 5.0 & 0.97 \\
            \hline
        \end{tabular}
    \end{minipage}%
    \hfill
    \begin{minipage}{0.48\textwidth}
        \centering
        \caption{Measured eclipse times for TIC 352830705 binary B. Table available in full as a machine-readable online supplement.}
        \begin{tabular}{@{}lcc@{}}
            \hline
            \hline
            Time & Cycle & std. dev. \\
            (BJD - 2,457,000) & & (min) \\
            \hline
            1714.3773 & 0.5 & 1.07 \\
            1716.1038 & 1.0 & 3.52 \\
            1717.8277 & 1.5 & 4.00 \\
            1719.5545 & 2.0 & 1.51 \\
            1721.2788 & 2.5 & 3.61 \\
            1723.0028 & 3.0 & 1.22 \\
            1726.4425 & 4.0 & 2.14 \\
            1728.1665 & 4.5 & 3.03 \\
            1729.8924 & 5.0 & 3.33 \\
            1733.3280 & 6.0 & 1.12 \\
            \hline
        \end{tabular}
    \end{minipage}
\end{table}

\begin{table}[htbp]
    \centering
    \begin{minipage}{0.48\textwidth}
        \centering
        \caption{Measured eclipse times for TIC 130946041 binary A. Table available in full as a machine-readable online supplement.}
        \begin{tabular}{@{}lcc@{}}
            \hline
            \hline
            Time & Cycle & std. dev.\footnote{Stellar variability likely affects the precision of the measurements.} \\
            (BJD - 2,457,000) & & (min) \\
            \hline
            1573.0010 & 2.5 & 0.32 \\
            1573.3968 & 3.0 & 1.26 \\
            1573.7934 & 3.5 & 0.35 \\
            1574.1888 & 4.0 & 0.87 \\
            1574.5861 & 4.5 & 0.25 \\
            1574.9818 & 5.0 & 0.31 \\
            1575.3789 & 5.5 & 0.47 \\
            1575.7742 & 6.0 & 1.30 \\
            1576.1712 & 6.5 & 0.23 \\
            1576.5673 & 7.0 & 0.67 \\
            \hline
        \end{tabular}
    \end{minipage}%
    \hfill
\end{table}

\begin{table}[htbp]
    \centering
    \begin{minipage}{0.48\textwidth}
        \centering
        \caption{Measured eclipse times for TIC 289822938 binary A. Table available in full as a machine-readable online supplement.}
        \begin{tabular}{@{}lcc@{}}
            \hline
            \hline
            Time & Cycle & std. dev.\footnote{Stellar variability likely affects the precision of the measurements.} \\
            (BJD - 2,457,000) & & (min) \\
            \hline
            1544.5516 & 0.0 & 0.58 \\
            1545.0107 & 0.5 & 2.70 \\
            1545.4721 & 1.0 & 0.39 \\
            1545.9281 & 1.5 & 2.85 \\
            1546.3943 & 2.0 & 0.48 \\
            1546.8536 & 2.5 & 7.47 \\
            1547.3182 & 3.0 & 1.79 \\
            1548.2399 & 4.0 & 1.03 \\
            1549.1595 & 5.0 & 0.07 \\
            1549.6125 & 5.5 & 1.03 \\
            \hline
        \end{tabular}
    \end{minipage}%
    \hfill
    \begin{minipage}{0.48\textwidth}
        \centering
        \caption{Measured eclipse times for TIC 289822938 binary B.}
        \begin{tabular}{@{}lcc@{}}
            \hline
            \hline
            Time & Cycle & std. dev.\footnote{Stellar variability likely affects the precision of the measurements.} \\
            (BJD - 2,457,000) & & (min) \\
            \hline
            1549.4723 & -143.0 & 0.73 \\
            1564.7166 & -142.0 & 0.63 \\
            2296.0871 & -94.0 & 0.49 \\
            2997.0176 & -48.0 & 0.48 \\
            3012.2538 & -47.0 & 0.56 \\
            3718.9009 & -0.5 & 0.92 \\
            3728.4096 & 0.0 & 0.35 \\
            3743.6455 & 1.0 & 0.44 \\
            \hline
        \end{tabular}
    \end{minipage}
\end{table}

\newpage
\clearpage

\end{document}